\documentclass[a4paper,fleqn,usenatbib]{mnras}

\usepackage{newtxtext,newtxmath}
\usepackage[T1]{fontenc}
\usepackage{ae,aecompl}

\usepackage{graphicx}	% Including figure files
\usepackage{amsmath}	% Advanced maths commands
\usepackage{bm} 

\input{pgfgraphicsmacros.tex}
\input{reviewmacros.tex}
\newcommand*\lbd{$(\partial \Omega_\mathrm{pr}/\partial{L_z})_{J_f}$}

\title[Positive Lynden-Bell derivative as a ticket to the bar trap?]{\textcolor{black}{Positive Lynden-Bell derivative as a ticket to the bar trap?}}

\author[Viktor D. Zozulia, Anton A. Smirnov, Natalia Ya. Sotnikova]{Viktor D. Zozulia$^{1,2}$, Anton A. Smirnov$^{2}$, Natalia Ya. Sotnikova$^{1,2}$\\
$^{1}$St. Petersburg State University,
Universitetskij pr.~28, 198504 St. Petersburg, Stary Peterhof, Russia\\
$^{2}$Central (Pulkovo) Astronomical Observatory of RAS, Pulkovskoye Chaussee 65/1, 196140 St. Petersburg, Russia\\
}
\date{Accepted XXX. Received YYY; in original form ZZZ}

\pubyear{2023}

\begin{document}
\label{firstpage}
%-----------------------------------------------------%
%-----------------------------------------------------%
\pagerange{\pageref{firstpage}--\pageref{lastpage}}
\maketitle

%-----------------------------------------------------%
%-----------------------------------------------------%
% Abstract of the paper
%-----------------------------------------------------%
%-----------------------------------------------------%
%%%%%%%%%%%%%%%%%%%%%%%%%%%%%%%%%%%%%%%%%%%%%%%%%%%%%%%
%%%%%%%%%%%%%%%%%%%%%%%%%%%%%%%%%%%%%%%%%%%%%%%%%%%%%%%
\begin{abstract}
\textcolor{black}{We have translated the results of $N$-body simulations of one barred model into the language of action variables and frequencies. Using this language, we analysed the behaviour of all orbits in the model on a large time scale at the stage of a mature bar. We show that the orbits join the bar while preserving their adiabatic invariant, \textcolor{black}{which takes into account the 3D structure of the orbits}. This allows us to apply the concept of the Lynden-Bell derivative for each of these orbits and trace how the sign of the derivative changes, i.e. how asynchronous changes in angular momentum $L_z$ and orbital precession rate $\Omega_\mathrm{pr}$ (normal orbital mode) change to synchronous (abnormal mode). The transition to the abnormal mode occurs when $\Omega_\mathrm{pr}$ reaches the angular velocity of the pattern $\Omega_\mathrm{p}$, after which the orbit becomes stuck in the bar trap. All this happens against the background of secular changes in actions ($L_z$ decreases, $J_\mathrm{R}$ and $J_z$ increase). At the same time, corotation particles near two stable Lagrange points are also subject to secular changes in their actions. They increase $L_z$ and drift to the periphery, shifting corotation outwards. We also show that a change in the orbital mode from normal to abnormal and the trapping of orbits in a bar is possible only when the bar speed decreases with time, regardless of what is causing the bar to slow down. Our findings clarify and expand the picture of bar formation and evolution in numerical models.}
\end{abstract}
%%%%%%%%%%%%%%%%%%%%%%%%%%%%%%%%%%%%%%%%%%%%%%%%%%%%%%%
%%%%%%%%%%%%%%%%%%%%%%%%%%%%%%%%%%%%%%%%%%%%%%%%%%%%%%%

% Select between one and six entries from the list of approved keywords.
% Don't make up new ones.
%%%%%%%%%%%%%%%%%%%%%%%%%%%%%%%%%%%%%%%%%%%%%%%%%%%%%%%
%%%%%%%%%%%%%%%%%%%%%%%%%%%%%%%%%%%%%%%%%%%%%%%%%%%%%%%
%%%%%%%%%%%%%%%%%%%%%%%%%%%%%%%%%%%%%%%%%%%%%%%%%%%%%%%
%%%%%%%%%%%%%%%%%%%%%%%%%%%%%%%%%%%%%%%%%%%%%%%%%%%%%%%

%%%%%%%%%%%%%%%%% BODY OF PAPER %%%%%%%%%%%%%%%%%%

%-----------------------------------------------------%
%-----------------------------------------------------%
\section{Introduction}
\label{sec:intro}
% Sec. 1.
%-----------------------------------------------------%
%-----------------------------------------------------%
The question of the formation of bars in galaxies is usually reduced to the analysis of a bar-forming mode instability. 
%This is reminiscent of Stanislav Lem's explanation of the term ``sepulka'', which was defined through the term ``sepulkaria'', which, in turn, referred to the word ``sepuling'', after which everything returned to ``sepulka''. 
But how exactly the bars form in real galaxies is an unresolved problem in galactic dynamics \citep{Sellwood2013a,Sellwood2014}. This is surprising in the era of large (all sky) and deep (high redshifts) surveys and a large suit of cosmological simulations that attempt to replicate the evolution of the observable Universe. However, the ``quantity'' further confuses the issue, since barred galaxies are not much different from their almost equally numerous unbarred counterparts (see, for example, \citealp{Courteau_etal2003,Sellwood2013a}).  In particular, the location of a galaxy in the Tully--Fisher plane is
independent of barredness. Moreover, barred and unbarred galaxies 
have, on average, comparable fractions of luminous and dark matter at a given radius. But even at the level of individual galaxies, the question turns out to be no less difficult. 
\par 
One can guess what factors can stop the bar formation (see below). However, even for one of the typical spiral galaxies in the local Universe, namely M~33, there is still no explanation for why this particular galaxy is stable to the bar formation~\citep{Sellwood_etal2019}. All reasonable $N$-body models of this galaxy, compatible with the mass distribution and those taking into account the morphological features of this galaxy, lead to the formation of a bar, which the real galaxy does not have~\citep{Sellwood_etal2019}. The inability to explain why some galaxies do not have bars actually means that we still do not understand all the details of the bar formation and do not see the whole picture of the galaxy secular evolution. \textcolor{black}{At the same time, numerical simulations of the last years gradually expanded the list of factors that can stop or slow down the formation of bars. For example, \cite{Collier_etal2018} found that a high halo spin $\lambda$ can prevent the bar formation. Nevertheless, a general principle of why a bar either forms or does not form in a particular simulation has yet to be understood. There is also an interesting discrepancy between the number of barred galaxies in the local Universe and cosmological up-to-date simulations~\citep{Reddish_etal2022}.} 
%\textcolor{black}{This is despite the fact that the list of factors that can stop or slow down the formation of bars is gradually expanding in theoretical works (see, for example, \citealp{Collier_etal2018}). And there is also an interesting discrepancy between the number of barred galaxies in the local Universe and in cosmological up-to-date simulations~\cite{Reddish_etal2022}.}
\par
If one goes down to the level of specific mechanisms, two mechanisms for the bar formation in isolated galaxies are usually discussed, namely: the swing amplification~\citep{Toomre1981} and the Lynden-Bell mechanism~\citep{Lynden-Bell1979}.
\par
The swing amplification mechanism connects the formation and growth of the bar with a special type of interaction between the wave packets and the corotation (CR) region. At the corotation, outgoing leading spiral waves can be super-reflected into amplified ingoing trailing waves. The ingoing trailing wave, in turn, can be reflected from the disc centre, transforming into a leading wave that propagates outward until it reaches the corotation. The cycle of wave travelling and repetitive transformations creates a positive feedback loop. This, in turn, leads to an exponential increase in the amplitude of the perturbation in time. 
\par  
The most valuable practical yields of the swing amplification theory are feedback-breaking factors. 
%First of all, the one, that should be discussed, is the inner Lindblad resonance (ILR). ILR has the ability to absorb waves that try to pass through it. Thus, the strong ILR halts the amplification cycle and prevents the bar formation~\citep{Sellwood1985,Sellwood_Moore1999,Sellwood_Evans2001}. 
\textcolor{black}{Specifically, \citet{Toomre1981} believed that a dense bulge-like mass component, that could induce inner Lindblad resonance (ILR), might be the main stabilising factor for the cold massive disc. However, up-to-date $N$-body simulations showed that the disc is stable against the formation of a bar only with a very dense and very massive central component \citep{Fujii_etal2018,Saha_Elmegreen2018,Kataria_Das2018}. As argued by \citet{Pol2016}, compact, but less massive bulges do not prevent bar formation due to the finite thickness of the disc. 
%\citet{Sellwood1989} showed that in early $N$-body simulations with low resolution and dense bulges, bars were formed due to high levels of shot noise. Shot noise created high-amplitude disturbances that easily passed through the ILR. However, $N$-body models with a large number of particles (up to $\sim 10^{7}-10^8$), and, hence, with a low level of shot noise, turn out to be stable against the formation of a bar only with a very dense and very massive central component \citep{Fujii_etal2018,Saha_Elmegreen2018,Kataria_Das2018}. 
%Compact, but less massive bulges do not prevent bar formation, even when a very large number of particles is employed. 
%\citet{Pol2016} explained this result by the absence of the ILR in discs of finite thickness even in the presence of a very cuspy bulge. 
In such discs, the precession profile when averaged over the height (``effective'' precession profile) turns into a curve with a rather low maximum.} Moreover, compact bulges, as well as the presence of gaseous central mass concentration, do not prevent the bars from appearing but lead to the formation of another subsystem, the so-called barlens \citep{Athanassoula_etal2015,Salo_Laurikainen2017,Smirnov_etal2021}. The destruction of the bar itself does not occur in this case. Observations show that galaxies with barlenses are a large class of objects \citep{Laurikainen_etal2011}. In their structure, they often exhibit compact classic bulges \citep{Laurikainen_etal2018}. But, despite the presence of a compact bulge, such galaxies also have a long elongated bar. This shows that the wave picture drawn by \citet{Toomre1981} is actually much more complex than originally thought.
\par
The mechanism that was suggested by~\citet{Lynden-Bell1979} is similar to the radial orbits instability (ROI) arising in three-dimensional systems
\textcolor{black}{~\citep{Polyachenko_Shukhman1972,Polyachenko_Shukhman1981,Antonov_1973}. \citet{Merritt_1987,Merritt_1999}, \citet{Saha_1991}, \citet{Weinberg_1991}, \citet{Palmer_1994} interpret the ROI precisely in terms of the Lynden-Bell's mechanism.}
The \textcolor{black}{Lynden-Bell's} mechanism qualitatively describes how the \textcolor{black}{planar eccentric} orbits line up in the direction of a rather slowly rotating bar-like perturbation and gradually become trapped that leads to an increase of the potential perturbation. The key idea is that the \textcolor{black}{eccentric} orbits in the central regions of the disc precess slowly and usually satisfy the following condition: $\Omega_\mathrm{pr}/\Omega \ll 1$, where $\Omega$ is an angular velocity and $\Omega_\mathrm{pr}$ is a precession rate. In this case, the orbital motion can be averaged over time, and one can consider not the individual stars moving along the corresponding orbits, but slowly precessing orbital ellipses. One can study how these ellipses interact with the gravitational field  (and the bar ``well'' or ``trap'') then. 
\textcolor{black}{Studying the normal modes of the stellar disc,~\cite{Pol2004a, Pol2013} argued that, while the orbits themselves precess slowly, the capturing of such orbits in the bar trap can occur relatively fast, i.e. on a dynamical timescale~\citep{Sellwood2014}}.
\par 
%If there is a weak oval distortion of the potential (bar) rotating with pattern speed $\Omega_\mathrm{p}$, then the low speed of precession of the orbits with respect to the bar $|\Omega_\mathrm{pr}-\Omega_\mathrm{p}|/\Omega \ll 1$ becomes important.
Assuming that $|\Omega_\mathrm{pr}-\Omega_\mathrm{p}|/\Omega \ll 1$ \textcolor{black}{(where $\Omega_\mathrm{p}$ is the pattern speed of the bar)}, the evolution of the orbit under the influence of the bar-like potential proceeds adiabatically, that is, with the conservation of a specific integral of motion \textcolor{black}{$J_f= J_\mathrm{R} + L_z / 2$,
%\int \mathbf{p \cdot dq}$, 
where $L_z$ is the vertical projection of the angular momentum, $J_\mathrm{R} = \int v_\mathrm{R}  dR$ is the radial action} (the integration is carried out over the trajectory of a star). 
%and $\mathbf{p}$ and $\mathbf{q}$ are vectors of the generalised momenta and coordinates, respectively. 
\citet{Lynden-Bell1979} explicitly showed that, depending on the sign of the partial derivative $(\partial \Omega_\mathrm{pr}/ \partial L_z)_{J_f}$ (hereinafter, LB-derivative) calculated at the fixed value of the invariant $J_f$, the \textcolor{black}{eccentric} orbit will be trapped by the bar potential (the positive sign of the LB-derivative) or moves out of it (negative sign). The region of the galaxy, where the mentioned derivative 
\textcolor{black}{has the negative sign, is called ``normal'' (usually the disc region), and the region, which has the negative sign of the derivative, is called ``abnormal''.}
%has the positive sign, is usually called an ``abnormal'' one since it is believed to be a rare occurrence in the central region of the galaxies. 
%\par 
%\textcolor{black}{Add intermediate phrase}
\par 
\textcolor{black}{The formalism of \cite{Lynden-Bell1979} was extended by \cite{Pol2020a,Pol2020b}. These authors tried to connect the ideas of Lynden-Bell and the modern view of the bar as a structure consisting of various orbital families~\citep{Contopoulos1980}, like orbits elongated along the bar major axis (the so-called $x1$ family) or elongated along the bar minor axis (the $x2$ family). For several models of the potential (including the Milky Way-like potential), \cite{Pol2020a,Pol2020b} studied stationary points of the Hamiltonian where the parenting orbits of the mentioned orbital families reside. Based on their findings, the authors concluded that, in addition to LB-derivative, two other parameters, the precession rate itself and ``bar responsiveness'' ($b$ in their notation), determine whether a particular orbital family can support the bar structure or not. Given that the orbital dynamics of the bar is found to be rich in different families in recent years~\citep{Patsis_Athanassoula2019,Smirnov_etal2021}, the studies by~\cite{Pol2020a,Pol2020b} highlight the importance of considering  the individual orbits (bar ``building blocks'') to understand the process of bar formation/evolution and not the regions or areas of the disc as was done by \cite{Lynden-Bell1979}.}
\par 
The orbital approach was partially adopted in~\cite{Petersen_etal2019}, where the authors specifically divided the orbital types on $x1$ orbits and ``other'' bar orbits. In particular, \cite{Petersen_etal2019} studied the evolution of angular momenta separately for the $x1$ orbits and other orbital types. Their findings indicate that the evolution indeed proceeds differently for different orbital types and, at the same time, bars in different models have different fractions of them (see also~\citealt{Smirnov_etal2021}). \textcolor{black}{The change and evolution of orbital morphology and characteristics are very important during the deeply nonlinear stages of a bar's life. In this case, it is important not only to know about the presence of different types of orbits, but also about the quantitative relationship between them and how they transform when interacting with the bar and falling into its trap. At the same time, the Lynden-Bell language (action variables, the LB-derivative), although it was initially applied to the early linear stages of bar formation, can also be useful for describing the later stages, if the focus is on the analysis of secular orbital transformation.}
\par
%Theoretical findings by~\citet{Pol2020a,Pol2020b}, which expanded the ideas of Lynden-Bell~(\citeyear{Lynden-Bell1979}) for up-today galactic models, can be tested on $N$-body models if one translates the results of simulations into the language of the action variables. 
In the present work, we want to extend the results of previous studies on the orbital evolution of the bar using up-to-date galactic models. Specifically, using the language of variables originally suggested by Lynden-Bell, namely $L_z$, $J_\mathrm{R}$, the precession rate $\Omega_\mathrm{pr}$, the invariant $J_f$, and partial derivatives $(\partial\Omega_\mathrm{pr}/\partial L_z)_{J_f}$ \textcolor{black}{expanded to account for the 3D structure of the orbits (see Sec.~\ref{sec:different_time_scales})}, we strive to describe the dynamical structure of a ``live'' $N$-body bar, which forms, grows, and changes its pattern speed, while gradually trapping orbits that experience a secular evolution of actions. This is the main goal of the present work. To do this, we developed a method for calculating action variables and frequencies directly from the orbits in the non-axisymmetric $N$-body potential of a numerical model at any time moment, \textcolor{black}{with an emphasis on the late stages of the model's evolution.}
%The main goal of the present work is to describe the dynamic structure of a ``live'' $N$-body bar that forms, grows, and changes its pattern speed, while gradually trapping orbits, in variables originally used by Lynden-Bell, 
%to formulate his idea, 
%namely the invariants $J_f$ and partial derivatives $(\partial\Omega_\mathrm{pr}/\partial L_z)_{J_f}$. 
\par
\textcolor{black}{Studying the structure of a ``live'' bar,} we want to verify whether the abnormal region, where the LB-derivative is positive, exists for a typical barred $N$-body model, and, if it exists, to what degree the region is connected with the bar. 
%The second 
\textcolor{black}{We also want to study the orbital trapping at the level of individual orbits, which is rarely done due to how complex this problem is. Specifically,  we want to check how the actions of individual orbits change when the orbits experience trapping and how the overall behaviour of the orbits changes in terms of actions and precession rates in this case.
%what role the sign of their Lynden-Bell's derivative plays in this case. 
In this way, we hope to obtain a more detailed picture of the orbital capturing by a bar, which presumably occurs or has occurred in our and other galaxies.}
%arguments to support 
%or refute 
%the Lynden-Bell theory.
\par
The paper is organised as follows.
In Section~\ref{sec:sim}, we present our $N$-body model and its global properties, including the evolution of the amplitude and pattern speed of the bar. We describe how we calculate actions and orbital frequencies of all particles from the $N$-body model in Section~\ref{sec:measuring_actions}. In Section~\ref{sec:maps}, we show how the system under consideration evolves in terms of $L_z, J_f,$ and $\Omega_\mathrm{pr}$. In Section~\ref{sec:orbits_trapping}, we identify the particles that are trapped by the bar during the particular time interval and check how their actions and frequencies evolve during this process. \textcolor{black}{In Section~\ref{sec:discussion}, we discuss the 
%distinctions between the Lynden-Bell's 
picture of the bar formation/growth 
%and the picture, which 
that stems from our analysis.} 
We summarise our conclusions in Section~\ref{sec:conclusions}.

\section{Simulations}
\label{sec:sim}
% Sec. 2.
%-----------------------------------------------------%
%-----------------------------------------------------%

%%%%%%%%%%%%%%%%%%%%%%%%%%%%%%%%%%%%%%%%%%%%%%%%%%%%%%%
% Fig. 1
%%%%%%%%%%%%%%%%%%%%%%%%%%%%%%%%%%%%%%%%%%%%%%%%%%%%%%%
\begin{figure*}
\centering
\begin{minipage}[t]{0.43\linewidth}
\includegraphics[width=\linewidth]{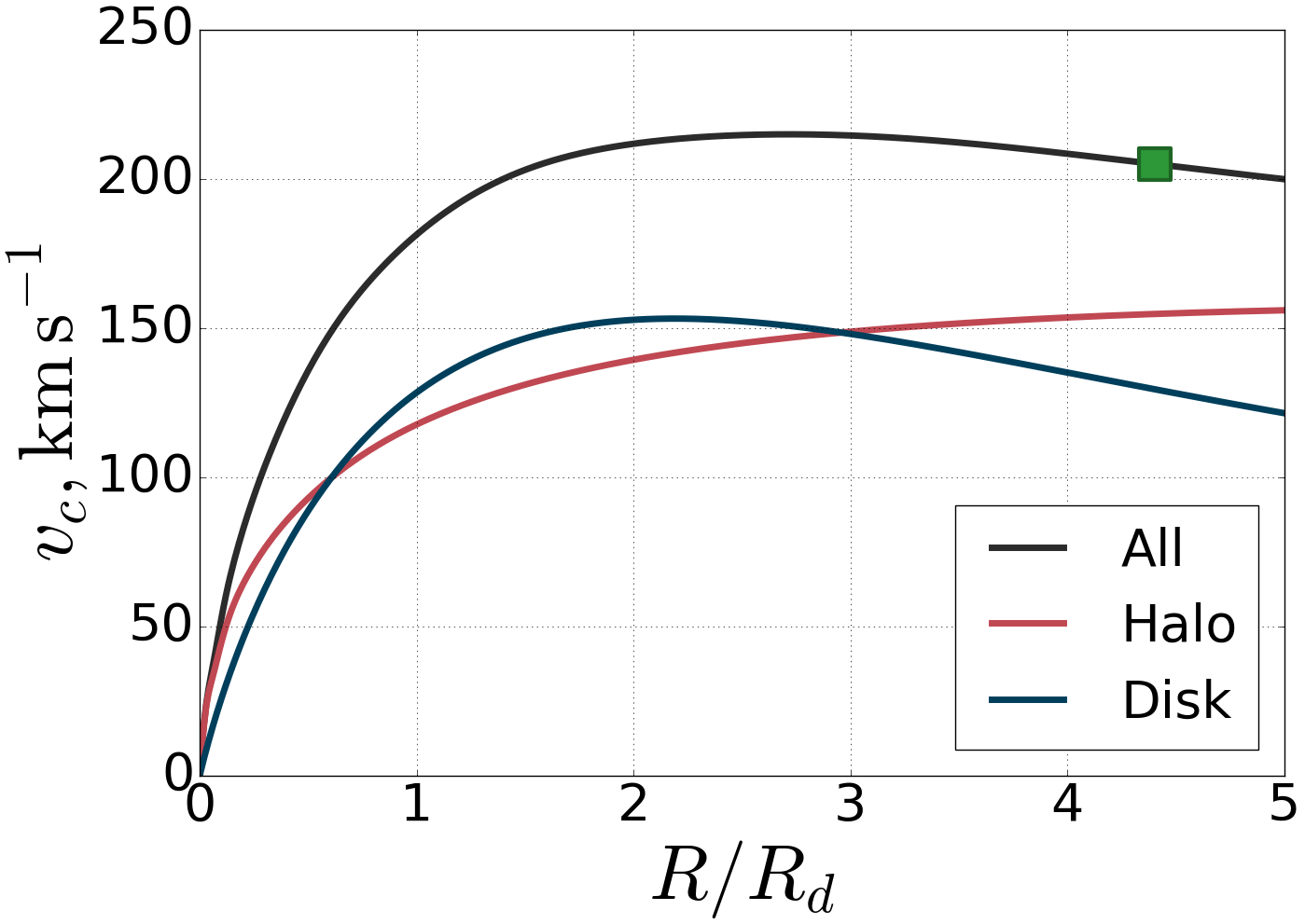}
\end{minipage}
\begin{minipage}[t]{0.485\linewidth}
\includegraphics[width=1\linewidth]{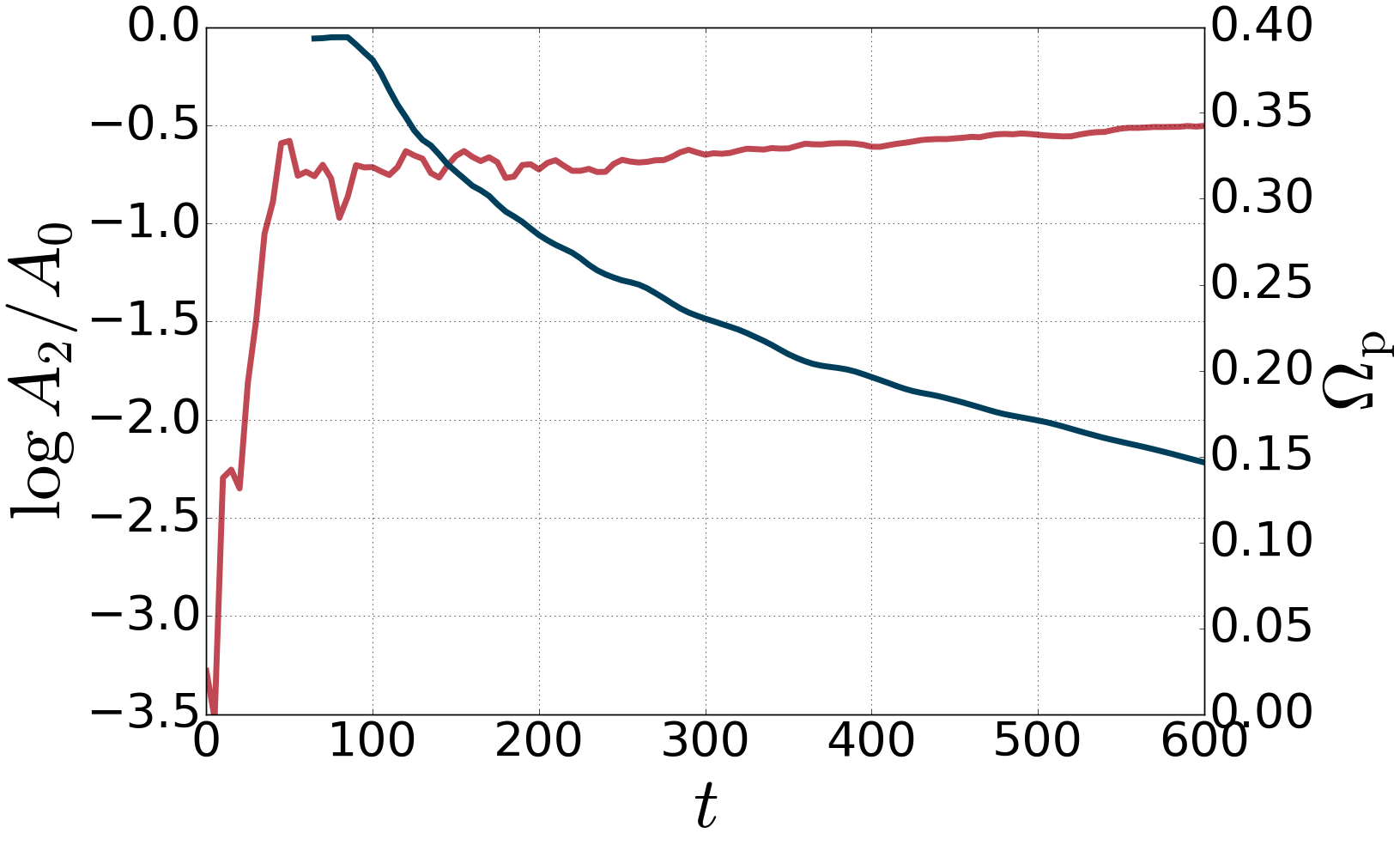}
\end{minipage}

\caption{
Properties of the $N$-body model. \textit{Left}: the rotation curve with disc and dark halo contributions for $t=0$. The green square marks the location of the zero value of \lbd for circular orbits (see Section~\ref{sec:maps}). \textit{Right}: the decimal logarithm of the normalized amplitude of the bar and the pattern speed of the bar for different time moments.}
\label{fig:Nbodymodel}
\end{figure*}
%%%%%%%%%%%%%%%%%%%%%%%%%%%%%%%%%%%%%%%%%%%%%%%%%%%%%%%
%%%%%%%%%%%%%%%%%%%%%%%%%%%%%%%%%%%%%%%%%%%%%%%%%%%%%%%

%%%%%%%%%%%%%%%%%%%%%%%%%%%%%%%%%%%%%%%%%%%%%%%%%%%%%%%
% Fig. 2
%%%%%%%%%%%%%%%%%%%%%%%%%%%%%%%%%%%%%%%%%%%%%%%%%%%%%%%
\begin{figure}
\centering
\includegraphics[width=0.95\linewidth]{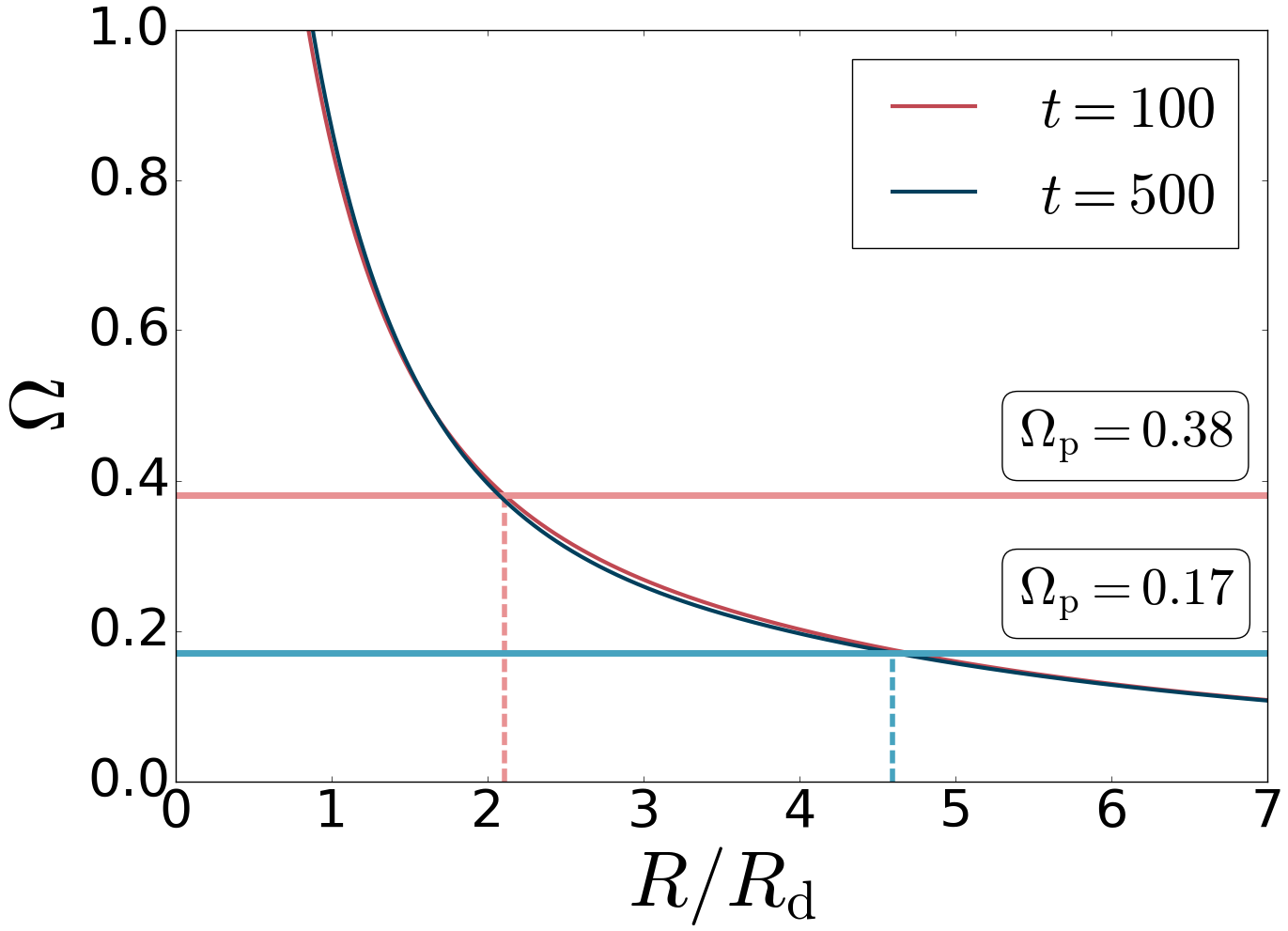}
\caption{Profiles of angular speed for time moments $t=100$ and $t=500$. Horizontal lines define pattern speeds for these time moments. Vertical lines indicate the location of the CR.}
\label{fig:Nbodymodel_bar}
\end{figure}
%%%%%%%%%%%%%%%%%%%%%%%%%%%%%%%%%%%%%%%%%%%%%%%%%%%%%%%
%%%%%%%%%%%%%%%%%%%%%%%%%%%%%%%%%%%%%%%%%%%%%%%%%%%%%%%

Our numerical model is taken from~\citet{Smirnov_Sotnikova2018}. At the start of the simulations, the model consists of the axisymmetric exponential disc (in the radial direction), which is isothermal along the vertical axis:
%%%%%%%%%%%%%%%%%%%%%%%%%%%%%%%%%%%%%%%%%%%%%%%%%%%%%%%
\begin{equation}
%%%%%%%%%%%%%%%%%%%%%%%%%%%%%%%%%%%%%%%%%%%%%%%%%%%%%%%
\rho(r,z) = \frac{M_\mathrm{d}}{4\pi R_\mathrm{d}^2 z_\mathrm{d}} \exp(-R/R_\mathrm{d}) \cdot \mathrm{sech}^2(z/z_\mathrm{d}) \,,
\label{eq:sigma_disk} 
\end{equation}
%%%%%%%%%%%%%%%%%%%%%%%%%%%%%%%%%%%%%%%%%%%%%%%%%%%%%%%
where $M_\mathrm{d}$ --- mass of the disc, $R_\mathrm{d}$  --- its radial scale, $z_\mathrm{d}$ --- the vertical scale height. We assume that the disc scale length $R_\mathrm{d}=3.5$~kpc, the scale height $z_\mathrm{d}=0.05R_\mathrm{d}$, and the disc mass $M_\mathrm{d}=5\times10^{10} M_{\sun}$. The dimensionless system of units is: $M_\mathrm{u}=M_\mathrm{d}=1$, $R_\mathrm{u}=R_\mathrm{d}=1$, $G=1$. Then the time unit $t_\mathrm{u}$ will correspond to 13.8~Myr, \textcolor{black}{the velocity unit will be about 252 km s$^{-1}$, and the angular velocity \mbox{unit is 71 km} s$^{-1}$ kpc$^{-1}$.}
\par 
The disc is embedded into a spherical dark halo of the Navarro-Frank-White-like profile~\citep{NFW} with a slightly steeper inner slope of the density profile (see~\citealt{Smirnov_Sotnikova2018} for details). We do not consider here the model with the gaseous component or the initial classical bulge to simplify the present analysis and avoid possible complex interactions between the subsystems. The physical parameters of the halo are related to the initial parameters of the disc in the following way: $M_\mathrm{h} (r<4R_\mathrm{d})/M_\mathrm{d}=1.5$, where $M_\mathrm{h}(r<4R_\mathrm{d})$ is the halo mass within the sphere of the radius of $4R_\mathrm{d}$; $r_h=6R_\mathrm{d}$, where $r_h$ is the halo scale length. An idea of the mass model used in this study is given by its rotation curve (Fig.~\ref{fig:Nbodymodel}, left plot).
\par
The kinematics of the components is typical for $N$-body calculations of this kind. The disc has an exponential profile of the radial velocity dispersion:
%%%%%%%%%%%%%%%%%%%%%%%%%%%%%%%%%%%%%%%%%%%%%%%%%%%%%%%
\begin{equation}
%%%%%%%%%%%%%%%%%%%%%%%%%%%%%%%%%%%%%%%%%%%%%%%%%%%%%%%
\sigma_\mathrm{R} = \sigma_0 \cdot \exp(-R/2 R_\mathrm{d}) \,.
\end{equation}
%%%%%%%%%%%%%%%%%%%%%%%%%%%%%%%%%%%%%%%%%%%%%%%%%%%%%%%

In practice, we set the Toomre parameter value $Q_0$ at $R=R_\mathrm{d}$ to $1.2$ and from this obtain the normalization value of $\sigma_0$. The halo has an isotropic velocity dispersion profile of Cuddeford type~\citep{Cuddeford_1991}.
\par
Numerical procedures like populating the model with particles and integration of the equation of motions were carried out via \texttt{mkgalaxy}~\citep{McMillan_Dehnen2007} and \texttt{gyrfalcON}~\citep{Dehnen2002} program utilities, respectively, incorporated in the toolbox for $N$-body simulations~\texttt{NEMO}~\citep{Teuben_1995}. We used 4$kk$ particles for the disc, 4.5$kk$ for the halo and followed the evolution of the system up to $\approx 8$~Gyr ($t\approx600$) with the adaptive time step. The smallest value of the time step is about $\approx 0.1$~Myr.
\par
After the start of the simulation, the disc quickly evolves ($\approx 1$~Gyr) to a non-axisymmetric state forming the bar. The bar starts small and fast-rotating but rapidly grows in size and gradually slows down (Fig.~\ref{fig:Nbodymodel}, right plot). 
\par 
Fig.~\ref{fig:Nbodymodel_bar} shows the profile of the angular speed of the model for two specific time moments, $t=100$ and $t=500$. Both curves are calculated for the spherically symmetric part of the potential and differ a little. The horizontal lines define the bar pattern speed. The figure also indicates where the corotation resonance is located for the bar in our model, $R_\mathrm{CR}\approx2$ and $R_\mathrm{CR}\approx4.5$ for  $t=100$ and $t=500$, respectively. We estimate the bar size as about $a_\mathrm{b}\approx1.5$ and $a_\mathrm{b}\approx3.0$ for these time moments using the location of the drop in the radial dependence of the bar amplitude $A_2$, which gives $\mathcal{R}=R_\mathrm{CR}/a_\mathrm{b}\approx1.3$ and $\mathcal{R}\approx1.5$, respectively. This means that the model bar marginally fits the definition of a fast bar~\citep{Debattista2000, Rautiainen2008, Aguerri2015,Geron2023}, although it becomes slower at later time moments.

%-----------------------------------------------------%
%-----------------------------------------------------%
\section{Measuring the actions}
\label{sec:measuring_actions}
% Sec. 3.
%-----------------------------------------------------%
%-----------------------------------------------------%

%%%%%%%%%%%%%%%%%%%%%%%%%%%%%%%%%%%%%%%%%%%%%%%%%%%%%%%
% Fig. 3
%%%%%%%%%%%%%%%%%%%%%%%%%%%%%%%%%%%%%%%%%%%%%%%%%%%%%%%
\begin{figure}
\centering
\includegraphics[width=\linewidth]{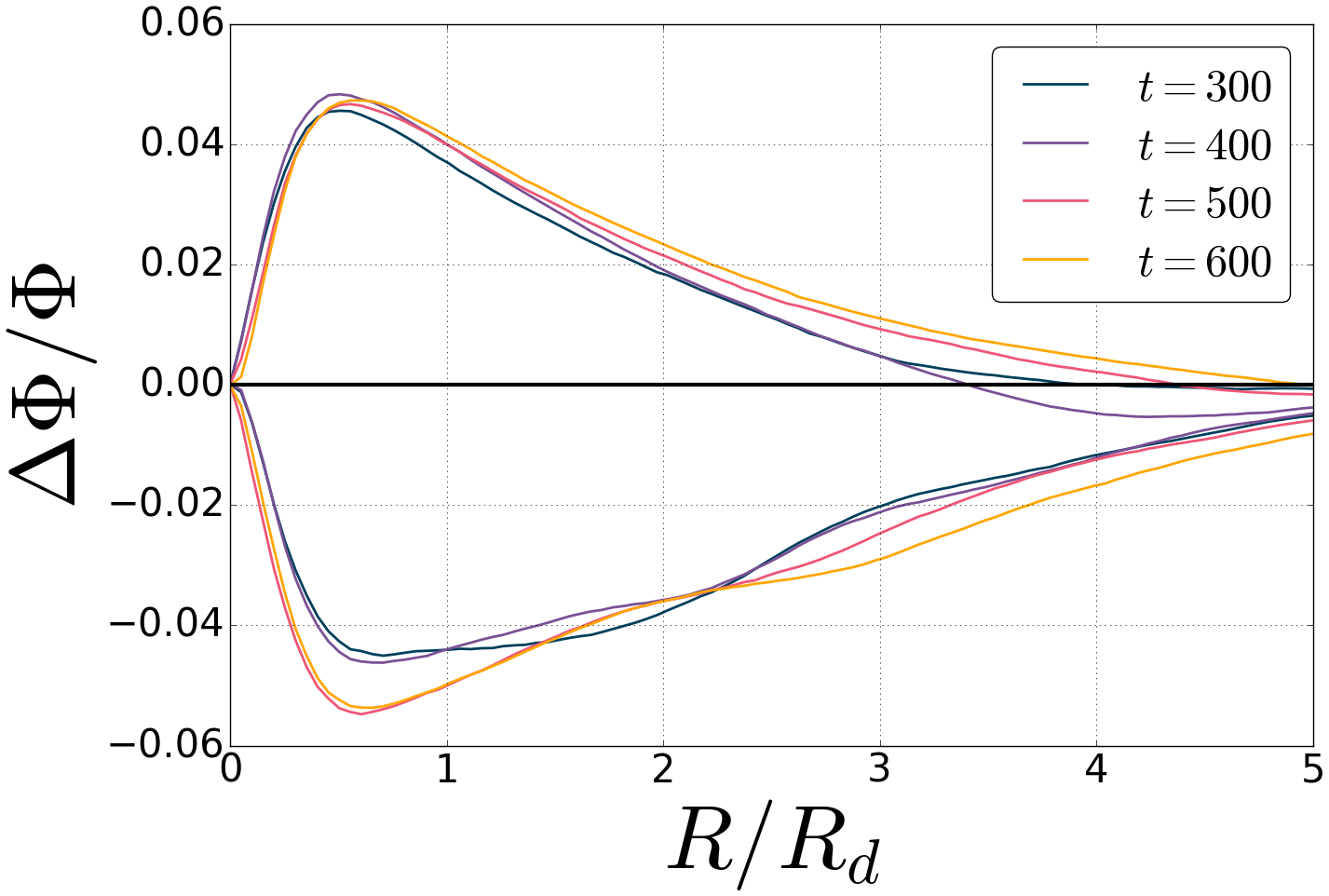}
\caption{The relative difference between the actual potential of the $N$-body model used and its approximation, Eq.\eqref{eq:pot}, along the major (upper curves) and minor (lower curves) axes of the bar. The radial distance is given in dimensionless units of simulations.}
\label{fig:pot_diff}
\end{figure}
%%%%%%%%%%%%%%%%%%%%%%%%%%%%%%%%%%%%%%%%%%%%%%%%%%%%%%%
%%%%%%%%%%%%%%%%%%%%%%%%%%%%%%%%%%%%%%%%%%%%%%%%%%%%%%%

%-----------------------------------------------------%
%-----------------------------------------------------%

% Sec. 3.1.
%-----------------------------------------------------%
%-----------------------------------------------------%
Action variables are a good tool for studying the dynamics of complex systems such as galaxies with bar and spirals. Recently, this approach has been widely used to study the resonance interactions of stars and non-axisymmetric features in the Milky Way, e.g., \cite{Trick_2021, Kawata2021, Trick_2022, Drimmel2023}. 
An extensive growth of this kind of studies is partially explained by the development of numerical methods for calculating the actions~\citep{Binney2012, Sanders_Binney2015}. The usage of these methods also made it possible to study the action space of the $N$-body models, where up to several millions particles are present~\citep{Debattista_etal2020,Mikkola2020}. Our work is in league with the mentioned works and further tests what fruitful results can be obtained if one analyses the whole action space of the system.

%\subsection{Actions in the non-axisymmetric potential}
\subsection{Different time scales in action measurement}
\label{sec:different_time_scales}
There are several pitfalls in the measurement of actions in a non-axisymmetric potential that we should mention.
\par
\begin{figure}
\begin{minipage}[h]{0.37\linewidth}
\includegraphics[width=\linewidth]{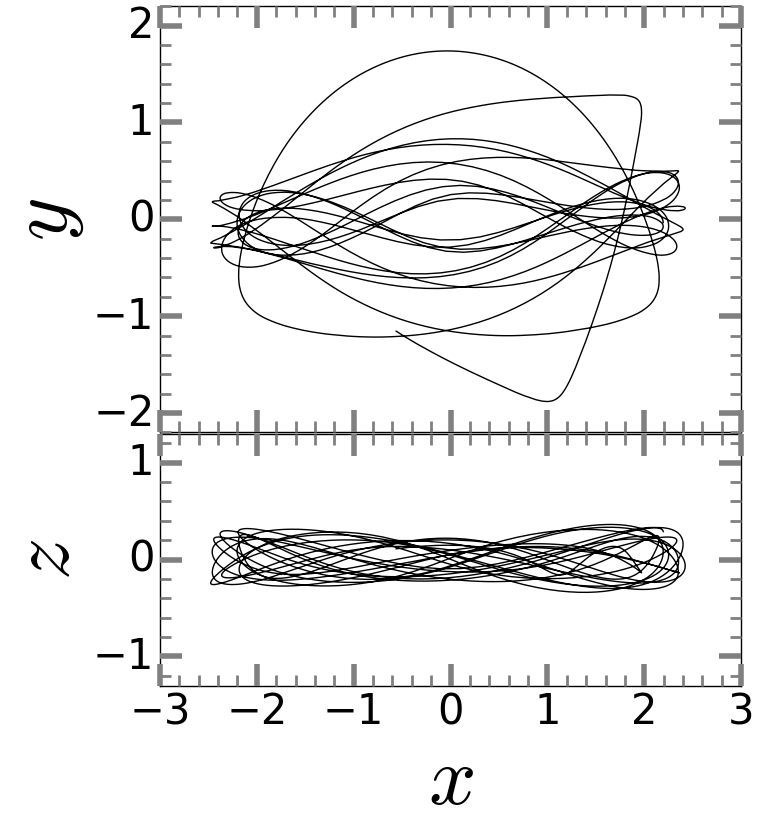}
\end{minipage}
\begin{minipage}[h]{0.62\linewidth}
\includegraphics[width=\linewidth]{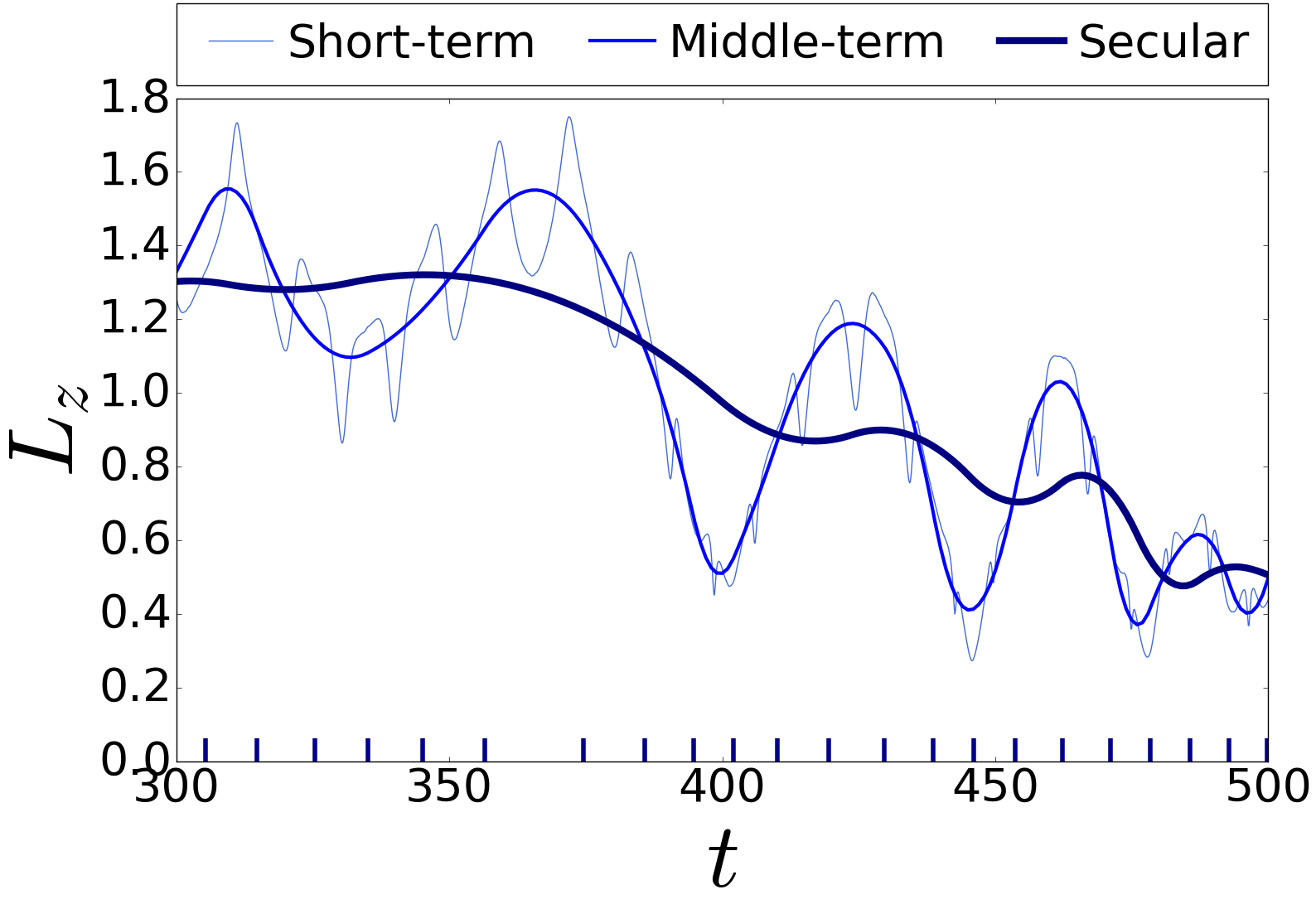}
\end{minipage}
\caption{\textcolor{black}{\textit{Left}: An orbit on $xy$- and $xz$-planes over the time interval from $t=250$ to $t=550$ ($x$-coordinate goes along the major axis of the bar). \textit{Right}: The evolution of $L_z$ on different time scales (short-term, middle-term and secular). The short marks at the bottom show the moments of passage of the apocentres of the orbit.}}
\label{fig:Lz_evolution}
\end{figure}

The first problem is that in the case of non-axisymmetric potential, the momentum and the radial action $J_\mathrm{R}$ cease to be ``true'' integrals of motion for orbits locked in a bar, and the actions oscillate around some specific values~\citep{Binney_Tremaine2008, Binney_2018, Trick_2021}. \textcolor{black}{In the non-axisymmetric rotating potential, the orbits librate or circulate around the orbits that satisfy the exact resonance condition: $n(\Omega-\Omega_\mathrm{p})=m\kappa$~\citep{Sellwood_2010,Binney_2018, Binney_2020}. If the orbit is being trapped by the bar, these librations occur simultaneously with the secular changes in the actions. For example, the momentum of the particle can periodically change and, at the same time, show a clear tendency towards a decreasing average value (see Fig.~\ref{fig:Lz_evolution}). This trend is usually observed when the almost circular orbit becomes an elongated one. The existence of middle-term librations significantly complicates the analysis of orbital trapping, since they are always present and one should distinguish between them and secular evolution of actions. Overall, we distinguish three different types of oscillations in the course of the present article:
\begin{enumerate}
    \item Short-term oscillations, i.e. fluctuations. They appear as bumps and troughs in Fig.~\ref{fig:Lz_evolution} and have a time scale of about one or less radial periods. We smooth out oscillations of this type and do not consider them further in the present analysis.
    \item Middle-term oscillations, i.e. librations, that occur on a longer time scale than fluctuations. As was already mentioned, they appear when the orbit under study is not in the exact resonance but nevertheless experiences the bar gravitational influence.
    \item Long-term changes (i.~e. the decrease or increase of the actions) that happen as a consequence of the orbit secular evolution.
\end{enumerate}}
%we address this issue by averaging these middle-term{\footnote{\textcolor{black}{Comparing to the capturing time, see Fig.~\ref{fig:}.}}} periodic changes over time for each particular orbit~\textcolor{black}{
\textcolor{black}{To study the secular evolution of the orbits in the present study, we average actions over the corresponding libration periods. ~\cite{Trick_2021, Trick_2022} also considered librations of the orbits around the average resonance line (ARL in their notation), see figure 3 in~\cite{Trick_2021}.}% In our case, we chose the averaging time interval large enough to include several periods of librations but at the same time short enough to study the secular evolution of the orbit (see subsection~\ref{sec:mean_actions_librations}).

\par 
\textcolor{black}{The second problem concerns the chaotic orbits and their impact on the analysis of the ensemble of orbits}. Strictly speaking, the actions can be defined only for regular orbits, chaotic orbits cannot be characterized this way~\citep{Binney_Tremaine2008}. \textcolor{black}{Here we do not carry out a throughout analysis of the chaoticity of the orbits in our model, \textcolor{black}{since it would require a separate study.} However, various studies of bar orbits in $N$-body models (e.g., \citealp{Wang_etal2016,Machado_Manos2016}) showed that most of the orbits supporting the bar should be regular, and their fraction should increase with increasing the bar strength. \citet{Voglis_etal2007}\footnote{While there are many chaotic orbits in the model, considered by~\cite{Voglis_etal2007}, the bar itself has a small fraction of chaotic orbits, see their figure 25, 2:1 resonance graph.} also showed that the fraction of chaotic orbits is rather small in the case of a strong bar and at the late stages of its evolution.
% Hence, this problem \textcolor{black}{should not be} severe for studies of a strong bar such as the present one.
%\textcolor{black}{First of all, the action variables can be defined only for regular orbits, chaotic orbits cannot be characterized this way~\citep{Binney_Tremaine2008}. However, at least we will use the action variables that are defined for the axisymmetric potential. They usually oscillate around the mean value for regular orbits in the non-axisymmetric potential.}
%\textcolor{black}{Here we do not carry out a throughout analysis of the chaoticity of the orbits in our model, \textcolor{black}{since it would require a separate study. Nevertheless}, based on the results of previous works~\citep{Voglis_etal2007,Valluri_etal2016,Machado_Manos2016}~, where it was shown that the fraction of chaotic orbits is rather small in the case of a strong bar and at the late stages of its evolution. 
Moreover, in \citet{Parul_etal2020}, we touched upon the issue of chaotic orbits in the $N$-body model, which is also used in the present work. We proceeded from the fact that the regular orbits are quasi-periodic, thus the Fourier spectra should consist of discrete lines. Many orbits in the bar of our model are like that. A lot of the orbits have only two frequencies of comparable amplitude in the spectrum. 
%The amplitudes of frequencies of a higher order, which could introduce chaos in the motion of a particle (Wang et al., 2016), are negligible.
Another important feature of the orbits that assembled into a bar in the model under consideration is their alignment along bright stripes and inside bright spots on the frequency maps (see figure~8, upper plot in \citealp{Smirnov_etal2021}). This suggests that the frequencies of most orbits are expressed as integer linear combinations of $N$ fundamental frequencies, which is true for regular orbits \citep{Wang_etal2016}. In \citet{Parul_etal2020}, we also estimated the frequency shifts, which were small for most bar orbits. According to \citet{Wang_etal2016,Contopoulos2002}, this also indicates the dominance of regular orbits in the model. We further support our statements regarding \textcolor{black}{the dominance of the regular orbits in Appendix~\ref{sec:appendix_LE}, where we study Lyapunov exponents for bar orbits.}} 
%Here we do not carry out an analysis of the chaoticity of the orbits. 
%Nevertheless, based on the results of previous works~\citep{Machado_Manos2016}, we assume that the fraction of chaotic orbits is small in the case of a strong bar and at the late stages of its evolution.
\par 
% The action space of $N$-body and theoretical models with a fixed potential (with the bar component included) have already been explored in~\citet{Debattista_etal2020, Trick_2021}. However, the action space in these works was studied only in terms of the usual actions: $J_\mathrm{R}$ (radial action), $L_z$ ($z$-projection of angular momentum), and $J_z$ (vertical action), but not in terms of \textcolor{black}{the adiabatic invariant $J_f=J_\mathrm{R}+L_z/2$, the complete adiabatic invariant $J_v=J_\mathrm{R}+L_z/2+J_z$} and the precession rates $\Omega_\mathrm{pr}=\Omega-\kappa/2$, where $\kappa$ is the frequency of radial oscillations. In addition, the potentials used in the mentioned works correspond to the initial axisymmetric state and do not take into account the mass redistribution induced by the bar.

The action space of $N$-body and theoretical models with a fixed potential (with the bar component included) have already been explored in~\citet{Debattista_etal2020, Trick_2021}. However, the action space in these works was studied only in terms of the usual actions: $J_\mathrm{R}$ (radial action), $L_z$ ($z$-projection of angular momentum), and $J_z$ (vertical action) and, moreover, the potentials correspond to the initial axisymmetric state and do not take into account the mass redistribution induced by the bar.
\textcolor{black}{In this work, we investigate the evolution of actions and frequencies in the changing $N$-body potential. In addition, we work in terms of the complete adiabatic invariant $J_v = J_\mathrm{R} + J_z + L_z/2$, which should be preserved in the three-dimensional bar of our model (see the Appendix \ref{sec:appendix_invariant}), the precession rates $\Omega_\mathrm{pr}=\Omega-\kappa/2$ and LB-derivative \lbd, which we calculate as the total derivative of the precession frequency with respect to the angular momentum $d \Omega_\mathrm{pr} / d L_z$ (a change of this derivative already occurs while preserving the invariant $J_v$).}

%%%%%%%%%%%%%%%%%%%%%%%%%%%%%%%%%%%%%%%%%%%%%%%%%%%%%%%
% Fig. 4
%%%%%%%%%%%%%%%%%%%%%%%%%%%%%%%%%%%%%%%%%%%%%%%%%%%%%%%
%%%%%%%%%%%%%%%%%%%%%%%%%%%%%%%%%%%%%%%%%%%%%%%%%%%%%%%
%%%%%%%%%%%%%%%%%%%%%%%%%%%%%%%%%%%%%%%%%%%%%%%%%%%%%%%

%-----------------------------------------------------%
%-----------------------------------------------------%
\subsection{Instantaneous actions}
\label{sec:axisym_potentia}
% Sec. 3.2.
%-----------------------------------------------------%
%-----------------------------------------------------%
In the present work, we study the bar structure in terms of \textcolor{black}{$J_v$} and $\Omega_\mathrm{pr}$ besides the usual actions, that is, in terms \textcolor{black}{close to} suggested by~\citet{Lynden-Bell1979}. We also verify how orbits evolve and are trapped in the bar in the overall changing potential of the underlying system. To this aim, we calculate all three action variables for each orbit of the system: the radial action $J_\mathrm{R}$, the azimuthal action or the angular momentum projection onto the $z$-axis $L_z$, the vertical action $J_z$, and the triplet of orbital frequencies $\mathbf{\Omega} = \partial H/ \partial \mathbf{J}$: frequency of radial oscillations $\kappa$, frequency of azimuthal oscillations $\Omega$, and frequency of vertical oscillations $\omega_z$. \textcolor{black}{We use several methods for calculating these quantities. One of them is implemented in the {\tt{AGAMA}} software package \citep{agama} and allows one to find approximate values of actions and frequencies directly from coordinates and velocities at any particular moment in time}.
%The calculations were carried out using the {\tt{AGAMA}} software package \citep{agama}. 
The utilities provided by the package are based on the so-called St\"ackel fudge method \citep{Binney2012,Sanders_Binney2016}, which evaluates action values  analytically if the considered potential is close enough to the St\"ackel potential (locally), which is defined as follows:
%%%%%%%%%%%%%%%%%%%%%%%%%%%%%%%%%%%%%%%%%%%%%%%%%%%%%%% 
\begin{equation}
%%%%%%%%%%%%%%%%%%%%%%%%%%%%%%%%%%%%%%%%%%%%%%%%%%%%%%%
\Phi(u, v) = \dfrac{U(u) - V(v)}{\sinh^2{u} + \sin^2{v}} \,, 
\end{equation}
%%%%%%%%%%%%%%%%%%%%%%%%%%%%%%%%%%%%%%%%%%%%%%%%%%%%%%%
where $u$ and $v$ are hyperbolic coordinates related to Cartesian coordinates as follows:
%%%%%%%%%%%%%%%%%%%%%%%%%%%%%%%%%%%%%%%%%%%%%%%%%%%%%%%
\begin{equation}
%%%%%%%%%%%%%%%%%%%%%%%%%%%%%%%%%%%%%%%%%%%%%%%%%%%%%%%
R = \Delta \sinh{u}\sin{v}\;\;;\;\;z = \Delta \cosh{u}\cos{v} \, .
\end{equation}
%%%%%%%%%%%%%%%%%%%%%%%%%%%%%%%%%%%%%%%%%%%%%%%%%%%%%%%
% has the form  
%In this case, the Hamilton - Jacobi equation gives:
%\begin{equation*}
% \dfrac{p_u^2}{2\Delta^2} = E \sinh^2{u} - I_3 - U(u) - \dfrac{L_z^2}{2\Delta^2\sinh^2{u}}
%\end{equation*}
%\begin{equation}
% \dfrac{p_v^2}{2\Delta^2} = E \sin^2{v} + I_3 - U(v) - \dfrac{L_z^2}{2\Delta^2\sin^2{v}}
%\end{equation}
%where $E$ is the orbit’s energy and $I_3$ is a constant of separation, $p_u$ and $p_v$ are canonical momenta. Moreover, they depend only on their conjugate coordinates, which allows us to introduce action variables:
%%%%%%%%%%%%%%%%%%%%%%%%%%%%%%%%%%%%%%%%%%%%%%%%%%%%%%%
%%%%%%%%%%%%%%%%%%%%%%%%%%%%%%%%%%%%%%%%%%%%%%%%%%%%%%%
%\begin{equation}
%J_\mathrm{R} = \dfrac{1}{\pi}\int_{u_{min}}^{u_{max}} p_u(u) du \;\; ;\;
%J_z = \dfrac{1}{\pi} \int_{v_{min}}^{v_{max}}  p_v(v) dv
%\end{equation}
%%%%%%%%%%%%%%%%%%%%%%%%%%%%%%%%%%%%%%%%%%%%%%%%%%%%%%%
%%%%%%%%%%%%%%%%%%%%%%%%%%%%%%%%%%%%%%%%%%%%%%%%%%%%%%%

\textcolor{black}{ The Stäckel fudge method from} {\tt{AGAMA}}  utilities can only be applied to an axisymmetric type of potential. The potential of the $N$-body model with the bar is not axisymmetric, of course. We obtain its axisymmetric approximation by expanding the potential into a series of multipoles and selecting only specific harmonics, namely those of 0-th order in azimuth and up to \textcolor{black}{12}-th order in $\cos{\theta}$:
%%%%%%%%%%%%%%%%%%%%%%%%%%%%%%%%%%%%%%%%%%%%%%%%%%%%%%%
\begin{equation}
%%%%%%%%%%%%%%%%%%%%%%%%%%%%%%%%%%%%%%%%%%%%%%%%%%%%%%%
\Phi_{0} = \Sigma_{0 \leq l \leq 12}\Phi_{l,0} \,,
\label{eq:pot}
\end{equation}
%%%%%%%%%%%%%%%%%%%%%%%%%%%%%%%%%%%%%%%%%%%%%%%%%%%%%%%
where $\Phi_{l,0}$ are taken from the series expansion:
%%%%%%%%%%%%%%%%%%%%%%%%%%%%%%%%%%%%%%%%%%%%%%%%%%%%%%%
\begin{equation}
%%%%%%%%%%%%%%%%%%%%%%%%%%%%%%%%%%%%%%%%%%%%%%%%%%%%%%%
\Phi(r, \theta, \varphi)=\Sigma_{l,m} \Phi_{l,m}(r)Y_l^m(\theta, \varphi) \,,
\end{equation}
%%%%%%%%%%%%%%%%%%%%%%%%%%%%%%%%%%%%%%%%%%%%%%%%%%%%%%%
where $Y_{l}^m$ are spherical functions of degree $l$ and order $m$. \textcolor{black}{The St\"ackel fudge method can be applied to this axisymmetric potential $\Phi_0$ to obtain the approximate values of actions and frequencies. 
To make sure that the axisymmetric approximation of the potential of our model differs slightly from the non-axisymmetric one, we compared both potentials at several time moments in Fig.~\ref{fig:pot_diff}. As can be seen, there is a good agreement between the approximation of the potential and the actual potential. The maximum value of the relative error ($\sim 5\%$) is achieved in the inner parts of the bar (the bar size is about three length units in this model). } 
%\textcolor{black}{To make sure that the replacement of a non-axisymmetric potential by its axisymmetric approximation does not introduce a significant error, we compared both potentials at several time moments.}

%and the order of the azimuthal Fourier expansion is always 0 for axisymmetric).

%Then we need to measure action-angle variables and orbital frequencies to this end using  In this method, it is necessary to fit the potential of the region where the orbit is located by the St\"ackel potential, which in coordinates $(u,v)$
 
%Also, we can obtain the angles and frequencies using the canonical transformation (see more details in Binney 2012).
\par
%{Clearly, approximating initially non-axisymmetric potential via axisymmetric component introduces some hard-to-estimate errors.}
%{However,}

%\textcolor{black}{Thus, we were convinced that the replacement does not introduce a significant error. In Sect.~\ref{sec:mean_actions_oscillations} below, we also directly compare the results of actions' calculations for the $N$-body potential and its axisymmetric approximation we obtain. As an additional bonus,}
\textcolor{black}{One can assume that the actions obtained by the St\"ackel fudge method for the axisymmetric approximate potential will differ slightly from their real values given that the axisymmetric and non-axisymmetric (original) potentials differ only slightly.
%However, we still carry out a more thorough analysis regarding this matter, i.~e. the comparison of different methods of calculation actions in paragraph \ref{sec:mean_actions_oscillations}.} 
A comparison of different methods for the same bar orbit is presented in Section~\ref{sec:mean_actions_oscillations}.}

As an additional bonus, the St\"ackel fudge method \textcolor{black}{operating with an axisymmetric potential} is much faster than \textcolor{black}{others methods used in this article and allows one to obtain estimates of actions and frequencies, if coordinates and velocities are known only at one point in time. We also want to point out that the axisymmetric approximation refers to the evolving potential, and not to the initial one, as, for example, in \citet{Debattista_etal2020}.} 
% directly integrating the orbit and deriving its actions numerically. 
% The use of the method significantly reduces the computation time. %\textcolor{black}{This is important in our case, when one calculates actions and their evolution for all orbits in the model (4$kk$), and not for selected particles, as is usually done. This is what makes our work different from others.}

%{, especially in the case one analyses the actions and their evolution for a large sample of orbits (4$kk$), as we do in the present work.}
%{We consider the approximation of the potential by its axisymmetric part as a necessary trade-off between accuracy and speed of calculations.}
\par 
%This matter was discussed in~\citet{Binney_2018} and~\citet{Trick_2021}. Since, by definition, the actions are integral of the motions Below we further improve the method of angle-actions calcualte% The system may not have three independent integrals of motion at all. However, we can still use them as an approximation for axisymmetric potential. And even so, they can tell a great deal about the dynamics of the system.
We should note the Stäckel fudge method itself is not ideal per se. In addition to errors arising from the approximation of the non-axisymmetric potential by its axisymmetric part discussed above, there would be errors in the estimates of actions even if the real underlying potential were hypothetically axisymmetric. This is especially true for orbits with significant eccentricity, which populate the bar. \textcolor{black}{However, as we will show in Section~\ref{sec:maps}, this method still allows one to quickly separate bar particles from the disc and obtain a rough view of the bar structure in terms of action variables.}

%-----------------------------------------------------%
%-----------------------------------------------------%
%\subsection{Actions averaged over time}
\subsection{Averaging actions}
\label{sec:mean_actions}
% Sec. 3.3.
%-----------------------------------------------------%
%-----------------------------------------------------%
\textcolor{black}{%Previously, we have examined that in a non-axisymmetric potential, actions could change with various timescales. Short-term timescale is associated with the motion of a particle in its orbit and has a characteristic scale equal to the period of radial oscillation for $L_z$ and $J_\mathrm{R}$ or vertical oscillation for $J_z$. Medium-term timescale is connected by a longer periodic process, namely a libration. And long-term timescale is associated with the secular evolution of the orbit. Therefore, in order to gain a comprehensive understanding of the evolution of orbital actions and frequencies, it is imperative to examine them at various time scales. 
In this subsection, we present a routine that we use to smooth out short-term fluctuations first (Sec.~\ref{sec:mean_actions_oscillations})  and then medium-term oscillations or libration (Sec. \ref{sec:mean_actions_librations}). %In this section, we demonstrate effective techniques for eliminating short-term fluctuations and medium-term libration . 
In the last part of this Section~(Sec.~\ref{sec:LB_derivative}), we show how one can estimate the value of the LB-derivative on the libration half-period scale.} 
\par
%-----------------------------------------------------%
%-----------------------------------------------------%
\subsubsection{Actions and frequencies averaged over short-time scale}
\label{sec:mean_actions_oscillations}
% Sec. 3.3.1
%-----------------------------------------------------%
%-----------------------------------------------------%
%In general, the actions are expected to oscillate for orbits that are heavily affected by the bar~(see, for example, a recent work by~\citealt{Binney_2020}). However, actions, calculated by {\tt{AGAMA}}, show sudden ups and downs on a very small timescale of several time units. 
\textcolor{black}{To begin with, we focus on calculating actions. To account for short-term fluctuations, but still calculate all actions and frequencies over relatively short time intervals, we use two independent approaches. One of them involves calculating actions using the {\tt{AGAMA}} package first and then averaging the obtained values between apocentres ($r_\mathrm{max}$) or maximums of the vertical excursions of the orbit ($z_\mathrm{max}$). Another method boils down to finding the actions directly from the orbit of a particle in an $N$-body model using the definition of radial and vertical actions in a cylindrical potential:}
%%%%%%%%%%%%%%%%%%%%%%%%%%%%%%%%%%%%%%%%%%%%%%%%%%%%%%%
% \begin{equation}
% %%%%%%%%%%%%%%%%%%%%%%%%%%%%%%%%%%%%%%%%%%%%%%%%%%%%%%%
% J_\mathrm{R} = \dfrac{1}{\pi} \int_{r_\mathrm{min}}^{r_\mathrm{max}} \upsilon_\mathrm{R} dr;\; J_z = \dfrac{1}{\pi} \int_{z_\mathrm{min}}^{z_\mathrm{max}} \upsilon_z dz\
% \end{equation}
%%%%%%%%%%%%%%%%%%%%%%%%%%%%%%%%%%%%%%%%%%%%%%%%%%%%%%%

%%%%%%%%%%%%%%%%%%%%%%%%%%%%%%%%%%%%%%%%%%%%%%%%%%%%%%%
\begin{equation}
%%%%%%%%%%%%%%%%%%%%%%%%%%%%%%%%%%%%%%%%%%%%%%%%%%%%%%%
\label{eq:int_JR_Jz}
J_\mathrm{R} = \dfrac{1}{2 \pi}  \oint  \upsilon_\mathrm{R} dr;\; J_z = \dfrac{1}{2 \pi}  \oint  \upsilon_z dz
\end{equation}
%%%%%%%%%%%%%%%%%%%%%%%%%%%%%%%%%%%%%%%%%%%%%%%%%%%%%%%
\textcolor{black}{The actions can be found by integrating the radial or vertical velocities of the particle from apocentre to apocentre or from one maximum of $z$ to another, respectively.}
\par
\textcolor{black}{Regarding the angular momentum, its instantaneous value is obtained directly from the particle velocities and coordinates. The projection of angular momentum onto the $z$-axis $L_z$ undergoes short-period oscillations (fluctuations), which we smooth out applying the same procedure as we apply to instantaneous $J_\mathrm{R}$ and $J_z$ from {\tt{AGAMA}}. As it is done for $J_\mathrm{R}$, the averaging is carried out between two adjacent apocentres.}
%\par
%\textcolor{black}{The results of both described approaches applied to a typical studied orbit are shown in Fig. \ref{fig:actions_change}. Black thin vertical lines (the second plot from the top, $J_\mathrm{R}$ panel) indicate the time of passage of the orbital apocentres. Blue horizontal dashed lines correspond to the value of $J_\mathrm{R}$ from {\tt{AGAMA}} averaged between apocentres. Black horizontal dashed lines corresponds to the value of $J_\mathrm{R}$ found via Eq.~\eqref{eq:int_JR_Jz}. As can be seen from the figure, the values obtained via different approaches are quite close.}
\par
\textcolor{black}{The resulting $J_\mathrm{R}$ and $J_z$ averaged between apocenters are piecewise functions of time. However, real actions should continuously change over time. To account for this problem, we apply the numerical procedure described in~\cite{Ruiz_Jose2022}. \cite{Ruiz_Jose2022} used mean-preserving interpolation for averaged data series in solar radiation modelling. This method of data smoothing allows one to transform a piecewise-defined function into a continuous one and, at the same time, preserve either the value of the integral~(\ref{eq:int_JR_Jz}) or the mean value of $J_\mathrm{R}$ from {\tt{AGAMA}}. We use a publicly available software implementation of the described procedure (\url{https://github.com/jararias/mpsplines/}) for our calculations.}
\par
\begin{figure}
\centering
\includegraphics[width=\linewidth]{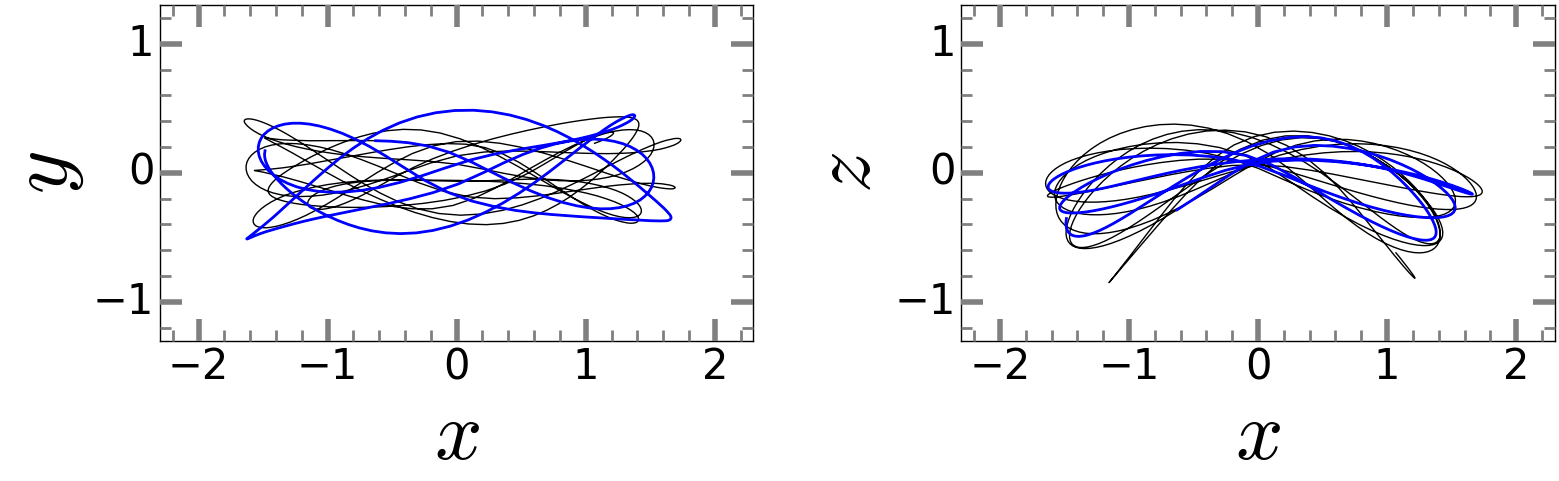}
\includegraphics[width=\linewidth]{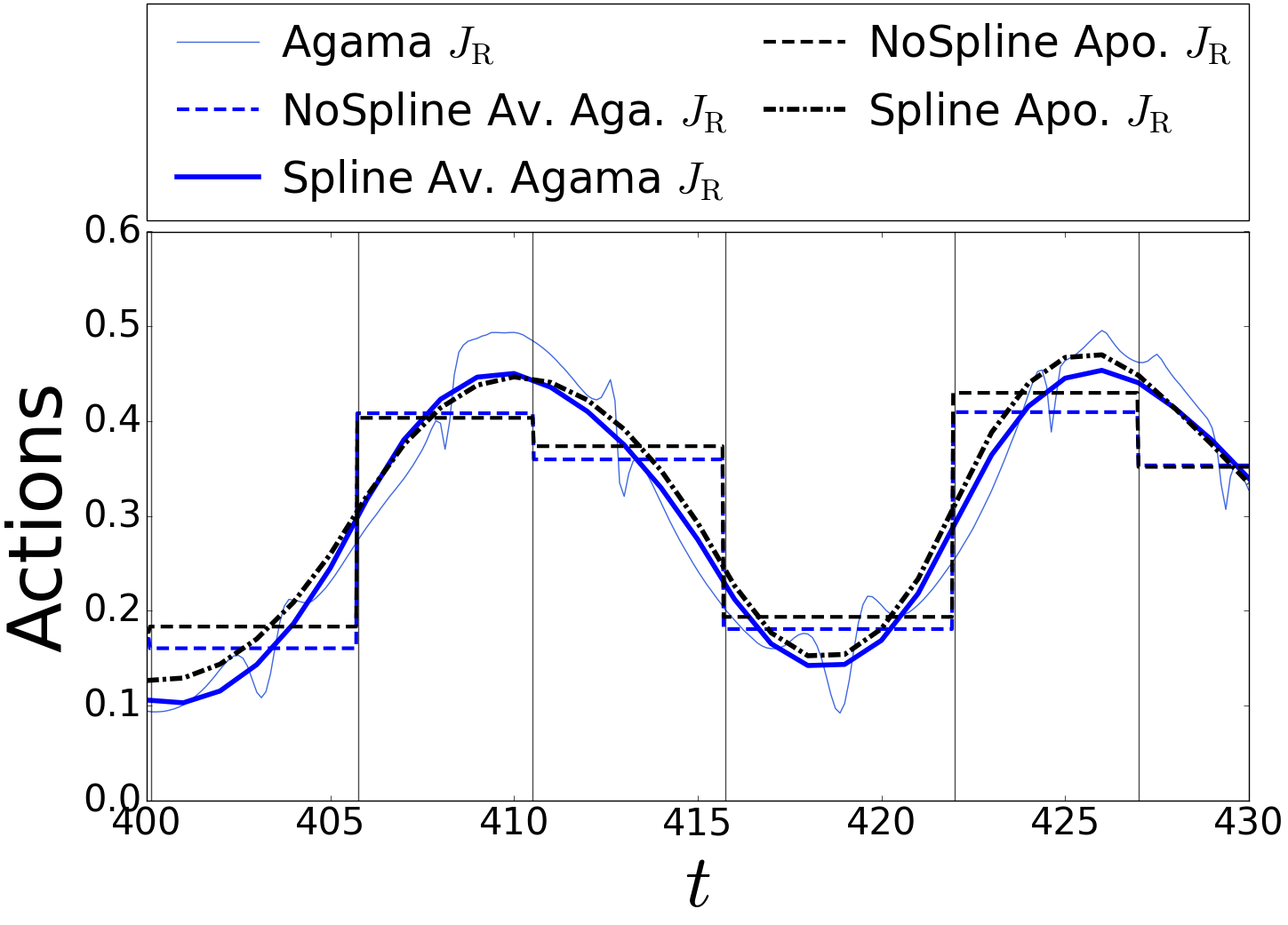}
\includegraphics[width=\linewidth]{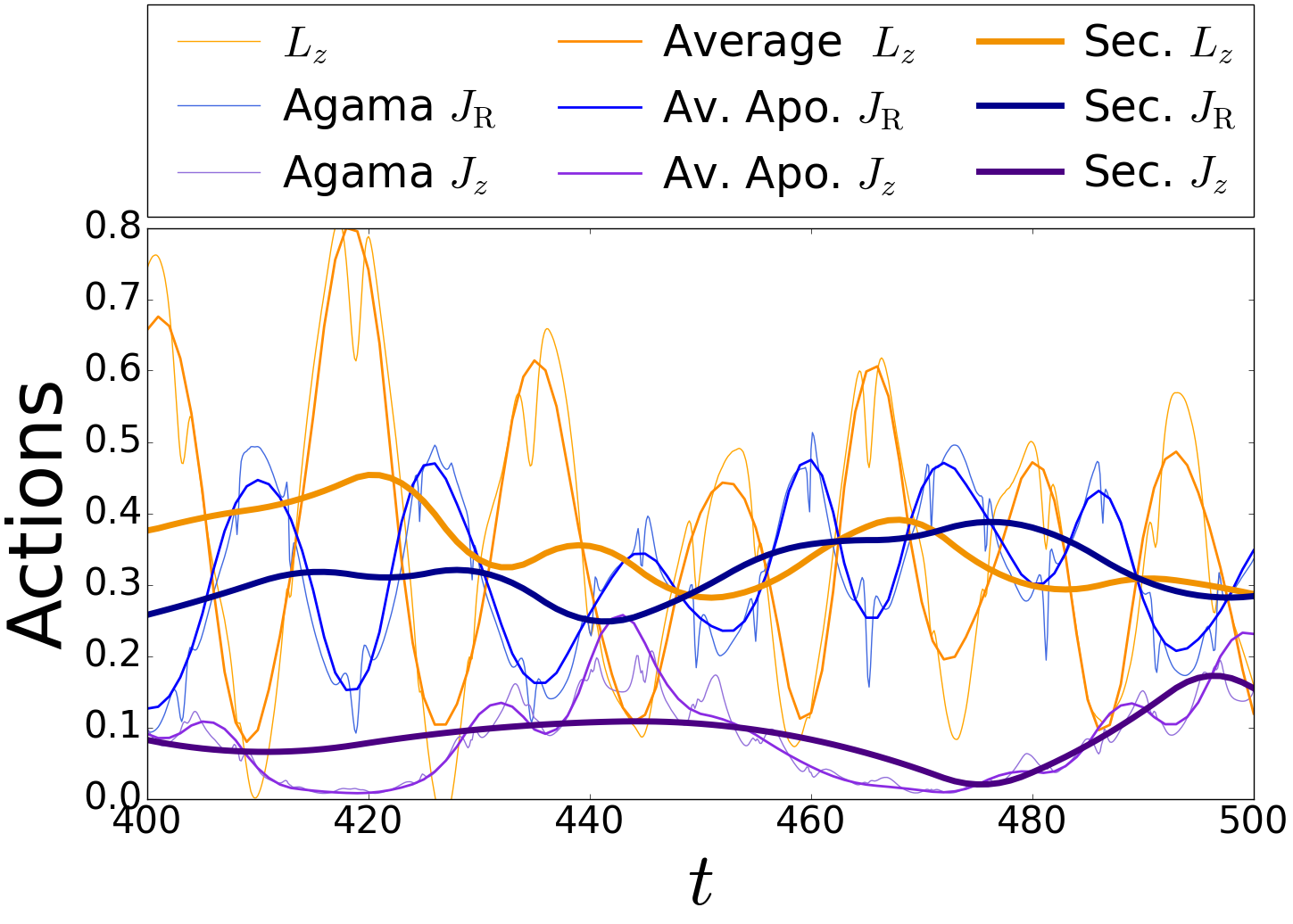}
\includegraphics[width=\linewidth]{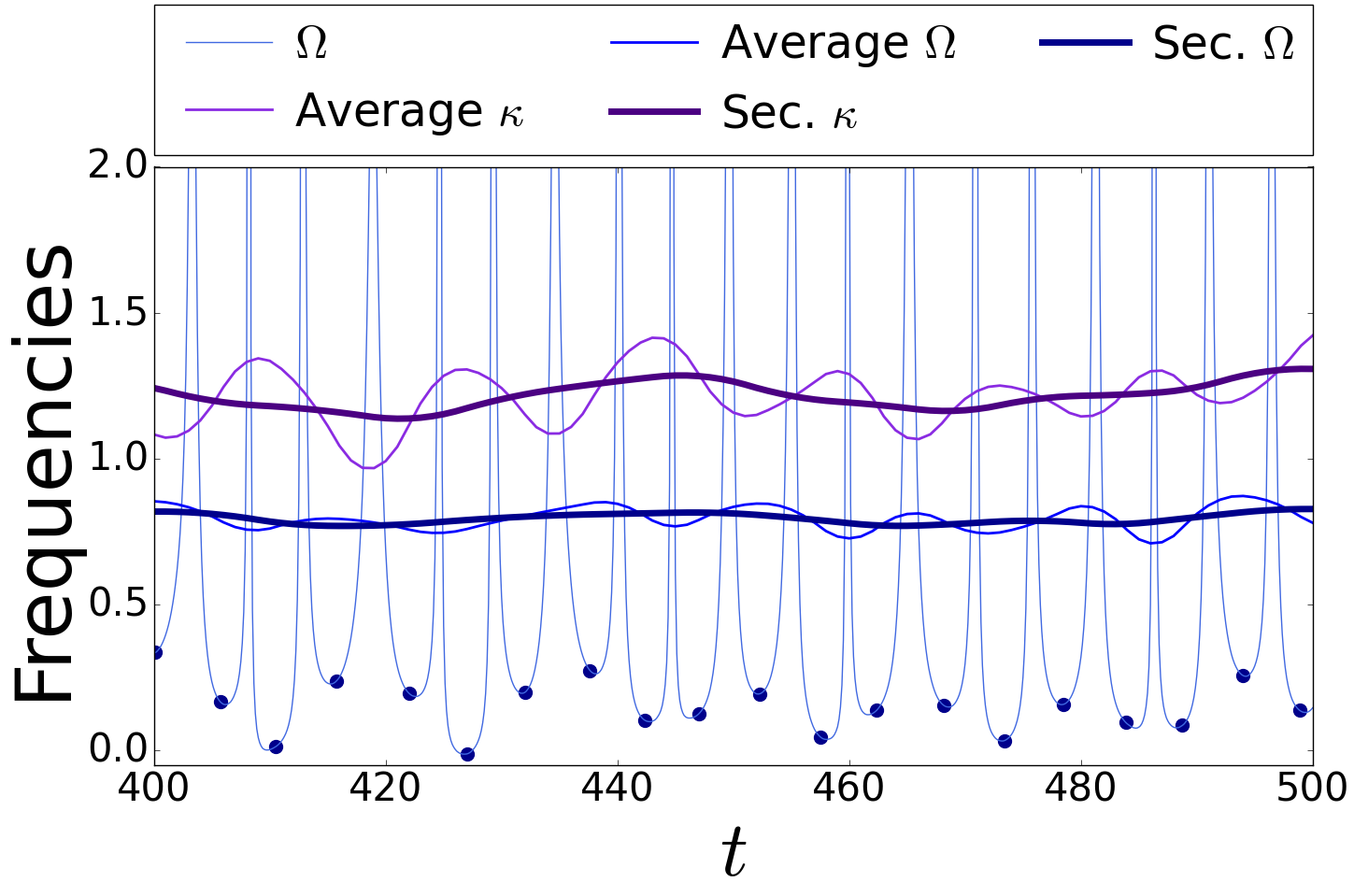}
\caption{\textcolor{black}{The bar particle orbit shown for the time intervals from $t = 400$ to $t = 430$ (\textit{blue}) and from $t = 430$ to $t = 500$ (\textit{black}) (\textit{two upper panels}; $x$, $y$ coordinates are given in dimensionless units of the simulations, $x$ coordinate goes along the major axis of the bar). The evolution of its actions $J_\mathrm{R}$, $J_z$, and $L_z$ \textit{(middle plots)} and frequencies $\kappa$ and $\Omega$ (\textit{bottom plot}), calculated using different procedures (see the main text).}}
\label{fig:actions_change}
\end{figure}

\textcolor{black}{
We demonstrate an example of this procedure application to the radial action of a typical bar orbit shown in the upper panel of Fig.~\ref{fig:actions_change} in the second panel of Fig.~\ref{fig:actions_change}. The blue dashed line in the second panel corresponds to $J_\mathrm{R}$ calculated as the average of instantaneous $J_\mathrm{R}$ from {\tt{AGAMA}}, while the black dashed line corresponds to $J_\mathrm{R}$ from Eq.~(\ref{eq:int_JR_Jz}). In turn, the blue solid and black dash-dotted lines represent the interpolation of these actions by the mean-preserving spline. We note that these two curves appear to be quite closely. Additionally, the curve of instantaneous actions from {\tt{AGAMA}} (thin light blue curve) also differs slightly from them, which indicates that the interpolation is carried out correctly.} 
\par 
\textcolor{black}{The interpolation results for all actions from Eq.~\ref{eq:int_JR_Jz} ($J_\mathrm{R}$ and $J_z$) and $L_z$ are shown in the third panel of Fig.~\ref{fig:actions_change} by lines of average thickness (``Av. Apo.'' prefix in the legend). As can be seen, the interpolated actions follow instantaneous {\tt{AGAMA}} actions (``Agama'' prefix) quite close.}
\par
\textcolor{black}{As for orbital frequencies, there are several methods for evaluating them suggested in the literature: by tracing time and angle of apocentre and $z$-maxima passages~\citep{Sellwood_Gerhard_2020}, by Fourier spectrum ~\citep{Smirnov_Sotnikova2018}, and by least squares approximation ~\citep{Ceverino_Klypin2007}. Each of these methods has its own advantages and disadvantages. Since we want to calculate frequencies both on small scales (of the order of one revolution) and on large time scales (in order to trace the secular evolution), we use the method based on apocentre tracing. Within this approach, the frequencies of radial and vertical oscillations are found using the following expressions:}

%Thus, mean frequencies of radial and vertical oscillations between apocentres are found from the
%%%%%%%%%%%%%%%%%%%%%%%%%%%%%%%%%%%%%%%%%%%%%%%%%%%%%%%
\begin{equation}
%%%%%%%%%%%%%%%%%%%%%%%%%%%%%%%%%%%%%%%%%%%%%%%%%%%%%%%
\label{eq:freq_apo}
\left< \kappa \right>= \dfrac{2 \pi}{T_\mathrm{R}};\;\;\; \left< \omega_z  \right> = \dfrac{2 \pi}{T_z};
\end{equation}
%%%%%%%%%%%%%%%%%%%%%%%%%%%%%%%%%%%%%%%%%%%%%%%%%%%%%%%
\textcolor{black}{where $T_\mathrm{R}$ and $T_z$ are the time intervals between two successive passages of apocentres or $z$ maxima, respectively.}
\par 
%\textcolor{black}{The instantaneous value of angular velocity $\Omega = L_z / R^2$ greatly changes greatly between two apocentres. To not take this into account, we define $\left<  \Omega \right>$ as the average value between two apocentres. Similarly,  
\textcolor{black}{The mean value of angular velocity $\left<  \Omega \right>$ can be obtained through the increase in the polar angle calculated between two adjacent apocentres:}
%%%%%%%%%%%%%%%%%%%%%%%%%%%%%%%%%%%%%%%%%%%%%%%%%%%%%%%
\begin{equation}
%%%%%%%%%%%%%%%%%%%%%%%%%%%%%%%%%%%%%%%%%%%%%%%%%%%%%%%
\label{eq:omega_apo}
\left< \Omega \right>  = \dfrac{1}{T_\mathrm{R}} \int \Omega dt =  \dfrac{1}{T_\mathrm{R}} \int d \phi = \dfrac{\Delta \phi}{T_\mathrm{R}},
\end{equation}
%%%%%%%%%%%%%%%%%%%%%%%%%%%%%%%%%%%%%%%%%%%%%%%%%%%%%%%
\textcolor{black}{where $\Delta \phi$ is the change in the angle, where the orbital revolutions are taken into account. We apply this procedure for the bar orbit.}

\textcolor{black}{Since we calculate the mean values, the frequencies are piecewise-defined functions again. To obtain their continuous representations, we apply the same interpolation procedure to them as we apply to actions. %the spline approach~\citep{Ruiz_Jose2022} to smooth these functions $(\kappa_S(t),\,\omega_z_{S(t)},\,\Omega_S(t))$. 
Note that, in this case, the mean-preserving spline conserves the total amount of the angle changes that occur as the particle moves from one apocenter to another. The interpolated frequencies $\kappa$ and $\Omega$ are shown in the bottom panel of Fig.~\ref{fig:actions_change} by medium thickness line. In addition, the thin line demonstrates the instantaneous value of an angular velocity $L_z / R^2$, the dots mark the moments of orbit apocentres passage. %($\Delta \phi = \left< \Omega \right> T_R = \int \Omega_S dt$, $2 \pi = \left< \kappa \right> T_R = \int \kappa_S dt$)
}
%Fig.~\ref{fig} (figure with orbit actions, frequencies change) The method described above gives the actions and frequencies that oscillate about some average value. The same behaviour 
%Although the actions calculated using the St\"ackel fudge can be considered as a good initial approximation for the galaxy with the bar, they still differ from the actual action of stars moving in the potential with some non-axisymmetric part~\citet{}. In this subsection, we describe how we further improved the calculation of action variables. 

%%%%%%%%%%%%%%%%%%%%%%%%%%%%%%%%%%%%%%%%%%%%%%%%%%%%%%%
% Fig. 4
%%%%%%%%%%%%%%%%%%%%%%%%%%%%%%%%%%%%%%%%%%%%%%%%%%%%%%%
\subsubsection{Actions averaged over librations}
\label{sec:mean_actions_librations}
% Sec. 3.3.2
%-----------------------------------------------------%
%-----------------------------------------------------%
%\textcolor{black}{Orbits in non-axisymmetric potential can librate around or circulate around the orbits in perfect
%resonance. At the same time actions and frequencies librate too. It is clearly visible on Fig.~\ref{fig:actions_change}, which demonstrates the long-term evolution of actions ($L_z, J_R$ and $J_z$) and frequencies ($\kappa$ and $\Omega$). In present work, we want to investigate secular orbits evolution. For this aim, it is necessary to smooth out, including libration. }
\par
\textcolor{black}{As a second step of our analysis, we smooth out the obtained actions and frequencies over the libration period. %To this aim, we carry out a second smoothing of time series obtained by smoothing out short-term fluctuations. 
We use the same procedure as we applied in the previous subsection to approximate the instantaneous $L_z$ and $J_\mathrm{R}$ and $J_z$ from {\tt AGAMA}. We take a smoothed time series of actions and frequencies, find the extrema there, and then find the mean value between two adjacent extrema. Thus, as in the previous subsection, a piecewise-defined function is obtained, which can be smoothed with the mean-preserving spline. The only difference between the procedure carried out here and in the previous subsection is that we average the values not between the maxima of the actions and frequencies themselves, but between the extrema. This is due to the fact that the libration period can be quite long. The secular evolution of actions and frequencies (averaged over a half-period of libration) is shown by thick lines in two bottom panels of Fig.~\ref{fig:actions_change} (``Sec.'' prefix in the legend).}
\par
\textcolor{black}{%All temporary variations discussed in previous are observed on Fig.~\ref{fig:actions_change}: short-term oscillations of instance actions, middle-term librations and secular evolution. 
As a finishing remark of this subsection, we want to note the following. Here, we describe three different types of oscillation in actions and frequencies that occur when the particle moves along its orbit in the realistic $N$-body potential. In general, we smooth out short-term oscillations. The algorithm that we presented allows us to investigate the middle-term and secular evolution of actions and frequencies for each particle-``star'' in the $N$-body model. As we showed here, one can estimate actions either using {\texttt{AGAMA}} routines applying them to the axisymmetric approximation of the potential or directly integrating the velocities along the orbit in the $N$-body potential. In the course of the present work, we mostly analyse the actions obtained \textit{directly from the $N$-body evolving potential}. Actions from {\texttt{AGAMA}} are only used for the analysis of the model at $t=0$, when the model is axisymmetric, and for one later time moment for the preliminary analysis of the action space.}

%-----------------------------------------------------%
%-----------------------------------------------------%
\subsubsection{Estimation of Lynden-Bell derivative}
\label{sec:LB_derivative}
% Sec. 3.3.3
%-----------------------------------------------------%
%-----------------------------------------------------%

%In the previous subsections, we have developed a procedure, which allow us to investigate the middle-term and secular evolution of actions and frequencies. As we noted, these values librate around the average on a middle-time scale. This means that we can find the change in some quantity between the extremes and find something like the libration amplitude. But if we simultaneously monitor the change in two quantities, it will be possible to estimate their ratio during the time between extrema, thus estimating the value of the derivative of one quantity with respect to the other. For example, consider how we can evaluate the derivative: 
%In the previous subsections, we have developed a procedure, which allow us to investigate the middle-term and secular evolution of actions and frequencies. As we noted, these values librate around the average on a middle-time scale. This fact allows us to estimate the average value of the derivative over the libration half-period, and even the derivative of one quantity with respect to the other. For example, consider how we can evaluate the derivative: 
%The Lynden-Bell mechanism assumed the preserving $J_f$ and initially considered an axisymmetric isochrone  potential. In turn, we can estimate the LB-derivitive for each particle separately.
\textcolor{black}{
In the course of the present article, we also obtain an estimation of LB-derivative \lbd for individual orbits. We show below that this quantity turns out to be tightly connected with how the orbit behaves under the bar influence. \textcolor{black}{In the 2D case (i.~e. planar orbits),} the numerical estimate of \lbd can be obtained by replacing the partial derivative for particles that conserve their adiabatic invariant $J_f$ with the total derivative}:

%%%%%%%%%%%%%%%%%%%%%%%%%%%%%%%%%%%%%%%%%%%%%%%%%%%%%%%

\begin{equation}
%%%%%%%%%%%%%%%%%%%%%%%%%%%%%%%%%%%%%%%%%%%%%%%%%%%%%%%
\left( \dfrac{\partial \Omega_{\mathrm{pr}}}{\partial L_z} \right)_{J_f} =\dfrac{d \Omega_{\mathrm{pr}}}{d L_z} - \left( \dfrac{\partial \Omega_{\mathrm{pr}}}{\partial J_f} \right)_{L_z} \cdot \dfrac{d J_f}{d L_z} \approx \dfrac{\Delta \Omega_{\mathrm{pr}}}{\Delta L_z}=\dfrac{\Delta \Omega_\mathrm{pr}}{\Delta t} \cdot \dfrac{\Delta t}{\Delta L_z},
\end{equation}
%%%%%%%%%%%%%%%%%%%%%%%%%%%%%%%%%%%%%%%%%%%%%%%%%%%%%%%
%\begin{equation}
%\dfrac{d\Omega_\mathrm{pr}}{dL_z} \approx \dfrac{\Delta \Omega_\mathrm{pr}}{\Delta L_z}=\dfrac{\Delta \Omega_\mathrm{pr}}{\Delta t} \cdot \dfrac{\Delta t}{\Delta L_z},     
%\end{equation}
 \textcolor{black}{where $\Delta L_z$ is the difference between successive extrema of $L_z$ in the corresponding time series, where the short-term fluctuations are smoothed out, and $\Delta \Omega_\mathrm{pr}$ is calculated as the difference between $\Omega_\mathrm{pr} = \Omega - \kappa/2$ corresponding to the extrema of $L_z$. We choose the time between $L_z$ extrema as $\Delta t$ because the time-series of middle-term $L_z$ is regular enough, and the extrema can be identified relatively easily (see Fig.~\ref{fig:actions_change}). In contrast, $\Omega_\mathrm{pr}$ extrema are harder to find numerically, since $\Omega_\mathrm{pr}$ is the difference of two splines $\Omega$ and $\kappa/2$ and is not
as regular as the time series of $L_z$. As a consequence of irregularity, the middle-term time series of $\Omega_\mathrm{pr}$ may contain false extrema. For the same reason, we have not considered the derivative found directly from the time series of medium-term values ($\Delta t$ is much less than the libration period).} \textcolor{black}{As an additional note, we should emphasize that orbits in the $N$-body galaxy are 3D in nature, and one should estimate the total derivative of the precession frequency with respect to the angular momentum $\Delta \Omega_\mathrm{pr} / \Delta L_z$.}
% by the Lynden-Bell derivative \lbd 

%\textcolor{black}{%Note, we have applied this method to calculate the derivative, because we investigate the secular evolution of actions. In additional, our numerical experiments have shown that the derivative found directly from the time series of medium-term values is calculated with large scatters and errors.}

%%%%%%%%%%%%%%%%%%%%%%%%%%%%%%%%%%%%%%%%%%%%%%%%%%%%%%%%
%\begin{equation}
%%%%%%%%%%%%%%%%%%%%%%%%%%%%%%%%%%%%%%%%%%%%%%%%%%%%%%%
%\left( \dfrac{\partial \Omega_{\mathrm{pr}}}{\partial L_z} \right)_{J_f} =\dfrac{d \Omega_{\mathrm{pr}}}{d L_z} - \left( \dfrac{\partial \Omega_{\mathrm{pr}}}{\partial J_f} \right)_{L_z} \dfrac{d J_f}{d L_z} \approx \dfrac{\Delta \Omega_{\mathrm{pr}}}{\Delta L_z}
%\end{equation}
%%%%%%%%%%%%%%%%%%%%%%%%%%%%%%%%%%%%%%%%%%%%%%%%%%%%%%%
%\textcolor{black}{We know algorithm, which allow us to evaluate the full derivative $d\Omega_\mathrm{pr}/dL_z$ and, most importantly, its \textit{sign}. Note, the estimation of the LB-derivitive can be found without calculating $J_R$ and $J_z$.}

\textcolor{black}{To summarise, the algorithm was described in this subsection allows us to calculate the \textcolor{black}{value} of LB-derivative between two successive extrema of $L_z$, and, in particular, to estimate the \textit{sign} of LB-derivative. A more detailed analysis of the change in the sign of the LB-derivative for individual orbits will be carried out in Section~\ref{sec:LB_averaged} and for an ensemble of particles in Section~\ref{sec:trapping}.
}

%-----------------------------------------------------%
%-----------------------------------------------------%
\section{Precession rate maps}
\label{sec:maps}
% Sec. 4.
%-----------------------------------------------------%
%-----------------------------------------------------%

%%%%%%%%%%%%%%%%%%%%%%%%%%%%%%%%%%%%%%%%%%%%%%%%%%%%%%%
% Fig. 6
%%%%%%%%%%%%%%%%%%%%%%%%%%%%%%%%%%%%%%%%%%%%%%%%%%%%%%%
\begin{figure*}
\centering
\includegraphics[width=0.45\textwidth]{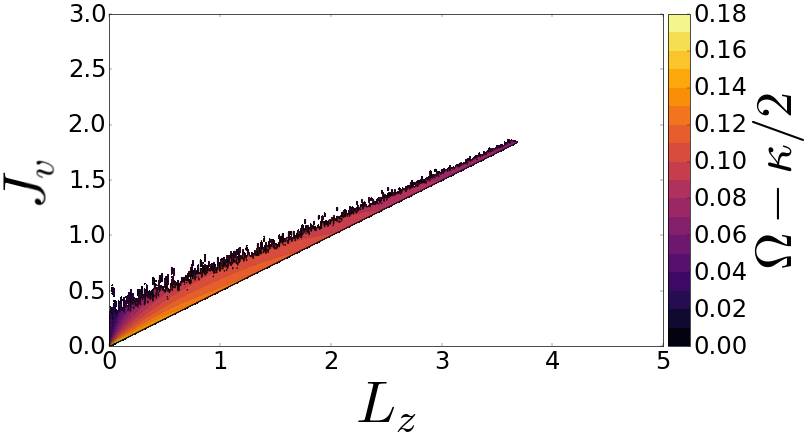}%
\includegraphics[width=0.45\textwidth]{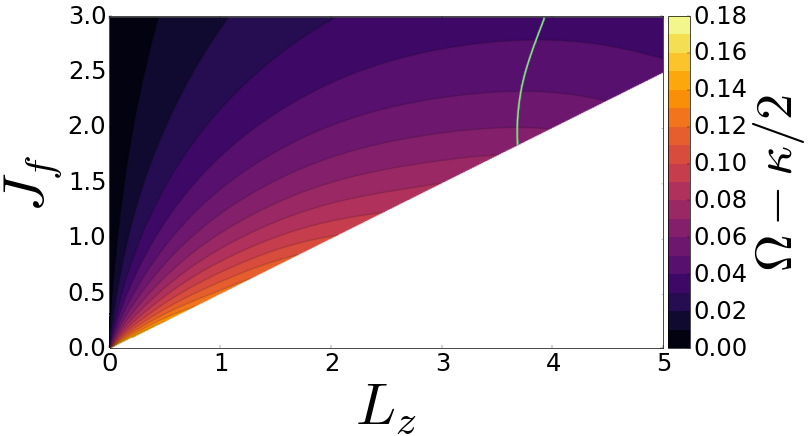}
\\
\includegraphics[width=0.45\textwidth]{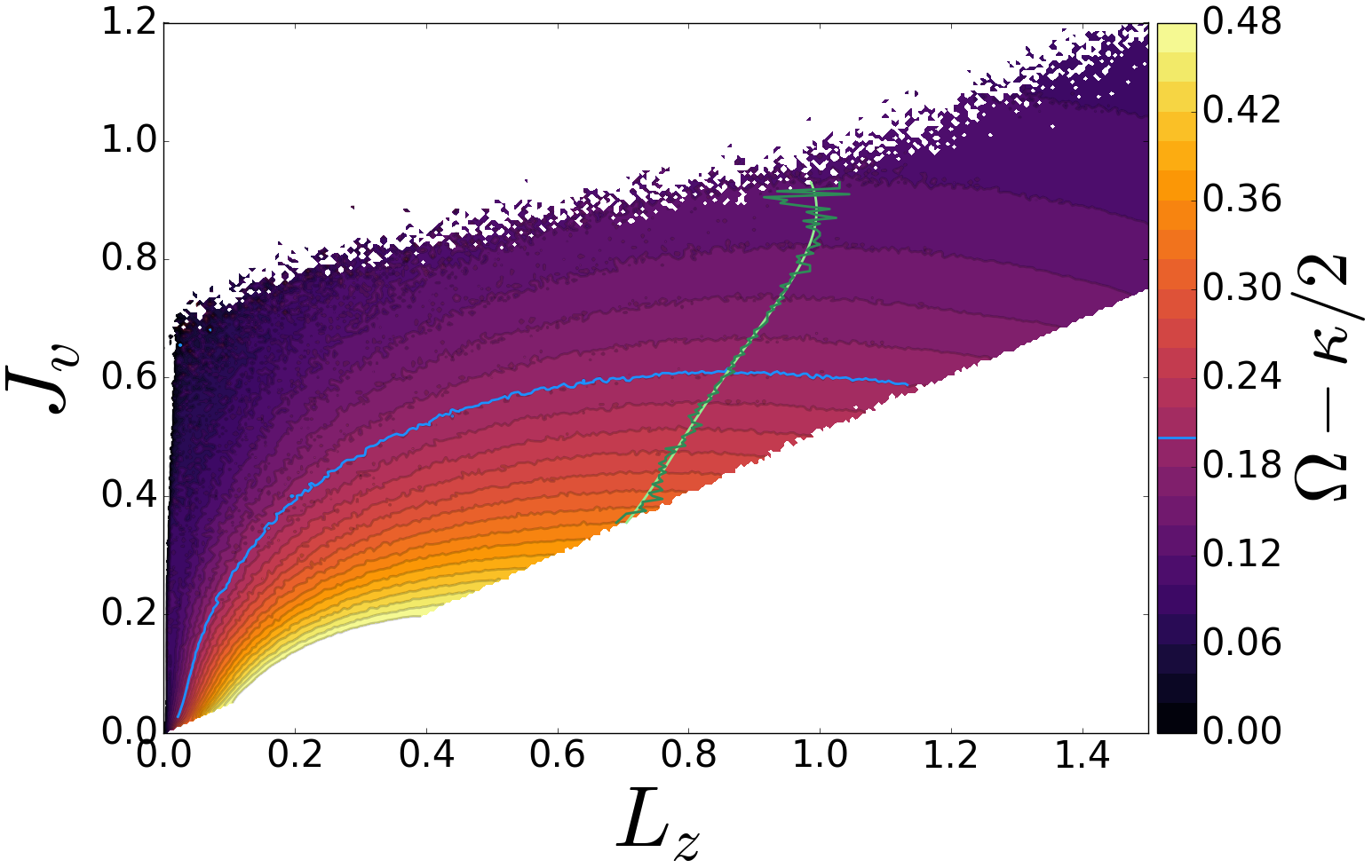}
\includegraphics[width=0.45\textwidth]{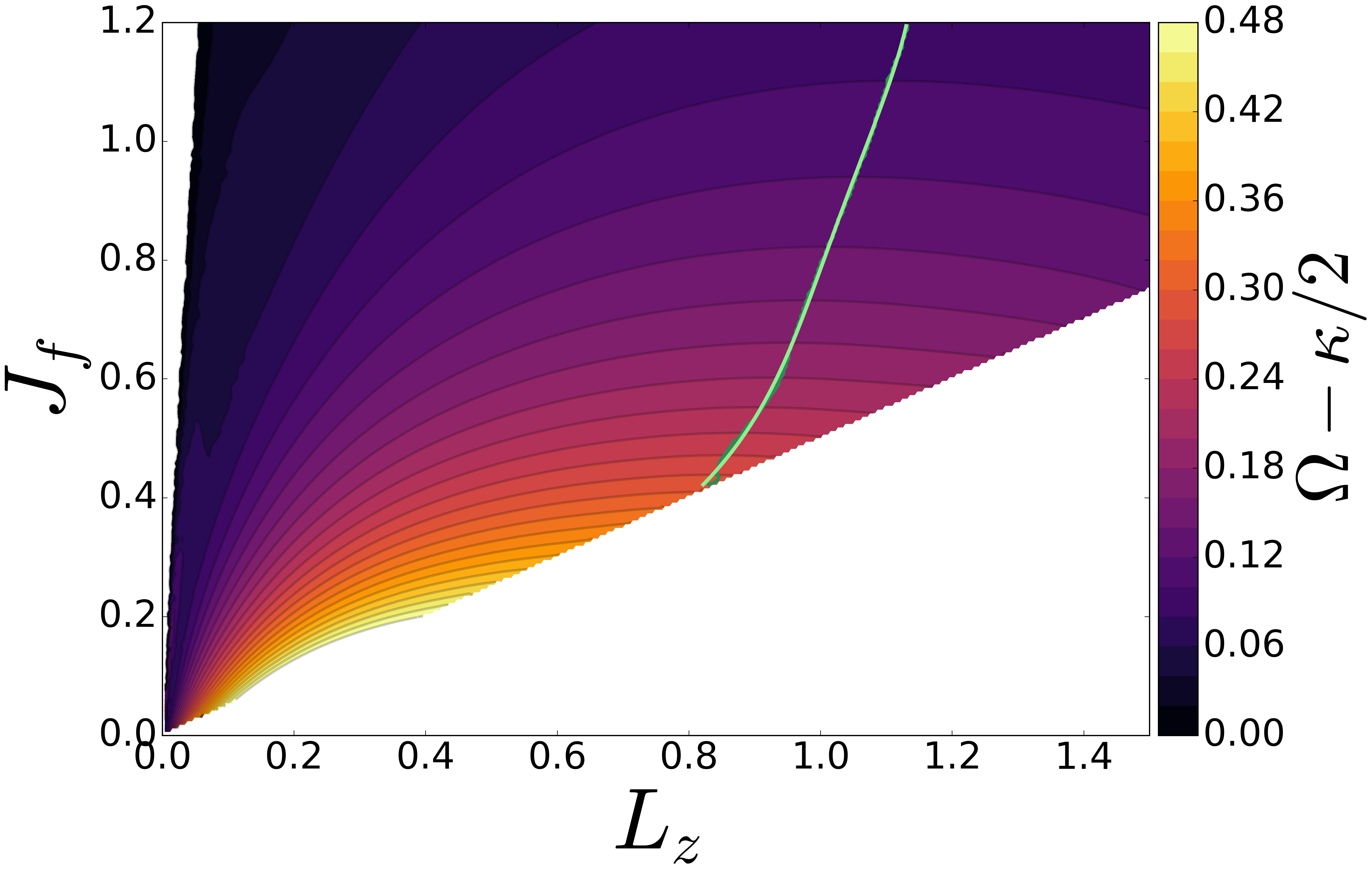}
\caption{Maps of the precession rates for $t=0$ (\textit{top}) and $t=400$ (\textit{bottom}) on the plane $L_z$--$J_f$. \textit{Left column}: maps, obtained by averaging the precession rate of the particles from the $N$-body model in bins of size~$0.005\times0.005$. \textit{Right column}: maps, obtained for the precession rate of the particles artificially populated in the phase space, with orbits of different radial actions and angular momentum. A dark green line corresponds to the condition \lbd$=0$. A light green line shows a spline approximation of the latter. Black lines are isolines of the precession rate.}
%Colored lines show tracks of some individual particles. \textit{Top right:} Distribution of ``slowness'' parameter $|\Omega - \kappa/2 - \Omega_p|/\Omega$. \textit{Bottom panels:} distribution of precession rates $\Omega-\kappa/2$ at times $t=10$ (\textit{left}) and $t=450$ (\textit{right}). In all panels, the dark green line shows the location of zero values of the partial derivative $\left( \partial \Omega_\mathrm{pr}/ \partial L_z \right)_{J_f}$, and the light green line shows its smoothed version.
%\textcolor{black}{Change order, bottom line should be the first one plus one scale, please.}}
\label{fig:Jf_Lz}
\end{figure*}
%%%%%%%%%%%%%%%%%%%%%%%%%%%%%%%%%%%%%%%%%%%%%%%%%%%%%%%
%%%%%%%%%%%%%%%%%%%%%%%%%%%%%%%%%%%%%%%%%%%%%%%%%%%%%%%

%%%%%%%%%%%%%%%%%%%%%%%%%%%%%%%%%%%%%%%%%%%%%%%%%%%%%%%
%\begin{figure*}
%\centering
%\includegraphics[width=\textwidth]{img/LB_map/Mb0_415_quart_resized.png}
%\caption{\textit{Top left}: 2D distribution of orbits over $L_z$ and $J_f = J_\mathrm{R} + J_z+ L_Z/2$ at time $t = 450$. Coloured lines show tracks of some individual particles. \textit{Top right:} Distribution of ``slowness'' parameter $|\Omega - \kappa/2 - \Omega_p|/\Omega$. \textit{Bottom panels:} distribution of precession rates $\Omega-\kappa/2$ at times $t=10$ (\textit{left}) and $t=450$ (\textit{right}). In all panels, the dark green line shows the location of zero values of the partial derivative $\left( \partial \Omega_\mathrm{pr}/ \partial L_z \right)_{J_f}$, and the light green line shows its smoothed version.
%\textcolor{black}{Change order, bottom line should be the first one plus one scale, please.}}
%\label{fig:Jf_Lz}
%\end{figure*}
%%%%%%%%%%%%%%%%%%%%%%%%%%%%%%%%%%%%%%%%%%%%%%%%%%%%%%%
%%%%%%%%%%%%%%%%%%%%%%%%%%%%%%%%%%%%%%%%%%%%%%%%%%%%%%%

%%%%%%%%%%%%%%%%%%%%%%%%%%%%%%%%%%%%%%%%%%%%%%%%%%%%%%%
% Fig. 7
%%%%%%%%%%%%%%%%%%%%%%%%%%%%%%%%%%%%%%%%%%%%%%%%%%%%%%%
\begin{figure*}
\centering
\includegraphics[width=\textwidth]{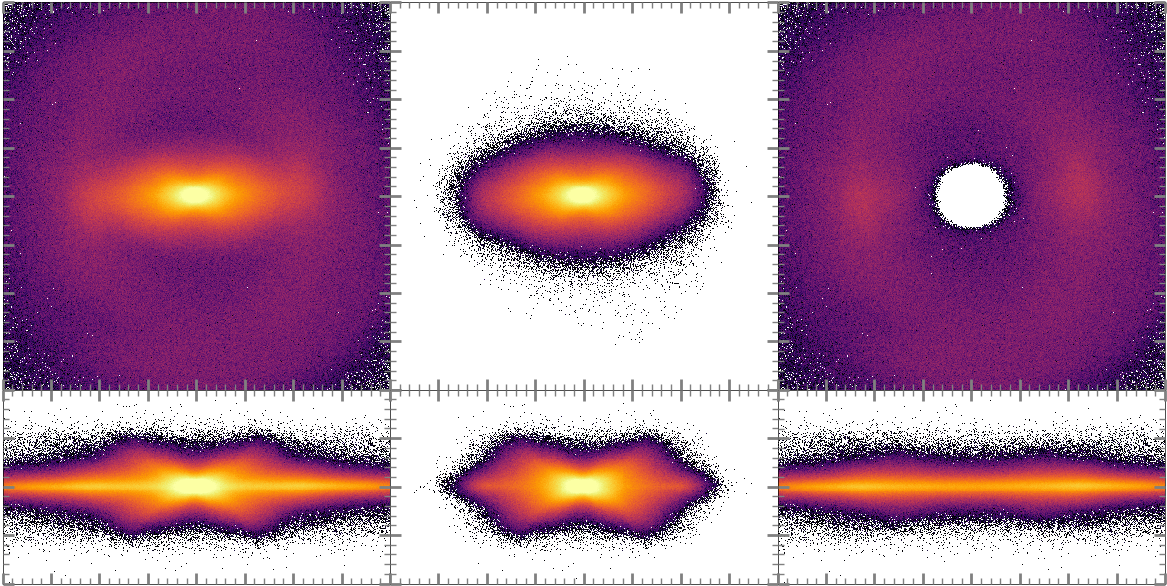}
\caption{The face-on (\textit{upper row}) and the side-on (\textit{bottom row}) density plots for all particles of the disc (\textit{left column}), those from the region where the LB-derivative has negative values (\textit{middle column}), and those from the region of the positive values (\textit{right} column). All plots are shown for the same time moment $t=400$. The face-on view is displayed in the square $(xy) = (-4,4) \times (-4,4)$ while the side-on view is displayed in the rectangle $(xz) = (-4,4)\times(-2,2)$.}
\label{fig:density_regions}
\end{figure*}
%%%%%%%%%%%%%%%%%%%%%%%%%%%%%%%%%%%%%%%%%%%%%%%%%%%%%%%
%%%%%%%%%%%%%%%%%%%%%%%%%%%%%%%%%%%%%%%%%%%%%%%%%%%%%%%
    
%\textcolor{black}{Do we believe in frequencies? If not, do we believe in isolines?} 
%\par
%\textcolor{black}{Need snapshots, with different zones marked}
%\par 
%\textcolor{black}{Instantaneous values of the actions!}
%\citet{Lynden-Bell1979} studied the ability of orbits to facilitate the bar growth analysing how the precession rate of orbits depends on the angular momentum projection $L_z$ and adiabatic invariant $J_f$. A key parameter, which was studied by \citet{Lynden-Bell1979} for the first time, is a sign of the partial derivative of precession rate $\Omega_\mathrm{pr}$ over $L_z$ at a constant value of $J_f$. Using qualitative arguments, \citet{Lynden-Bell1979} suggested that this parameter should determine the susceptibility of orbits for trapping. In particular, he argued that bars should form in the area where the derivative is positive, within the so-called abnormal region. Following~\citet{Lynden-Bell1979}, we study the dependence of precession rates on $L_z$ and $J_f$ for our numerical model in this section.
\par
\textcolor{black}{The secular evolution of orbits will be studied in the next Section. In this section, we show how the general structure of the $N$-body system with a bar looks like in the action space taken at a single moment of time. Following the Lynden-Bell~(\citeyear{Lynden-Bell1979}) approach, we present the `portrait' of a bar on the $(L_z, J_f)$ plane. Here, we show the distributions of non-averaged (instantaneous) actions and precession rates, i.e. those calculated by~{\tt AGAMA} without additional post-processing on our side. This is also to show that instantaneous values obtained from~{\tt AGAMA} can be directly used to identify the bar particles.} 
%In general, we show the distributions of precession rates on $(L_z, J_f)$ plane following~\cite{Lynden-Bell1979}'s approach.}
%In practice, the analysis boils down to the overplotting of the precession rate $\Omega_{pr}$ isolines over the 2D distribution of orbits by $L_z$ and $J_f$ and distinguishing the areas of. Doing so, distinguishing the area where partial derivative o. We call the maps obtained in this way ``Lynden-Bell maps''.
\par 
One of the important notions concerning the maps of precession rate in numerical models is that the maps can be constructed using two different ways. One way is to calculate the actions and precession rates of the particles constituting the particular model and produce the maps from the obtained values by averaging the values of $\Omega-\kappa/2$ over small bins in the plane $L_z-J_f$. The problem, which is implicitly introduced in this approach, is that the particles constituting the self-consistent model occupy only a small part of the phase space. In practice, most of the orbits are circular (especially at the start of the simulation) and, therefore, the maps obtained in this way are not representative, i.e. they do not allow one to picture how the evolution of orbits from circular to more radially elongated can proceed in the given potential. The second approach, which we use here along with the first one, is to artificially populate a wide region of the phase space $L_z-J_f$ by orbits and calculate the actions and precession rates using a frozen potential of $N$-body model. The maps obtained in this way are more representative compared to the first case since they show where in the phase space the circular orbits can evolve under the influence of perturbation if they conserve their adiabatic invariant.
\par 
%\todo{NS}{Fig.~\ref{fig:Jf_Lz}. To explain why non-averaged maps were used.} \todo{AS}{Added some notes above, check, please.} 
Maps of the precession rates in $(L_z, J_f)$ coordinates for $t=0$ and $t=400$ are shown in~Fig.~\ref{fig:Jf_Lz}. Each map is obtained by averaging the precession rate value over the bins of width 0.005 by $L_z$ and the same width of 0.005 by $J_f$. The line $J_f=L_z/2$ corresponds to the line of circular orbits. As discussed above, we analyse two types of maps. For the first type, we consider instantaneous actions and frequencies from {\texttt{AGAMA}} (left plots). %The corresponding values are obtained from the $N$-body model ($\Omega$~also found using {\tt AGAMA}). 
Maps of the second type are presented in the right panels of Fig.~\ref{fig:Jf_Lz}. Here we also consider the instantaneous values of the precession rate but the maps are constructed analytically using an axisymmetric approximation of the potential at the appropriate time moments (see~\ref{sec:axisym_potentia}). In this case, we consider only in-plane motions and the adiabatic invariant is calculated simply as $J_f=J_\mathrm{ R}+L_z/2$ (without the addition of the vertical component). In general, the maps in the right panel of Fig.~\ref{fig:Jf_Lz} were built in the same way as it was done by~\citet{Lynden-Bell1979} for the isochrone model, but all integrals and derivatives are calculated numerically here.
\par 
In all subpanels of Fig.~\ref{fig:Jf_Lz}, we also show the location of the line \lbd$=0$ (green colour). The line was found via the spline fitting of $\Omega-\kappa/2$ dependence on $L_z$ considered for a fixed value of $J_f$. In practice, we take a horizontal slice of the map $(L_z, J_f)$ and identify the location of a zero value \lbd in each of the slices. The typical error resulting from this procedure is about $0.03$ in corresponding simulation units. 
%in $xy$ plane
\par
As the left top panel of Fig.~\ref{fig:Jf_Lz} shows, the most of orbits are circular at $t=0$, i.e. they adjoin to the line $J_f=L_z/2$ and have a relatively small value of the radial action. This is expected by the construction of the model. But the unexpected thing is that in contrast to the isochrone model built by~\citet{Lynden-Bell1979}, in which the abnormal region was constrained within the central part of the model, almost the entire exponential $N$-body disc lies in the abnormal region, where \lbd$>0$, as the top right panel of Fig.~\ref{fig:Jf_Lz} shows. There, it can be clearly seen that the line \lbd$=0$ is located at about $L_z\approx3.6$. Such a value corresponds to the disc boundary of the initial $N$-body system. The location of zero of \lbd for circular orbits is shown by the small green square in the initial rotation curve of the model (Fig.~\ref{fig:Nbodymodel}). Of course, we consider a potential that differs from the one considered by~\cite{Lynden-Bell1979} and this should explain the difference. Nevertheless, the fact that for a typical $N$-body model the normal region turns out to be almost non-existent is intriguing. In his short review of the dynamics of the Milky Way, Weinberg~(\citeyear{Weinberg2001}) gives a general rule stating when two orbits should precess toward or away from each other. They precess toward (away from) each other if the rotation curve is rising (flat). Such an interpretation of the Lynden-Bell's~\citeyear{Lynden-Bell1979} mechanism comes from the widespread belief that the sticking of elongated orbits occurs in the central regions of galaxies, and the mechanism itself generates the so-called slow bars. We see that in realistic $N$-body models of galaxies, the abnormal region, where orbits can be trapped in a bar-like perturbation, covers almost the entire initial disc and the zero of \lbd lies far from the maximum of the rotation curve (see Fig.~\ref{fig:Nbodymodel}, left plot).
\par
This is also consistent with the analytical result by \citet{Pol2003} who found that the normal region for an infinitely thin exponential disc lies outside four radial scales\footnote{The difference in the location of zero of \lbd for circular orbits for the exponential disc considered by \citet{Pol2003} and for our model is due to the presence of the dark matter halo in our model.}. Moreover, recently~\citet{Pol2020a,Pol2020b} have shown that, for the initial stages of evolution, the role of the normal orbits in the bar weakening/destruction is much smaller than previously believed, since the actual values of \lbd, even if they are negative over the disc, are actually close to zero for several types of galaxy models typically considered.
%In each individual panel, black lines are isolines of the precession rates. %The line corresponding to a zero value of the partial derivative $(\partial \Omega_\mathrm{pr}/ \partial L_z)_{J_f}$ is marked by green, while the green shaded area depicts the region, where the absolute value of the derivative is smaller than 0.05 in simulation units.
%Interestingly, isolines of the initial state of the system ($t=10$) show positive inclination for all values of $L_z$ meaning that the Lynden-Bell derivative has positive value almost everywhere.
\par
Lower panels of Fig.~\ref{fig:Jf_Lz} show what happens when the system evolves. The left panel shows the precession rates of the particles of $N$-body model at $t=400$, while the right panel demonstrates the theoretical map for the same time moment. Both panels show that the system is now divided into two distinct parts: one part, where \lbd$<0$, and the second one, where \lbd$>0$. It can be seen that the line \lbd$=0$ has a slightly different shape depending on the map, but in general, there is a match between the two maps in terms of how the line runs. 
%TODO ORBITS WITH THE NEGATIVE PRECESSION RATES
%We note that there is a number of orbits in the $N$-body model for which the calculated precession rates are actually negative. We purposefully exclude such orbits\footnote{We attribute the appearance of such orbits to an error in frequencies calculation for fast rotating and quite elongated orbits in the most central area of the bar. These particles are not relevant for further discussion on the orbital trapping, since such particles remain in the bar and in its most central area.} from the present analysis and, consequently, a cavity appears on the lower left map of Fig.~\ref{fig:Jf_Lz}. 
\par 
We present snapshots taken for the particles occupying the regions to the left and right from the line \lbd$=0$ and for the entire disc in Fig.~\ref{fig:density_regions} at $t=400$. It can be immediately seen that the region, where \lbd$>0$, consists of the particles supporting the bar. The second region is populated mostly by disc particles but also contains a portion of particles that form the so-called bar ``spurs'', or ansae. %Hereinafter, we refer to the regions as ``bar'' and ``disc'' regions, respectively. 
\par 
\par
%As a result of this Section, we conclude that the bar is contained within the region where \lbd$>0$, which is overall consistent with the idea of the bar formation suggested by~\citet{Lynden-Bell1979}. In addition, we found that for the initial (non-evolved) model the region, where \lbd$<0$, is almost non-existent in a typical $N$-body model considered here. %In any case, it is outside $4.5\mathrm{d}$. 
%All this indicates that ~\citet{Lynden-Bell1979}'s idea may have more to do with the fast bars than previously thought. 
\textcolor{black}{As a result of this Section, we conclude that the bar is mostly contained within the region where \lbd$>0$. In addition, we found that, for the initial (non-evolved) model, the region, where \lbd$<0$, is almost non-existent. 
%In any case, it is outside $4.5\mathrm{d}$. 
Although these facts can be considered as supporting the idea of orbital trapping suggested by~\citet{Lynden-Bell1979}, in the next Section we will show that the situation is actually much more complex if one considers individual orbits and how they are captured by the bar, rather than the maps obtained by averaging over some bins.} 
\par 
%The corresponding destitution of particles on $(xy)$ plane is presented in Fig.~\ref{fig:} with  

%%%%%%%%%%%%%%%%%%%%%%%%%%%%%%%%%%%%%%%%%%%%%%%%%%%%%%%
%%%%%%%%%%%%%%%%%%%%%%%%%%%%%%%%%%%%%%%%%%%%%%%%%%%%%%%
%\begin{figure*}
%\centering
%\includegraphics[width=\textwidth]{img/Mb0_LzJR_n_z.png}
%\caption{2D distribution of orbits over $L_z$ and $J_f = J_\mathrm{R} + J_z+ L_Z/2$ at times $t = 10$ (\textit{left} panel), 300 (\textit{middle} panel), and 600 (\textit{right} panel). Dark contours correspond to isolines of the quantity $|\Omega| - \kappa/2$, the blue ones mark the isoline corresponding to the bar pattern speed. The dark green line and the shaded green area shows where the value of the so-called Lynden-Bell derivative is zero and within the range from $-0.05$ to $0.05$, respectively.}
%\label{fig:Jf_Lz}
%\end{figure*}
%%%%%%%%%%%%%%%%%%%%%%%%%%%%%%%%%%%%%%%%%%%%%%%%%%%%%%%
%%%%%%%%%%%%%%%%%%%%%%%%%%%%%%%%%%%%%%%%%%%%%%%%%%%%%%%

%-----------------------------------------------------%
%-----------------------------------------------------%
\section{Orbital trapping}
\label{sec:orbits_trapping}
% Sec. 5.
%-----------------------------------------------------%
%-----------------------------------------------------%

%%%%%%%%%%%%%%%%%%%%%%%%%%%%%%%%%%%%%%%%%%%%%%%%%%%%%%%
% Fig. 8
%%%%%%%%%%%%%%%%%%%%%%%%%%%%%%%%%%%%%%%%%%%%%%%%%%%%%%%
\begin{figure*}
%\begin{minipage}[h]{0.45\linewidth}
%\center{\includegraphics[width=\linewidth]{img/LB_map/LB_map_Mb0_415_add_small.png}}
%\center{\includegraphics[width=\linewidth]{img/new/Mb0_415_new_LB_map_test.png}}
%\end{minipage}
%\begin{minipage}[h]{0.45\linewidth}
%\center{\includegraphics[width=\linewidth]{img/LB_map/Mb0_415_new_LzJR_mean_Wpr-Wp:W.png}}
%\center{\includegraphics[width=\linewidth]{img/new/Mb0_415_new_LB_map_wpr_blue0.1.png}}
%\end{minipage}
\begin{minipage}[h]{0.45\linewidth}
%\center{\includegraphics[width=\linewidth]{img/LB_map/LB_map_Mb0_415_add_small.png}}
\center{\includegraphics[width=\linewidth]{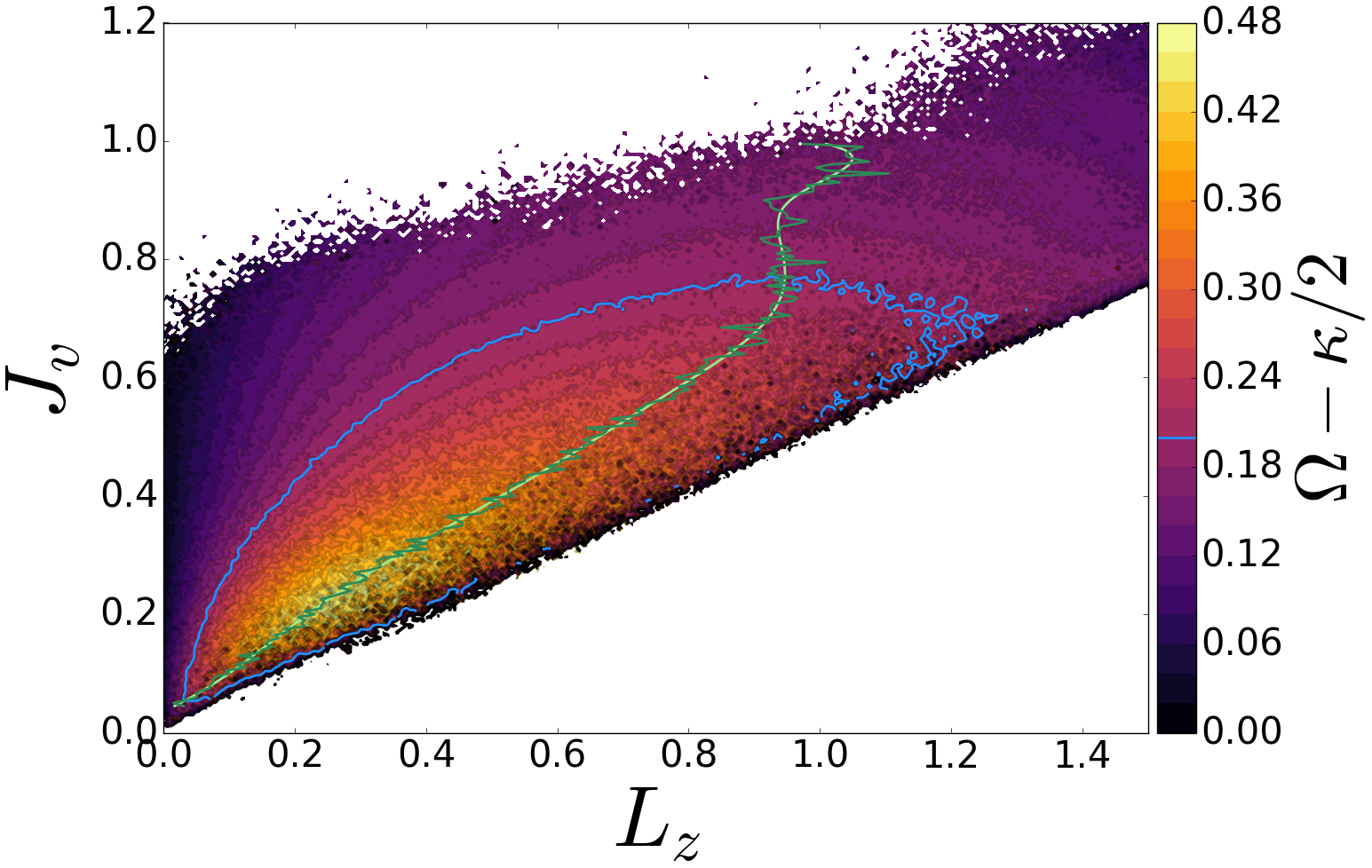}}
\end{minipage}
\begin{minipage}[h]{0.45\linewidth}
%\center{\includegraphics[width=\linewidth]{img/LB_map/Mb0_415_new_LzJR_mean_Wpr-Wp:W.png}}
\center{\includegraphics[width=\linewidth]{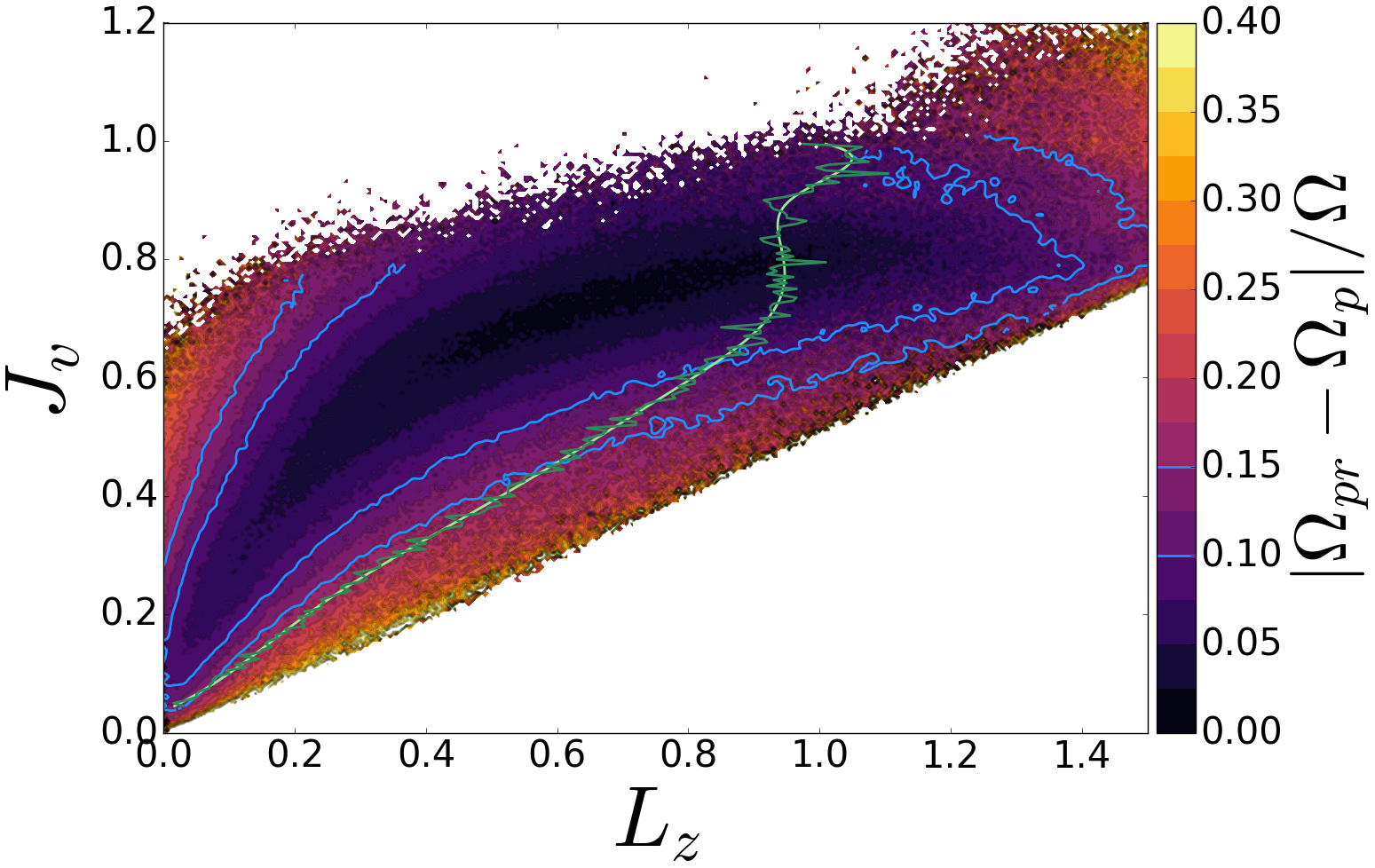}}
\end{minipage}
\caption{\textcolor{black}{The Lynden-Bell maps derived in the $N$-body potential for $t=400$. \textit{Left}: 2D distribution of the precession rate $\Omega_\mathrm{pr}=\Omega - \kappa/2$ on the plane $L_z$--\textcolor{black}{$J_v$}, where $J_v=J_\mathrm{R}+L_z/2+J_z$. The dark blue line is the line of $\Omega_\mathrm{pr}=\Omega_\mathrm{p}$. \textit{Right}: 2D distribution of $|\Omega_\mathrm{pr}-\Omega_\mathrm{p}|/\Omega$ on the same plane. \textcolor{black}{In both panels, the dark green line corresponds to the condition \lbd=0.} The light green line is the same line but smoothed via the spline approximation.}}
\label{fig:LB-map}
\end{figure*}
%%%%%%%%%%%%%%%%%%%%%%%%%%%%%%%%%%%%%%%%%%%%%%%%%%%%%%%
%%%%%%%%%%%%%%%%%%%%%%%%%%%%%%%%%%%%%%%%%%%%%%%%%%%%%%%

%%%%%%%%%%%%%%%%%%%%%%%%%%%%%%%%%%%%%%%%%%%%%%%%%%%%%%%
% Fig. 9
%%%%%%%%%%%%%%%%%%%%%%%%%%%%%%%%%%%%%%%%%%%%%%%%%%%%%%%
\begin{figure*}
\center{\includegraphics[width=\linewidth]{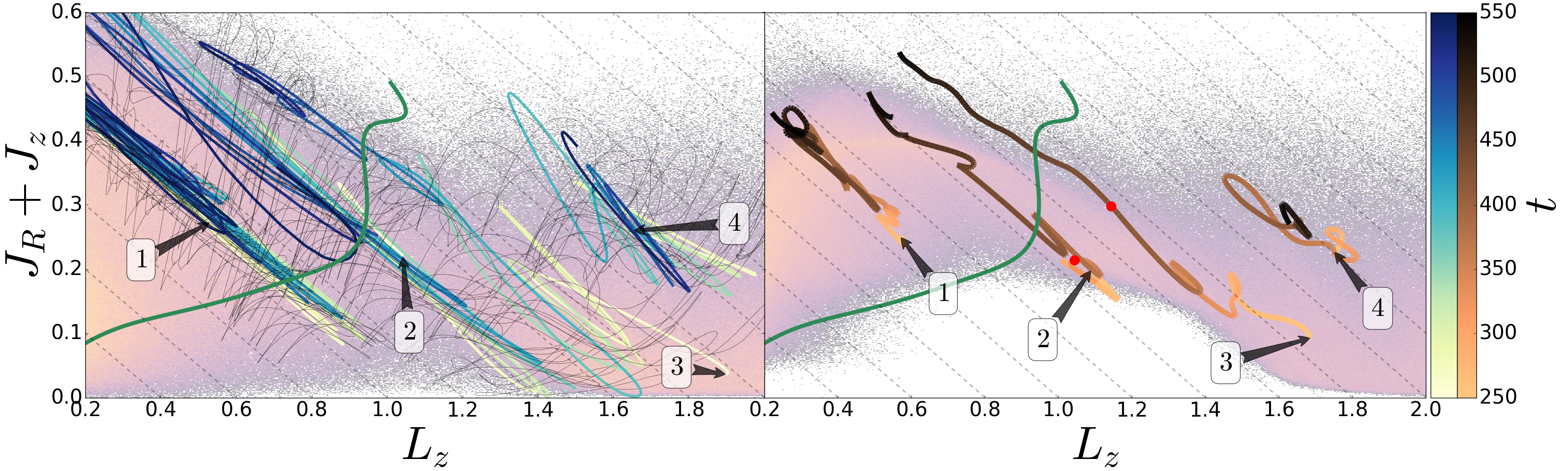}}
\caption{\textcolor{black}{Tracks of several orbits on the plane $(L_z, J_\mathrm{R}+J_z)$ (\textit{left}: short-term averaged actions; \textit{right}: actions averaged over librations). Arrows indicate the position of the beginning of the track. The background is the density distribution at $t=400$. The thick green line is the line of zero value of the LB-derivative smoothed via the spline approximation at $t=400$. Each track line is plotted in different colours (see the colour bar) to show how actions change with time. The thick red dot on the right plot indicates the moment of the orbital behaviour changes (see Section~\ref{sec:trapping}).}}
\label{fig:LB-map_density_tracks}
\end{figure*}
%%%%%%%%%%%%%%%%%%%%%%%%%%%%%%%%%%%%%%%%%%%%%%%%%%%%%%%
%%%%%%%%%%%%%%%%%%%%%%%%%%%%%%%%%%%%%%%%%%%%%%%%%%%%%%%

%%%%%%%%%%%%%%%%%%%%%%%%%%%%%%%%%%%%%%%%%%%%%%%%%%%%%%%
% Fig. 10
%%%%%%%%%%%%%%%%%%%%%%%%%%%%%%%%%%%%%%%%%%%%%%%%%%%%%%%
\begin{figure*}
\begin{center}
\centering
    \includegraphics[width=\textwidth]{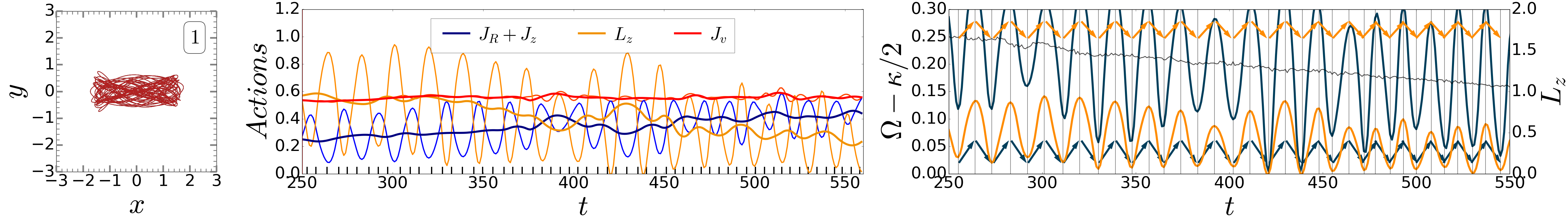}\\%\bigskip
    \includegraphics[width=\textwidth]{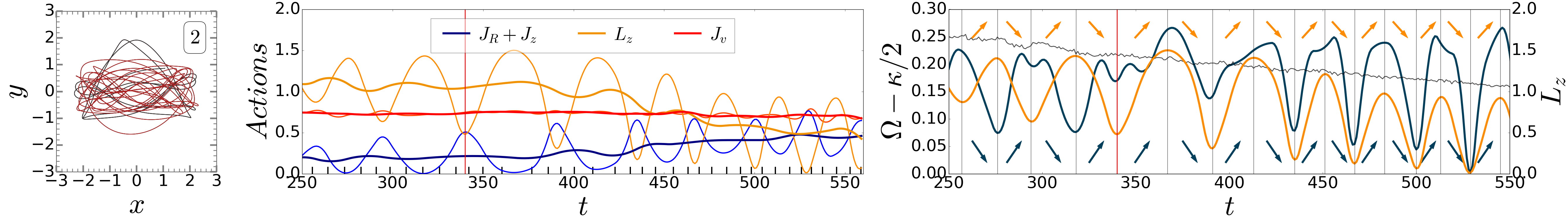}\\
    %\bigskip
    \includegraphics[width=\textwidth]{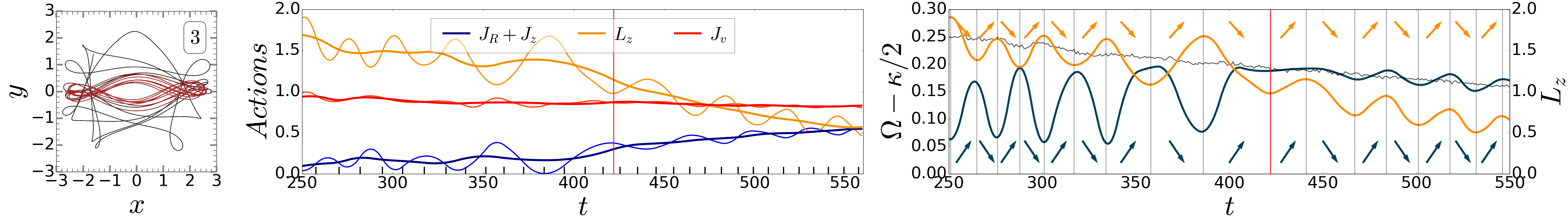}\\
    %\bigskip
    \includegraphics[width=\textwidth]{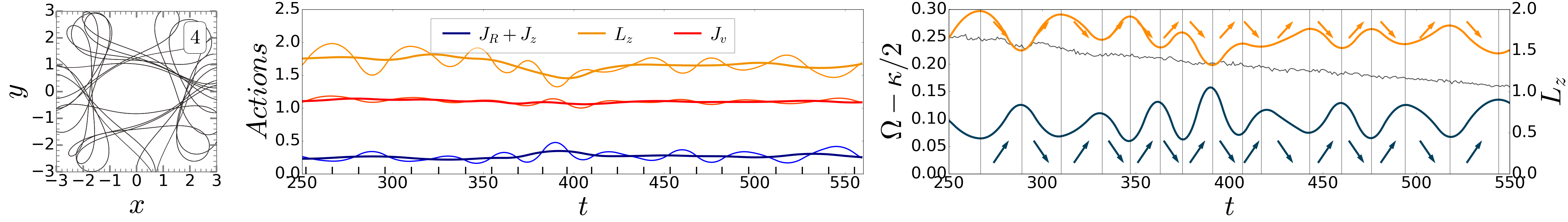}
\end{center}
%%%%%%%%%%%%%%%%%%%%%%%%%%%%%%%%%%%%%%%%%%%%%%%%%%%%%%%%%%%%%%%%%%%%%%%%%%%%
\caption{\textcolor{black}{\textit{Left:} Orbits on $xy$-plane over the time interval from $t=250$ to $t=550$. \textit{Black} and \textit{red} colours correspond to negative and positive values of LB-derivative, respectively. \textit{Centre:} Evolution of middle-term (thin lines) and secular (thick lines) orbital actions $(J_\mathrm{R}+J_z, L_z, J_f)$. 
%The short marks at the bottom show the moments of passage of the apocentres of the orbit. 
\textit{Right:} Coevolution of middle-term $L_z$ (\textit{orange} line) and $\Omega_\mathrm{pr}=\Omega-\kappa/2$ (\textit{blue} line). Coloured arrows indicate the direction of changes in the values during the time between two adjacent extrema of $L_z$. The red vertical line in the centre and right plots corresponds to the time when the sign of LB-derivative is changed.
%The evolution in a stationary rotating potential was separately traced for orbits number 2 and 3 before and after capture in the bar.
}}
\label{fig:orbits_evolution}
\end{figure*}
%%%%%%%%%%%%%%%%%%%%%%%%%%%%%%%%%%%%%%%%%%%%%%%%%%%%%%%
%%%%%%%%%%%%%%%%%%%%%%%%%%%%%%%%%%%%%%%%%%%%%%%%%%%%%%%

%%%%%%%%%%%%%%%%%%%%%%%%%%%%%%%%%%%%%%%%%%%%%%%%%%%%%%%
% Fig. 11
%%%%%%%%%%%%%%%%%%%%%%%%%%%%%%%%%%%%%%%%%%%%%%%%%%%%%%%
\begin{figure*}
\center{\includegraphics[width=\linewidth]
%{img/new/hist_Lz_Jf_Mb0_500_barnobar_LBder_lib_apo.png}}
{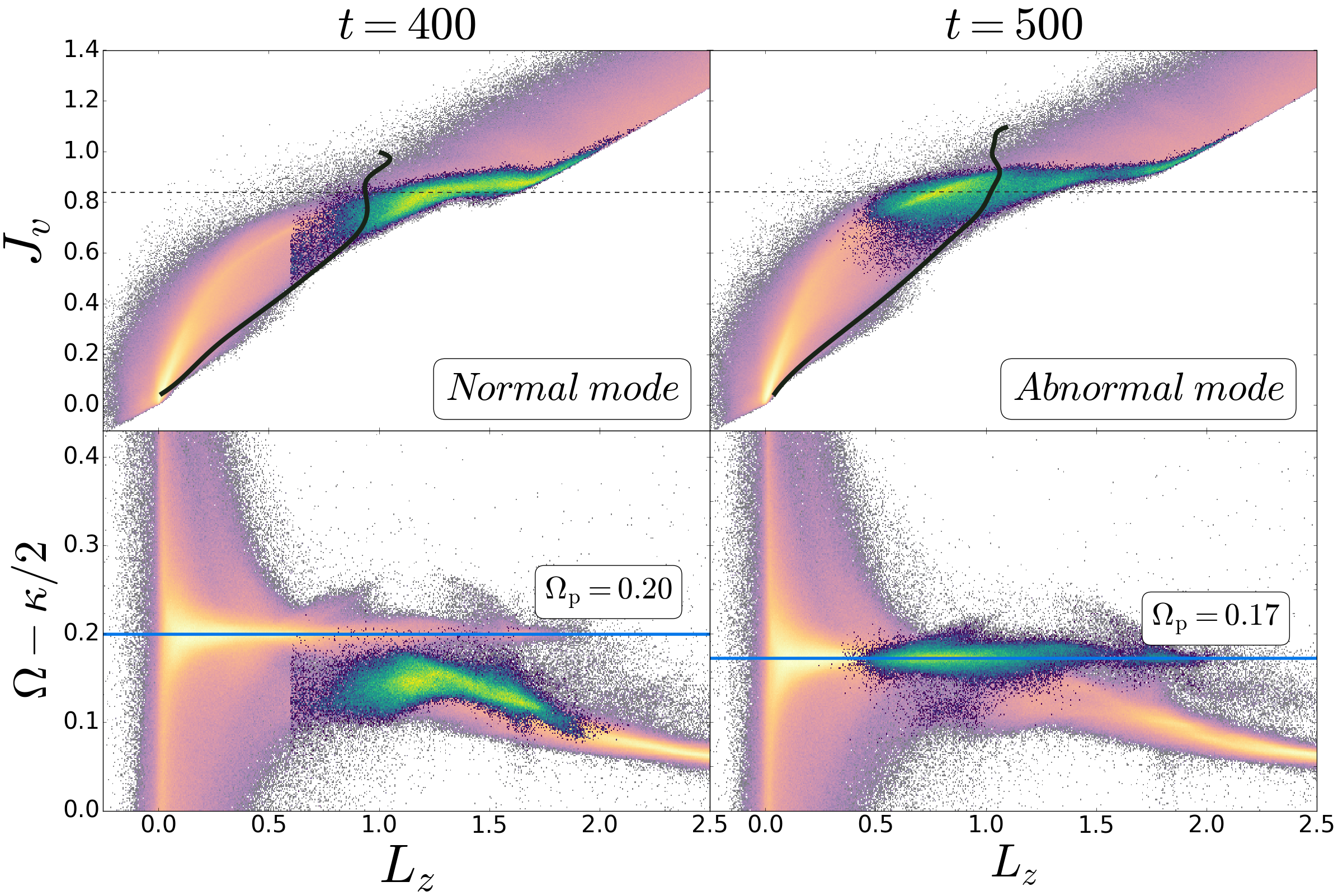}}
\caption{\textcolor{black}{\textit{Upper row}: 
The 2D density distribution of particles in $N$-body model over $L_z$--\textcolor{black}{$J_v$} plane for $t=400$ (\textit{left}) and $t=500$ (\textit{right}). The thick black line is the location of the line of \lbd$=0$ (smoothed via spline approximation) at the corresponding point in time. Actions are averaged over librations. The green areas show how particles changing the sign of their \lbd from negative to positive are distributed on the maps for times $t=400$ and $t=500$, respectively. The dashed horizontal line is the line of the median value of \textcolor{black}{$J_v$} for `green' particles.
\textit{Bottom row}: Distributions of particle precession rate depending on the angular momentum $z$-projection $L_z$. A straight blue line corresponds to the bar pattern speed with the value indicated in the annotating textbox. The green areas mean the same as in the top plots.}}
\label{fig:LB-map_density}
\end{figure*}
%%%%%%%%%%%%%%%%%%%%%%%%%%%%%%%%%%%%%%%%%%%%%%%%%%%%%%%
%%%%%%%%%%%%%%%%%%%%%%%%%%%%%%%%%%%%%%%%%%%%%%%%%%%%%%%

%%%%%%%%%%%%%%%%%%%%%%%%%%%%%%%%%%%%%%%%%%%%%%%%%%%%%%%
% Fig. 12
%%%%%%%%%%%%%%%%%%%%%%%%%%%%%%%%%%%%%%%%%%%%%%%%%%%%%%%
\begin{figure*}
\centering
%\center{\includegraphics[width=\linewidth]{img/Map_xy_xz/nobar415_bar485_galaxy_Mb0_z_mean_small.png}}
\center{\includegraphics[width=\linewidth]{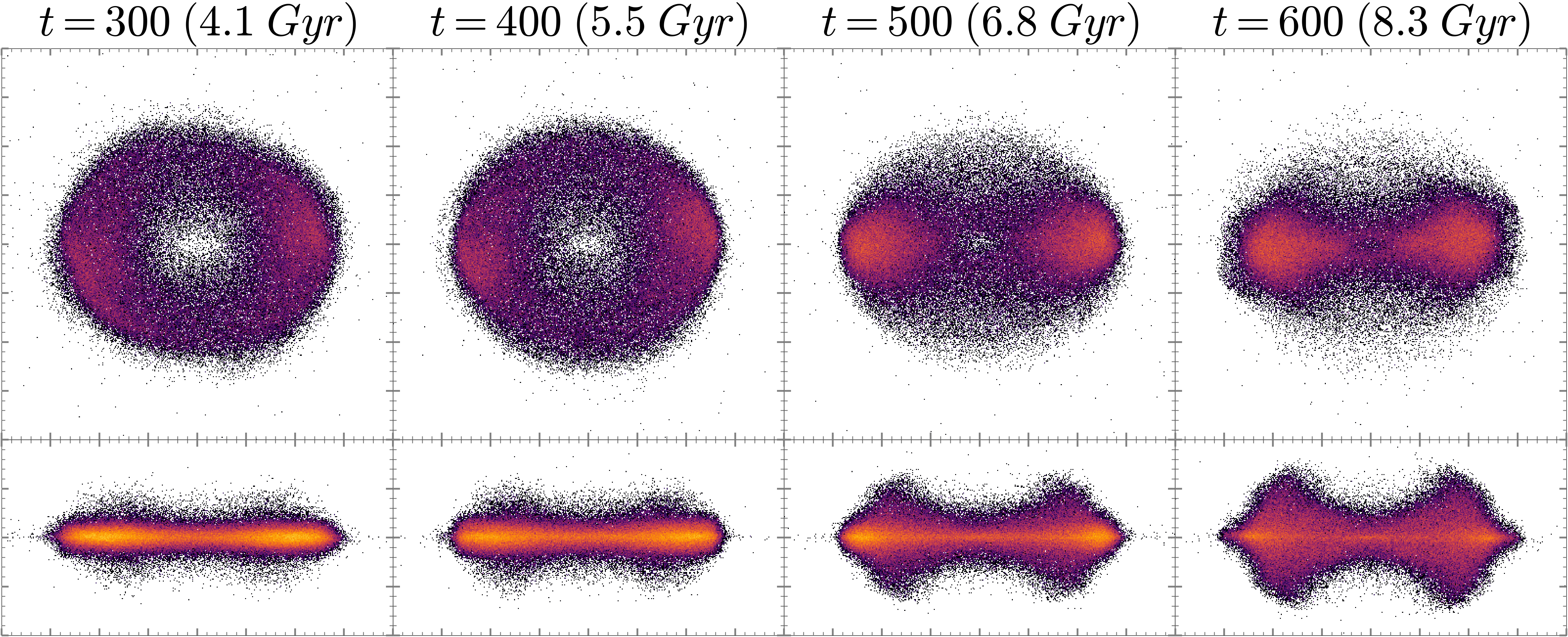}}
\caption{\textcolor{black}{The face-on (\textit{upper row}) and the side-on (\textit{bottom row}) density plots of particles whose orbits change their behaviour from normal to abnormal over a time interval of $t=400-500$. Face-on view is displayed in the square $(xy) = (-4,4) \times (-4,4)$ while side-on view is displayed in the rectangle $(xz) = (-4,4)\times(-2,2)$.}}
\label{fig:density}
\end{figure*}
%%%%%%%%%%%%%%%%%%%%%%%%%%%%%%%%%%%%%%%%%%%%%%%%%%%%%%%
%%%%%%%%%%%%%%%%%%%%%%%%%%%%%%%%%%%%%%%%%%%%%%%%%%%%%%%

%%%%%%%%%%%%%%%%%%%%%%%%%%%%%%%%%%%%%%%%%%%%%%%%%%%%%%%
%%%%%%%%%%%%%%%%%%%%%%%%%%%%%%%%%%%%%%%%%%%%%%%%%%%%%%%
%\begin{figure}
%\centering
%\includegraphics[width=\linewidth]{img/LB_map/Mb0_415_new_LzJR_mean_Wpr-Wp:W_zoom.png}
%\caption{The relative difference between the precession rate of the orbits and the bar pattern speed. The line of \lbd$=0$ is marked by green.}
%\label{fig:slowness}
%\end{figure}
%%%%%%%%%%%%%%%%%%%%%%%%%%%%%%%%%%%%%%%%%%%%%%%%%%%%%%%
%%%%%%%%%%%%%%%%%%%%%%%%%%%%%%%%%%%%%%%%%%%%%%%%%%%%%%%

%%%%%%%%%%%%%%%%%%%%%%%%%%%%%%%%%%%%%%%%%%%%%%%%%%%%%%%
% Fig. 13
%%%%%%%%%%%%%%%%%%%%%%%%%%%%%%%%%%%%%%%%%%%%%%%%%%%%%%%
\begin{figure}
\centering
%\center{\includegraphics[width=\linewidth]{img/Histograms/hist_part_nobar400_bar500_mean_sns_R_sm.png}}
\center{\includegraphics[width=\linewidth]{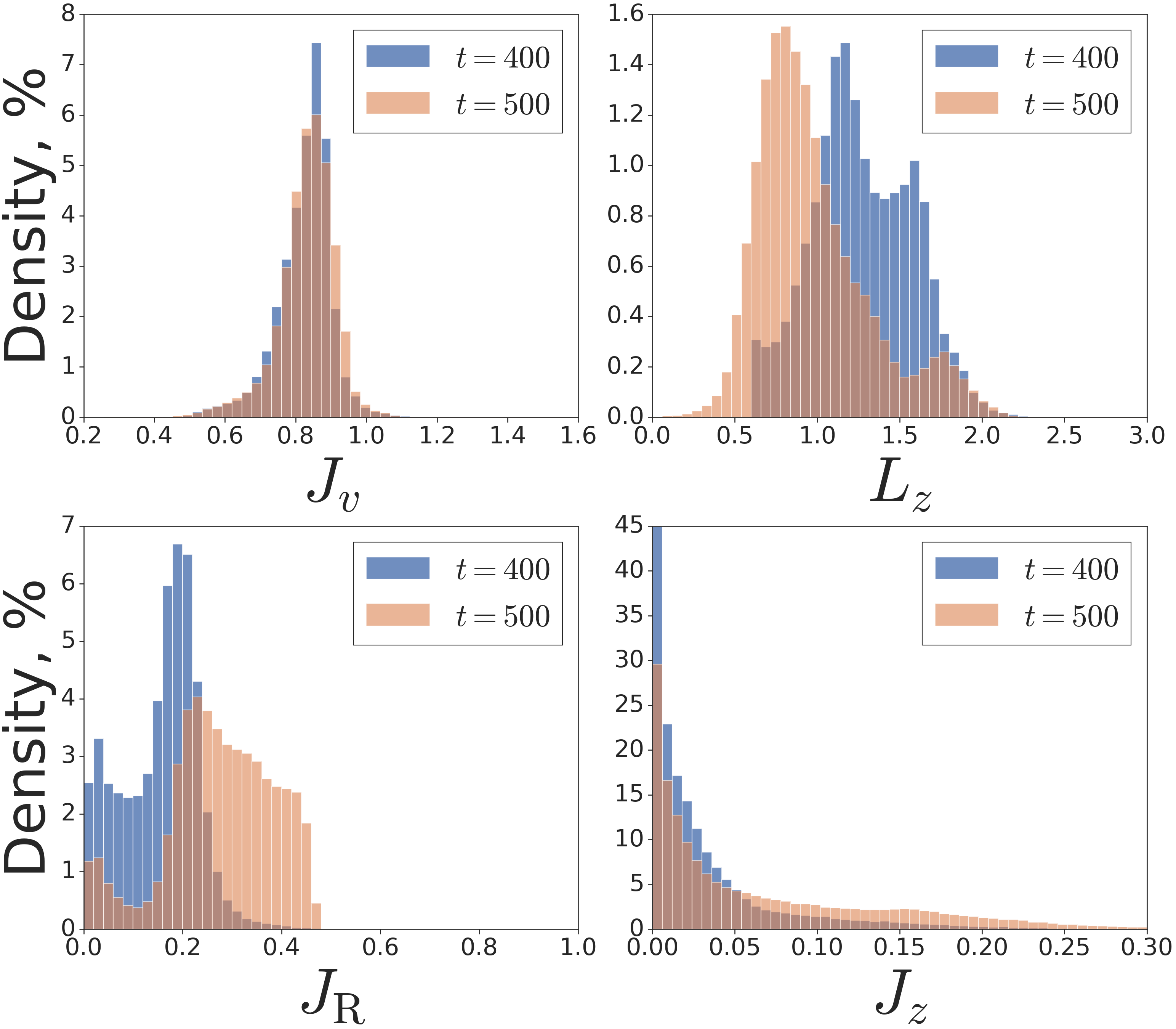}}
\caption{\textcolor{black}{Distributions of \textcolor{black}{$J_v$}, $L_z$, $J_\mathrm{R}$, and $J_z$  for orbits changing the sign of \lbd in the time interval $400-500$ at the beginning ($t=400$) and at the end ($t=500$) of particle path.}}
\label{fig:hist}
\end{figure}
%%%%%%%%%%%%%%%%%%%%%%%%%%%%%%%%%%%%%%%%%%%%%%%%%%%%%%%
%%%%%%%%%%%%%%%%%%%%%%%%%%%%%%%%%%%%%%%%%%%%%%%%%%%%%%%

%%%%%%%%%%%%%%%%%%%%%%%%%%%%%%%%%%%%%%%%%%%%%%%%%%%%%%%
% Fig. 14
%%%%%%%%%%%%%%%%%%%%%%%%%%%%%%%%%%%%%%%%%%%%%%%%%%%%%%%
\begin{figure}
\centering
%\center{\includegraphics[width=\linewidth]{img/Histograms/hist_part_nobar400_bar500_mean_sns_R_sm.png}}
\center{\includegraphics[width=\linewidth]{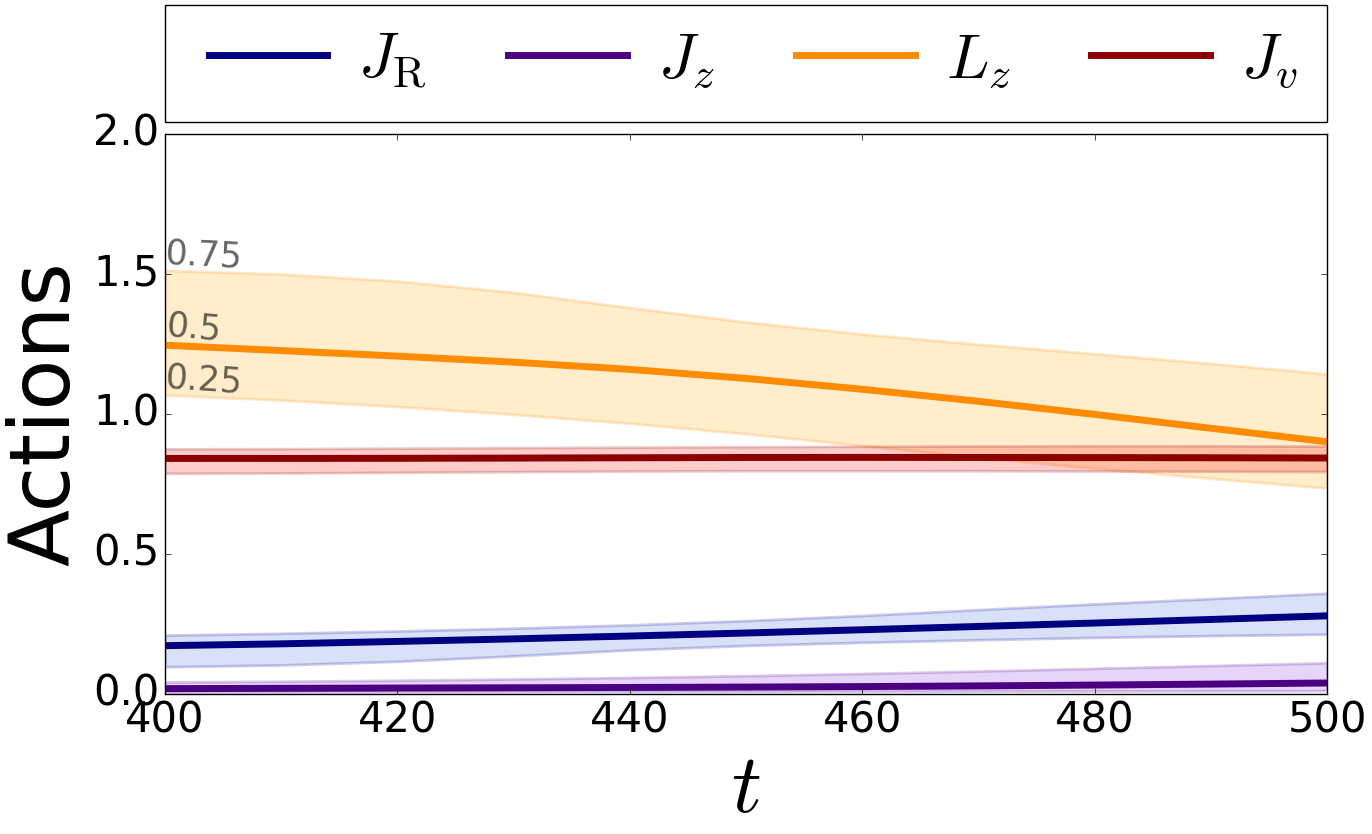}}
\caption{\textcolor{black}{Evolution of median action values (\textcolor{black}{$J_v$}, $L_z$, $J_\mathrm{R}$, and $J_z$) for orbits changing the sign of \lbd in the time interval $400-500$. Dashed areas indicate the first quartiles of the corresponding distributions.}}
\label{fig:mean_actions}
\end{figure}
%%%%%%%%%%%%%%%%%%%%%%%%%%%%%%%%%%%%%%%%%%%%%%%%%%%%%%%
%%%%%%%%%%%%%%%%%%%%%%%%%%%%%%%%%%%%%%%%%%%%%%%%%%%%%%%

%%%%%%%%%%%%%%%%%%%%%%%%%%%%%%%%%%%%%%%%%%%%%%%%%%%%%%%
% Fig. 13
%%%%%%%%%%%%%%%%%%%%%%%%%%%%%%%%%%%%%%%%%%%%%%%%%%%%%%%
% \begin{figure}
% \centering
% %\center{\includegraphics[width=\linewidth]{img/Histograms/kde2d_deltaJfJf_nobar400bar500_sns_R_sm.png}}
% \center{\includegraphics[width=\linewidth]{img/new/orbits/hist_Lz_Jf_Mb0_500_mean_barnobar_lib_apo.png}}
% \caption{Changes in the adiabatic invariant $J_f$ of the particles that are being trapped in the bar in the time interval $t=400-500$.}
% \label{fig:deltajf}
% \end{figure}
%%%%%%%%%%%%%%%%%%%%%%%%%%%%%%%%%%%%%%%%%%%%%%%%%%%%%%%
%%%%%%%%%%%%%%%%%%%%%%%%%%%%%%%%%%%%%%%%%%%%%%%%%%%%%%%

%-----------------------------------------------------%
%-----------------------------------------------------%
%\subsection{Different time scales during trapping}
%\label{sec:time_scales}
% Sec. 5.1.
%-----------------------------------------------------%
%-----------------------------------------------------%
\textcolor{black}{When analysing $N$-body simulations, we must distinguish between processes occurring on different time scales. The slow \textit{secular} changes in actions and frequencies occur simultaneously with the rapid oscillations of the corresponding actions and frequencies in the rotating barred potential, and \textit{also} simultaneously with longer-term librations. Therefore, to trace the longer-term evolution of the actions and precession rates, we should use their averaged-in-time values obtained via the procedure described in Sec.~\ref{sec:mean_actions_librations}. But first, let us analyse the Lynden-Bell maps, free from short-term oscillations, i.e. calculated according to the scheme presented in Sec.~\ref{sec:mean_actions_oscillations}. In this case, we investigate actions calculated directly in the $N$-body potential but averaged over small-amplitude oscillations. There will be inevitable librations in the dependence of such actions on time, i.e. periodic changes of a moderate amplitude on a time scale of the order of one revolution of the particle.}

%-----------------------------------------------------%
%-----------------------------------------------------%
\subsection{The Lynden-Bell maps averaged over short-term oscillations}
\label{sec:LB_averaged}
% Sec. 5.1.
%-----------------------------------------------------%
%-----------------------------------------------------%
The Lynden-Bell's~(\citeyear{Lynden-Bell1979}) idea of the bar formation due to the mutual attraction and alignment of precessing stellar orbits is based on two interconnected assumptions. The first one is that an orbit should precess relatively slowly with respect to the bar. And the second one, which stems from the first, is that \textcolor{black}{the orbit, falling into the trap, must retain} its adiabatic invariant $J_f$. 
\textcolor{black}{The last condition means that the capturing process is slow compared to other characteristic times. We understand the ``slowness'' of all processes precisely in this sense, namely the precession of the orbits that are being trapped is slow if the precession rate is close to the speed of the pattern, and the difference of these two quantities, divided by the angular velocity $\Omega$, is close to zero. In this case, even if the bar pattern speed slowly decreases, the adiabatic invariant will be preserved for the orbits that are being trapped. Changes in the action variables on the libration time scale have no relation to the characteristic time of orbital trapping in the bar. For capturing, what is important is the time of gradually accumulating secular changes rather than the time of periodic changes in actions, that occur with the preservation of $J_f$ throughout the entire period of capturing.}
\par
Here we investigate whether the above two assumptions hold true for the orbits in the `abnormal' region and, in particular, for those orbits that are being trapped in the bar in our $N$-body model.
\par
The left panels of Fig.~\ref{fig:LB-map} show how the \textcolor{black}{precession rate maps averaged over short-term oscillations (see Section~\ref{sec:mean_actions_oscillations}) look like. Building maps in Fig.~\ref{fig:LB-map}, we do not get rid of librations of actions and frequencies that are inevitably present for orbits that are far from circular or even regular. It is important that, unlike the maps in Fig.~\ref{fig:Jf_Lz}, maps in Fig.~\ref{fig:LB-map} use actions, frequencies, and precession rates determined directly in the \textit{non-axisymmetric} $N$-body potential.} 
%Compared to Fig.~\ref{fig:Jf_Lz}, there is now a cavity at about $L_z\approx0.3$. This cavity is occupied by orbits with negative averaged precession rates, which we do not depict here. The appearance of such orbits is rather controversial since the bar rotates as a whole in the ``positive'' direction, and the orbits should support this rotation. Since such orbits occupy the most central areas, which we are not interested in, we do not specifically study the orbits further here. We nevertheless note that there is a clear possibility that negative precession rates can arise due to the errors in frequency measurements, which become greater due to strong bar influence in the central areas, or too rough averaging. %todo{NS}{Anton, Vitya, explain, please, the shape of the map on the left and the shape of zero value of partial derivative \lbd. It will be useful to highlight the line $\Omega - \kappa/2 = \Omega_\mathrm{p}$ (=0.191).} 
Judging by the colour bar, it can be seen that the particles in the abnormal region and immediately beyond zero of \lbd in the normal region precess with a speed close to that speed of the bar (dark blue line). 
\par
The right plot of Fig.~\ref{fig:LB-map} shows how the difference between the bar pattern speed and the orbit precession rate normalised on the value of the angular frequency is distributed over $L_z$ and $J_f$. As can be seen, the precession rate of the orbits is very close to the bar pattern speed (up to $\sim10\%$, in the left plot of Fig.~\ref{fig:LB-map}, the blue lines show the contours of 10\% and 15\% differences from zero, see also the tips on the colour bar.) in the vicinity of the line \lbd$=0$, i.e., the condition $|\Omega_\mathrm{pr}-\Omega_\mathrm{p}|/\Omega \ll 1$ is well satisfied, which is actually equivalent to conservation of the integral of motion $J_f$. \textcolor{black}{This is true both for most of the orbits to the left of the zero LB-derivative line and for a number of orbits adjacent to the line to the right.} Thus, the orbits precess relatively slowly compared to the bar before trapping.
\par
%We should also note that the overall number of the particles to the left of the line \lbd$=0$ almost does no change compared to the case when we use \textcolor{black}{the map for instant values of actions as in Fig.~\ref{fig:Jf_Lz}}, i.e. the difference between two sets is about $6\times10^4$ particles (about 3\% of bar particles). 
\par
\textcolor{black}{Filling the maps with the values of $L_z$ and \textcolor{black}{$J_v$} in Fig.~~\ref{fig:Jf_Lz} and Fig.~\ref{fig:LB-map} is very similar, which is not surprising, since the smoothed values of actions determined via $r_\mathrm{max}$ and $z_\mathrm{max}$ in a non-axisymmetric potential well fit the dependence of instantaneous actions on time (Fig.~\ref{fig:actions_change}, second panel from the top) determined in an axisymmetric potential for given time moments. At the same time, the precession rates determined via the apocentres and the instantaneous precession rates, that are obtained by a routine from {\texttt{AGAMA}}, differ. This is precisely what is causing the difference in the position of the lines of zero partial derivative \lbd. On the maps presented in Fig.~\ref{fig:LB-map}, a long `corridor', that is formally a part of the normal region, appears. This `corridor' is adjacent to the line of circular velocities and stretches from $L_z\approx0.9$ towards very small $L_z$ values. However, there are no circular orbits in this region. The orbits there are all bar orbits that experience librations of actions. These orbits draw tracks on the plane ($L_z$; $J_\mathrm{R}+J_z$) with periodic changes in $L_z$, $J_\mathrm{R}$, and $J_z$, while preserving their \textcolor{black}{$J_v$}. 
Fig.~\ref{fig:LB-map_density_tracks} (left plot) shows a map of $L_z$ and $J_\mathrm{R}+J_z$ at $t=400$ with several representative orbital tracks from $t=250$ to $t=550$ plotted on it. The position of the line of the zero LB-derivative is also drawn for $t=400$. Particle~1 is already trapped in the bar by $t=400$, but, from time to time, it goes beyond the boundary of the zero derivative into the region where the derivative is formally negative due to librations.
Particles~2 and~3 are gradually drawn into the abnormal region, and, once there, continue to experience periodic changes in their actions. The dotted inclined lines in the plot are lines of equal values of \textcolor{black}{$J_v$}. It can be seen that particles 1, 2, and 3 move into the bar region, almost exactly following the lines of constant \textcolor{black}{$J_v$}. Particle 4 is in the disc and, during the time from $t=250$ to $t=550$, does not experience secular changes in its actions, but only periodic ones. In the next section, we will analyse the tracks of particles averaged over librations, i.e. we will trace the secular evolution of the actions of the particles that are trapped in the bar.}

%-----------------------------------------------------%
%-----------------------------------------------------%
\subsection{Trapping of particles}
\label{sec:trapping}
% Sec. 5.2.
%-----------------------------------------------------%
%-----------------------------------------------------%
\textcolor{black}{As can be seen from the right plot of Fig.~\ref{fig:LB-map_density_tracks}, if one averages the actions for all particles over librations, then orbits trapped in the bar by $t=400$ will lie completely in the abnormal region. Particle~1 is located in the abnormal region by $t=250$. Particle~2 will be there at $t\approx340$, and particle~3 at $t\approx420$, after they cross the zero LB-derivative line. At the same time, already trapped orbits or orbits that are being trapped demonstrate secular changes in their actions, despite the averaging over librations, and the trapping process of such orbits is rather slow (Fig.~\ref{fig:LB-map_density_tracks}, right plot, particle~3). Particle 4, which is not trapped in the bar, experiences periodic changes in actions (without secular evolution), and it does not yet have a chance of getting into the bar.}
\par
\textcolor{black}{Fig.~\ref{fig:orbits_evolution} shows the orbits of particles (left column), the tracks of which are presented on the maps in Fig.~\ref{fig:LB-map_density_tracks}. Fig.~\ref{fig:orbits_evolution} also demonstrates changes with time of actions for these particles, both non-averaged over librations and averaged. For particles 1, 2, and 3, the averaged values of $L_z$ (yellow lines) continue to systematically decrease even after $t=400$. This occurs not only due to an increase in $J_\mathrm{R}$ but also due to an increase in $J_z$, when the orbit, already trapped in the bar, begins to protrude from the mid-plane in the vertical direction. At the same time, for all particles the action \textcolor{black}{$J_v$} (red lines) is surprisingly preserved, i.e. the process of orbital evolution occurs adiabatically.}
\par
\textcolor{black}{Given the above, the criterion for orbital capture by a bar can be formulated in two ways. In the Lynden-Bell language, a particle can be considered to be trapped in a bar if it crosses the Lynden-Bell zero derivative line. Here, one needs to use actions averaged over librations. It is also possible to trace at what moment $L_z$ and $\Omega_\mathrm{pr}$ of a particular orbit begin to change synchronously in the same direction (increase or decrease simultaneously), without resorting to the procedure of averaging over librations. Since \textcolor{black}{$J_v$} is well preserved for orbits near the line of zero LB-derivative, such behaviour means that the orbit has individually acquired a positive value of the LB-derivative.}
\par
\textcolor{black}{The plots in the right column of Fig.~\ref{fig:orbits_evolution} demonstrate the changes with time of the $L_z$ (yellow lines) and $\Omega_\mathrm{pr}$ (grey lines) values non-averaged over librations. The evolution of $\Omega_\mathrm{p}$ (thin black line) is also plotted. The arrows (yellow and grey) point out the direction of $L_z$ and $\Omega_\mathrm{pr}$ changes. The red vertical line (middle and right plots in Fig.~\ref{fig:orbits_evolution}) marks the time moment when the changes of $L_z$ and $\Omega_\mathrm{pr}$ start to go in the same direction (to increase or to decrease simultaneously). In line with \citet{Lynden-Bell1979}, we can introduce the concepts of `normal' and `abnormal' orbital behaviour. To the left of the red vertical line, we observe normal behaviour, and to the right, there is a transition to abnormal behaviour\footnote{The moment when the transition from abnormal behaviour to normal occurs is indicated on the tracks in Fig.~\ref{fig:LB-map_density_tracks} (right plot) by a red bold dot.}}. 
\par
\textcolor{black}{The outline of the orbit in the abnormal mode is also marked in red (left column in Fig.~\ref{fig:orbits_evolution}). Orbit 1 by the time $t=250$ is already trapped in the bar (abnormal mode), so it is all drawn in red. Orbit 2, upon transition to the abnormal mode (practically coinciding with the moment of intersection of the line of zero LB-derivative), becomes elongated and lies completely inside a rectangle elongated along the major axis of the bar. At the same time, orbit 3, upon transition to the abnormal mode, acquires the shape of a regular bar orbit. All these changes correlate well with $\Omega_\mathrm{pr}$ behaviour in the right plots of Fig.~\ref{fig:orbits_evolution}. %For orbit 2, when $L_z$ starts to change synchronously with $\Omega_\mathrm{pr}$, $\Omega_\mathrm{pr}$, as if clinging to $\Omega_\mathrm{p}$, begins to oscillate around $\Omega_\mathrm{p}$. 
For orbit 2, when it goes into an abnormal mode, the oscillations of $\Omega_\mathrm{pr}$ are getting clung to $\Omega_\mathrm{p}$. For orbit 3, which acquires a regular shape, $\Omega_\mathrm{pr}$ literally sits on the line of changing $\Omega_\mathrm{p}$ values after the mode change. As for orbit 4, no changes occur in its behaviour during the time interval under consideration, and $\Omega_\mathrm{pr}$ always remains below the value of $\Omega_\mathrm{p}$.}
%-----------------------------------------------------%
%-----------------------------------------------------%
\subsection{Ensemble of trapped particles}
\label{sec:trapped_particles}
% Sec. 5.3.
%-----------------------------------------------------%
%-----------------------------------------------------%
\textcolor{black}{Now, using knowledge about the sign of \lbd for each orbit at any time moment, it is possible to identify an ensemble of particles trapped in the bar over a certain period of time and to monitor the evolution of these particles. Based on the results of the previous section, we identify such particles as those that change the sign of \lbd from negative to positive. Below we show that such particles are indeed those that are being trapped.}
%Now we can find the \textcolor{black}{the entire ensemble of} particles that experience trapping \textcolor{black}{over a certain period of time.} Based on the results of the previous section, we identify such particles as those that traverse the line of zero partial derivative \lbd. Below we show that such particles are indeed those that are being trapped. 
\par 
%\textcolor{black}{Recall that orbit trapping} is a continuous process, and the particles constantly traverse the line, \textcolor{black}{experiencing librations}.
We limit ourselves by selecting only two specific time moments $t=400$ and $t=500$, analysing the changes in particle actions that happen between these time moments.
\par 
\textcolor{black}{Fig.~\ref{fig:LB-map_density} (top row) shows the 2D density distribution of particles over $(L_z, J_f)$ plane. 
%with contours indicating the location of particles before and after capturing. 
The maps are obtained by averaging the number of particles in each bin of size of 0.005 both in $L_z$ and \textcolor{black}{$J_v$} at time moment $t=400$ (left plot) and $t=500$ (right plot). The action values were also averaged over librations. That is why the `corridor' of the normal region, adjacent to the line of circular orbits with small values of $L_z$, disappeared on the maps. The bar particles now lie almost entirely to the left of the Lynden-Bell zero derivative line, without going beyond it. 
%The exception is particles on the periphery of the bar with rather large values of $L_z$, which gradually join it. 
There is a narrow `bottleneck' near the value of $L_z\approx1.0$ with a small scatter in \textcolor{black}{$J_v$}, through which the disc particles fall into the trap of the bar.}
\par
\textcolor{black}{An additional, perhaps not so clear matter, is that the line of zero partial derivative changes its location with the evolution of the system potential. Therefore, we also mark the location of the line \lbd$=0$ at left and right plots for corresponding time moments\footnote{The position of the Lynden-Bell zero derivative line is calculated for $t=400$ (left plot) and $t=500$ (right plot) from the actions and precession rates averaged over librations.}. It can be seen that the line does not change significantly either its position or its shape in time.}
\par
\textcolor{black}{On the maps in Fig.~\ref{fig:LB-map_density} (top row), the particles that experience change from `normal' to `abnormal' mode from $t=400$ to $t=500$ are marked in green (superimposed areas with gradations of the green colour). These particles have a negative LB-derivative at the moment $t=400$ (left plot) and a positive derivative by the moment $t=500$ (right plot).}
\par
\textcolor{black}{Fig.~\ref{fig:LB-map_density} (upper left plot) shows that the particles marked in green (with \lbd$<0$) are located almost entirely to the right of the Lynden-Bell zero derivative line at $t=400$. A very small number of orbits with \lbd$>0$ that have not yet fallen into the trap of the bar is located on the left, in the abnormal region. We also want to note that we excluded a certain number of orbits from consideration. These are chaotic orbits in the central regions, which change the sign of the derivative several times and accidentally have a negative derivative at the time $t=400$, and a positive one at the time $t=500$. We formally cut off such orbits by the condition $L_z>0.6$ (see also Appendices~\ref{sec:appendix_LE} and \ref{sec:appendix_normal_abnormal} regarding chaotic orbits and their locations). We also would like to note that for an orbit to have a negative LB-derivative value and to be to the right of the zero LB-derivative line at a given time is not completely the same thing\footnote{We will return to this issue in Section~\ref{sec:normal_abnormal}}. However, the differences between these two properties are small and this mainly concerns the central chaotic orbits.}
\par
\textcolor{black}{We selected orbits marked in green as those that have \lbd$<0$ at $t=400$ and change the sign of the derivative before $t=500$ (\lbd$>0$). By the time $t=500$, most of these orbits have moved to the abnormal region (to the left of the line of the zero LB-derivative, Fig.~\ref{fig:LB-map_density}, upper right plot). A certain number of orbits continue to remain on the right, in the normal region. Among them, there are about 10\% of particles that have high values of $L_z>1.6$ and are adjacent to the line of circular orbits. These orbits have increased the rate of precession $\Omega_\mathrm{pr}$ to values close to $\Omega_\mathrm{p}$ during the studied time interval and are elliptical in shape, with the major axis elongated along the major axis of the bar. We will discuss these orbits in Section~\ref{sec:elliptical orbits}.}
\par
\textcolor{black}{In Fig.~\ref{fig:LB-map_density} (bottom row) we depict the dependence of the precession rate over the $z$-projection of the angular momentum. The bar forms a thick straight strip in this figure, while the not yet trapped particles (marked by green colour) lie on the descending branch of the precession rate curve. The latter is somewhat thick due to the differences in eccentricities of orbits. One can see that the bar perturbation rotates faster than the typical precession rate of the orbits it traps. By the time $t=500$ (bottom right plot), all trapped particles increase the precession rate to approximately the same value, the angular speed of the bar $\Omega_\mathrm{p} = 0.17$ at $t=500$. Even particles located in the region of almost circular orbits and having $L_z>1.6$ land on the `shelf' of the bar if their orbits manage to change their behaviour from normal to abnormal.}
\par
\textcolor{black}{Fig.~\ref{fig:density} presents the evolution of the subsystem of the particles whose orbits change their behaviour from normal to abnormal over the considered time interval and end up in the `bar' region.} Initially, these particles form a ring-like structure (although, with a clearly inhomogeneous density profile, the upper left plot of Fig.~\ref{fig:density}). In the course of the evolution, the structure gradually changes its shape to ellipsoidal, becoming part of the bar. Interestingly, the characteristic farfalle-like shape is formed simultaneously in a side-on view. This indicates that the particles increase their vertical excursions and start to support the so-called B/PS bulge (a vertically thick part of the bar).
\subsection{Secular evolution of actions}
\label{sec:LB_secular}
% Sec. 5.4.
%-----------------------------------------------------%
%-----------------------------------------------------%
The distributions of the actions and the adiabatic invariant \textcolor{black}{$J_v$} for the trapped particles for two time moments, $t=400$ and $t=500$, are shown in Fig.~\ref{fig:hist}. The histograms show that the particles lose their angular momentum while increasing their radial and vertical actions. This is consistent with an overall change in morphology \textcolor{black}{of snapshots} observed in Fig.~\ref{fig:density}: orbits become more elongated and acquire greater vertical excursions. \textcolor{black}{At the same time, \textcolor{black}{$J_v$} is very well preserved over this period of time, and the overall shape of the corresponding distribution does not change. Fig.~\ref{fig:mean_actions} shows the evolution of median action values over time and their deviations by one quartile. It is clearly seen that the median for the entire distribution over \textcolor{black}{$J_v$} practically does not change.} The trapped orbits keep their \textcolor{black}{$J_v$} with an accuracy of about \textcolor{black}{4\%-5\%}. \textcolor{black}{The preservation of \textcolor{black}{$J_v$} is in good agreement with the slow, secular decrease in $L_z$ and the increase in $J_\mathrm{R}$ and $J_z$.}
%Fig.~\ref{fig:deltajf} shows how $J_f$ changes for individual particles depending on the $J_f$ value itself. Compared to the histogram presented in Fig.~\ref{fig:hist}, the figure demonstrates that there are some outliers with a significant change in $J_f$, but it is important that their number is relatively small. However, if we take into account that there is also a change in $J_z$, and the change should be accounted for when one calculates $J_f$, then the preservation of $J_f$ turns out to be good (blue colour region). Thus, the trapped orbits keep their $J_f$ with an accuracy of about 2\%-3\%. 
%It is a good finding in $N$-body models in general, where various non-linear interactions are known to take a place.
\par
Summing up the results of the Section~\ref{sec:orbits_trapping}, we found that orbits from the normal region, where \lbd$<0$, eventually lose their angular momentum and become trapped by a bar. 
\textcolor{black}{To be more precise, those orbits that change their behaviour from normal to abnormal, fall into the trap of the bar. Before being trapped, such orbits systematically decrease $L_z$ and increase the precession rate $\Omega_\mathrm{pr}$. After falling into the trap, the orbits begin to change $L_z$ in accordance with $\Omega_\mathrm{pr}$. In the abnormal mode, $L_z$ increases or decreases simultaneously with $\Omega_\mathrm{pr}$, which oscillates around the value of the angular velocity of the pattern $\Omega_\mathrm{p}$.} 
In addition, we find that the trapped orbits initially precess slowly with respect to the bar, i.e. the Lynden-Bell condition for particle trapping is certainly satisfied. In this case, the trapping occurs with the preservation of the adiabatic invariant \textcolor{black}{$J_v$}.

%-----------------------------------------------------%
%-----------------------------------------------------%
\section{Discussion}
\label{sec:discussion}
% Sec. 6.
%-----------------------------------------------------%
%-----------------------------------------------------%

%-----------------------------------------------------%
%-----------------------------------------------------%
\subsection{Normal/abnormal regions versus normal/abnormal orbital modes}
\label{sec:normal_abnormal}
% Sec. 6.1.
%-----------------------------------------------------%
%-----------------------------------------------------%
\textcolor{black}{A crucial fact that we show for the first time is that the mature $N$-body bar is indeed separated from the disc by the Lynden-Bell zero derivative line (Fig.~\ref{fig:Jf_Lz} and Fig.~\ref{fig:density_regions}). The latter divides the model into two regions, the abnormal one (the bar region, where the partial derivative of angular momentum over the precession rate is positive) and the normal region (disc area with \lbd$<0$). However, Fig.~\ref{fig:orbits_evolution} and Fig.~\ref{fig:LB-map_density} demonstrate that, for understanding the process of adding new particles to the bar, the important concept is not the concept of a normal/abnormal region, but the behaviour of individual orbits, i.e. the mode in which the orbit is, normal/abnormal.}
\par
\textcolor{black}{In normal mode, the orbit, before joining the bar, decreases its angular momentum while simultaneously increasing its precession rate until the latter approaches the angular speed of the pattern. After this, the orbit switches to abnormal mode and begins to change its angular momentum in accordance with the precession rate (Fig.~\ref{fig:orbits_evolution}). In turn, the precession rate begins to oscillate around $\Omega_\mathrm{p}$ (orbit 2) or even literally gets glued to $\Omega_\mathrm{p}$ (orbit 3). This prevents the orbit from jumping out of the bar trap. Now the orbit is locked in it.}
%\par
%\textcolor{blue}{Chaotic with changing sign of \lbd.}
\par
\textcolor{black}{It is important to note that, at the same time, an orbit in the abnormal mode is not necessarily located to the left of the zero LB-derivative line (Fig.~\ref{fig:LB-map_density}, upper right plot, green area). It turns out that normal/abnormal regions and normal/abnormal orbital modes are close, but not the same concept. In the normal/abnormal region, pixel-by-pixel averaging of orbital characteristics occurs. The corresponding pixel shows typical orbital behaviour rather than individual behaviour. That is why, in Fig.~\ref{fig:LB-map_density}, we find orbits that reside in the normal region, have large values of $L_z$, and, at the same time, change the sign of \lbd from negative to positive over a time interval of $t=400$--$500$.}

%-----------------------------------------------------%
%-----------------------------------------------------%
\subsection{Bar orbits with large values of $L_z$}
\label{sec:elliptical orbits}
% Sec. 6.2.
%-----------------------------------------------------%
%-----------------------------------------------------%
%%%%%%%%%%%%%%%%%%%%%%%%%%%%%%%%%%%%%%%%%%%%%%%%%%%%%%%
% Fig. 15
%%%%%%%%%%%%%%%%%%%%%%%%%%%%%%%%%%%%%%%%%%%%%%%%%%%%%%%
\begin{figure*}
\begin{center}
\centering
\includegraphics[width=\textwidth]{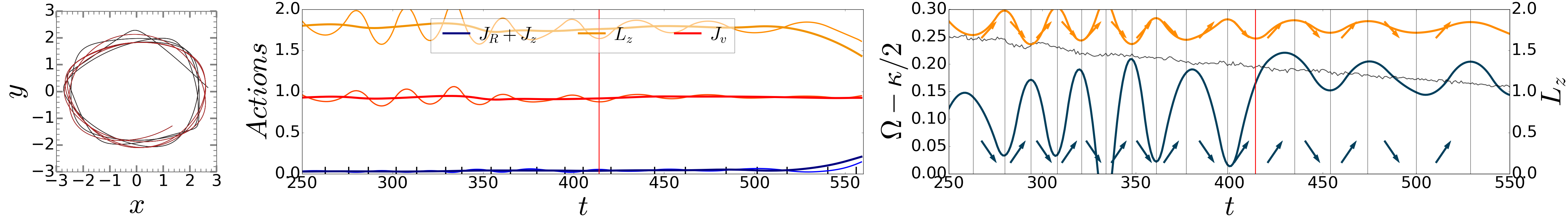}   
\bigskip
\includegraphics[width=\textwidth]{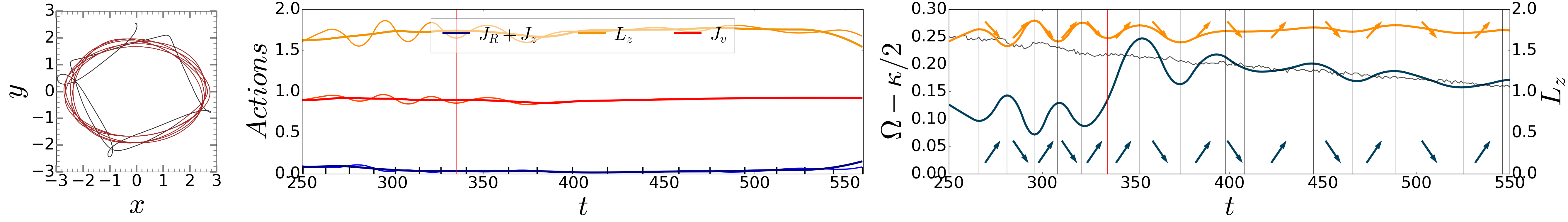}
\bigskip
\includegraphics[width=\textwidth]{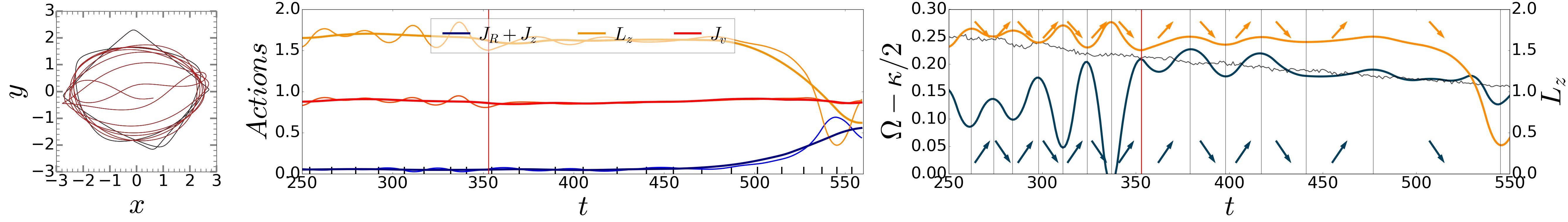}
\end{center}
%%%%%%%%%%%%%%%%%%%%%%%%%%%%%%%%%%%%%%%%%%%%%%%%%%%%%%%%%%%%%%%%%%%%%%%%%%%%
\caption{\textcolor{black}{The same as in Fig.~\ref{fig:orbits_evolution} but for orbits with initially large $L_z$.}}
\label{fig:orbits_largeLz}
\end{figure*}
%%%%%%%%%%%%%%%%%%%%%%%%%%%%%%%%%%%%%%%%%%%%%%%%%%%%%%%
%%%%%%%%%%%%%%%%%%%%%%%%%%%%%%%%%%%%%%%%%%%%%%%%%%%%%%%
\par
\textcolor{black}{Examples of three orbits with the abnormal mode of oscillations and large values of $L_z$ are shown in Fig.~\ref{fig:orbits_largeLz} Their shape is close to elliptical (left column of Fig.~\ref{fig:orbits_largeLz}). The orbit shown in the top panel changes the orbital mode immediately after $t=400$ (red vertical line in the middle and right subplots), while, for the orbits from the middle and lower panels, the change occurs near $t=350$. It is after the change of sign of the \lbd that the orbits acquire an elliptical shape, and their major axis becomes elongated along the major axis of the bar. For the upper and middle orbits, this shape practically does not change, just as their $L_z$, the value of which keeps these orbits to the right of the line of the zero LB-derivative, does not change. Until these orbits reduce $L_z$ and increase $J_\mathrm{R}$ and $J_z$, they support the thin peripheral part of the bar outside the B/PS bulge (see Fig.~\ref{fig:density_regions}, bottom middle plot).}
\par
\textcolor{black}{The example of the lower orbit shows that in the future the shape of the orbit may change. In this particular example, the orbit becomes more elongated and looks like candy in a candy wrapper, like orbit~3 in Fig.~\ref{fig:orbits_evolution} after falling in the bar trap. In this case, the $J_\mathrm{R}$ increases, and the value of $L_z$ drops below 1.2, after which the orbit, without changing $\Omega_\mathrm{pr} \approx \Omega_\mathrm{p}$ and maintaining \textcolor{black}{$J_v$}, moves to the region to the left of the line of zero LB-derivative. In the future, it can increase $J_z$, leaving the mid-plane and populating the B/PS bulge.}

%-----------------------------------------------------%
%-----------------------------------------------------%
\subsection{Expanded picture of orbital trapping by a bar}
\label{sec:how_trapping}
% Sec. 6.3.
%-----------------------------------------------------%
%-----------------------------------------------------%
%\textcolor{black}{Angular momentum exchange...}
%\textcolor{black}{What is the stopping factor of the bar growth then?}
%A crucial fact that we show for the first time is that the $N$-body bar growth occurs because the orbits are trapped from the normal region, where the partial derivative of angular momentum over the precession rate is negative. 
\textcolor{black}{The fact that the $N$-body bar growth occurs because the orbits are trapped mainly from the normal region contradicts Lynden-Bell's~(\citeyear{Lynden-Bell1979}) idea that the bar should form and grow from the slowly precessing orbits lying within the abnormal region.
We should note that this is not the first time that the role of the normal orbits in the building of the bar has been discussed. In particular, \citet{Pol2020a,Pol2020b} found that normal orbits can facilitate the bar growth under certain conditions.}
% Here we would like to provide some arguments that such behaviour can still be explained within the framework proposed by Lynden-Bell and developed by \citet{Pol2020a,Pol2020b}.}

%%%%%%%%%%%%%%%%%%%%%%%%%%%%%%%%%%%%%%%%%%%%%%%%%%%%%%%
% Fig. 14
%%%%%%%%%%%%%%%%%%%%%%%%%%%%%%%%%%%%%%%%%%%%%%%%%%%%%%%
%\begin{figure}
%\centering
%\center{\includegraphics[width=\linewidth]{img/Histograms/Lz_omega-kappa:2_mean.png}}
%\center{\includegraphics[width=\linewidth]{img/Histograms/Lz_omega-kappa:2_mean_barnobar_415485_sm.png}}
%\center{\includegraphics[width=\linewidth]{img/new/Lz_omega-kappa:2_barnobar_400500_lib_LBder_1_cmin3.png}}
%\caption{Distributions of particles' precession rate depending on the angular momentum $z$-projection $L_z$. A straight blue line corresponds to the bar pattern speed with the value indicated in the annotating textbox. Contours indicate the location of particles that are being captured by the bar, i.e. traverse the line of \lbd=0.}
%\label{fig:ILR}
%\end{figure}
%%%%%%%%%%%%%%%%%%%%%%%%%%%%%%%%%%%%%%%%%%%%%%%%%%%%%%%
%%%%%%%%%%%%%%%%%%%%%%%%%%%%%%%%%%%%%%%%%%%%%%%%%%%%%%%
The key difference between the case considered by \citet{Lynden-Bell1979} and the physical situation considered here is that the bar perturbation rotates faster than the typical precession rate of the orbits it traps. 
%This can be clearly seen in Fig.~\ref{fig:ILR}, where we depict the dependence of the precession rates over the $z$-projection of angular momentum. The bar forms a thick straight strip in this figure, while the not yet trapped particles (within the green contours) lie on the descending branch of the precession rate curve. The latter is somewhat thick due to the differences in orbits' eccentricities. 
As can be seen, the trapped particles from the disc area should increase their precession rate while decreasing their angular moment. But such behaviour is precisely what one should expect from particles inhabiting a normal region, i.e. such orbits behave like donkeys increasing their precession rate when slowed down and vice versa. %\textit{should} increase its precession rate if the angular momentum decreases and the adiabatic invariant is conserved. 
A decrease of angular momentum leads to an increase in orbit precession rate up until the moment its precession rate becomes close enough to the bar pattern speed and the particle falls into an abnormal region, \textcolor{black}{or, more precisely, switches to the abnormal mode}.  
\par
Naturally, orbits can facilitate the bar growth in the described way if and only if they initially precess slower than the bar. In the opposite case, when the bar is slower than the precession rate of orbits, an orbit will run away from the bar possibly weakening it. To sum up, if an orbit possesses some channel which can be used to transfer its angular momentum then such an orbit will end up in the bar since it will increase its precession rate. Thus, fast bars can be fed by orbits from the normal region.
\par
%We should note that it is not the first time that the idea of orbits from the normal region not contributing to the bar weakening or destruction is discussed. 
\citet{Pol2020a,Pol2020b} analysed the bar formation mechanism in the framework of the standard Hamiltonian approach and individual orbit families. These authors found that not only the partial derivative of angular momentum is the deciding factor in the response of orbits to the bar but also their precession rate in the rotating frame. \citet{Pol2020a,Pol2020b} also introduced a separate parameter (parameter $b$ in notation of~\citealt{Pol2020a,Pol2020b}), which describes the orbital responsiveness to the bar-like perturbation and depends on the properties of the bar itself (mass and size) and underlying potential. In their figure 6, the authors provided a scheme that sums up their results on which orbits should be trapped depending on the combinations of all three parameters, partial derivative, precession rate, and orbital responsiveness. In particular, ~\citet{Pol2020a,Pol2020b} found that the bar should trap the so-called long-axis orbits\footnote{Orbits that, after being trapped, become a part of the ``backbone'' of the bar.} from the normal region if the latter are slow (i.~e. precess slower than the bar rotates, \textcolor{black}{see also Fig.~\ref{fig:LB-map_density}, bottom left plot, green area}) and the orbital responsiveness is positive. We do not analyse the orbital responsiveness here. Nevertheless, we obtained here that orbits from the normal region can be trapped in the bar \textcolor{black}{changing their orbital behaviour (mode)}. Thus, our results support conclusions of~\citet{Pol2020a,Pol2020b} from the perspective of self-consistent $N$-body simulations.

%-----------------------------------------------------%
%-----------------------------------------------------%
\subsection{Angular momentum sink}
\label{sec:Lz_sink}
% Sec. 6.4.
%-----------------------------------------------------%
%-----------------------------------------------------%
%%%%%%%%%%%%%%%%%%%%%%%%%%%%%%%%%%%%%%%%%%%%%%%%%%%%%%%
% Fig. 16
%%%%%%%%%%%%%%%%%%%%%%%%%%%%%%%%%%%%%%%%%%%%%%%%%%%%%%%
\begin{figure*}
\includegraphics[width=0.48\textwidth]{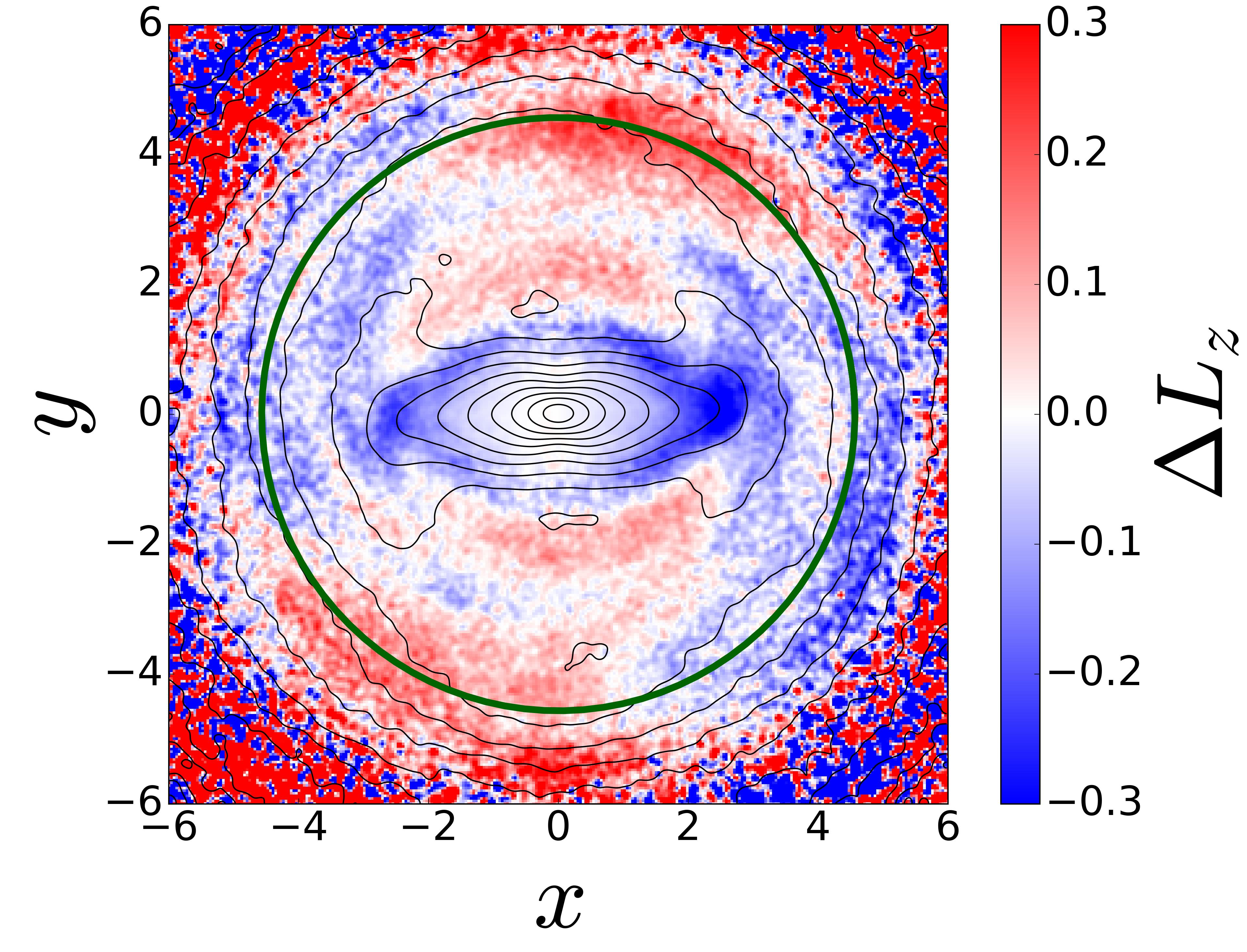}
\includegraphics[width=0.48\linewidth]{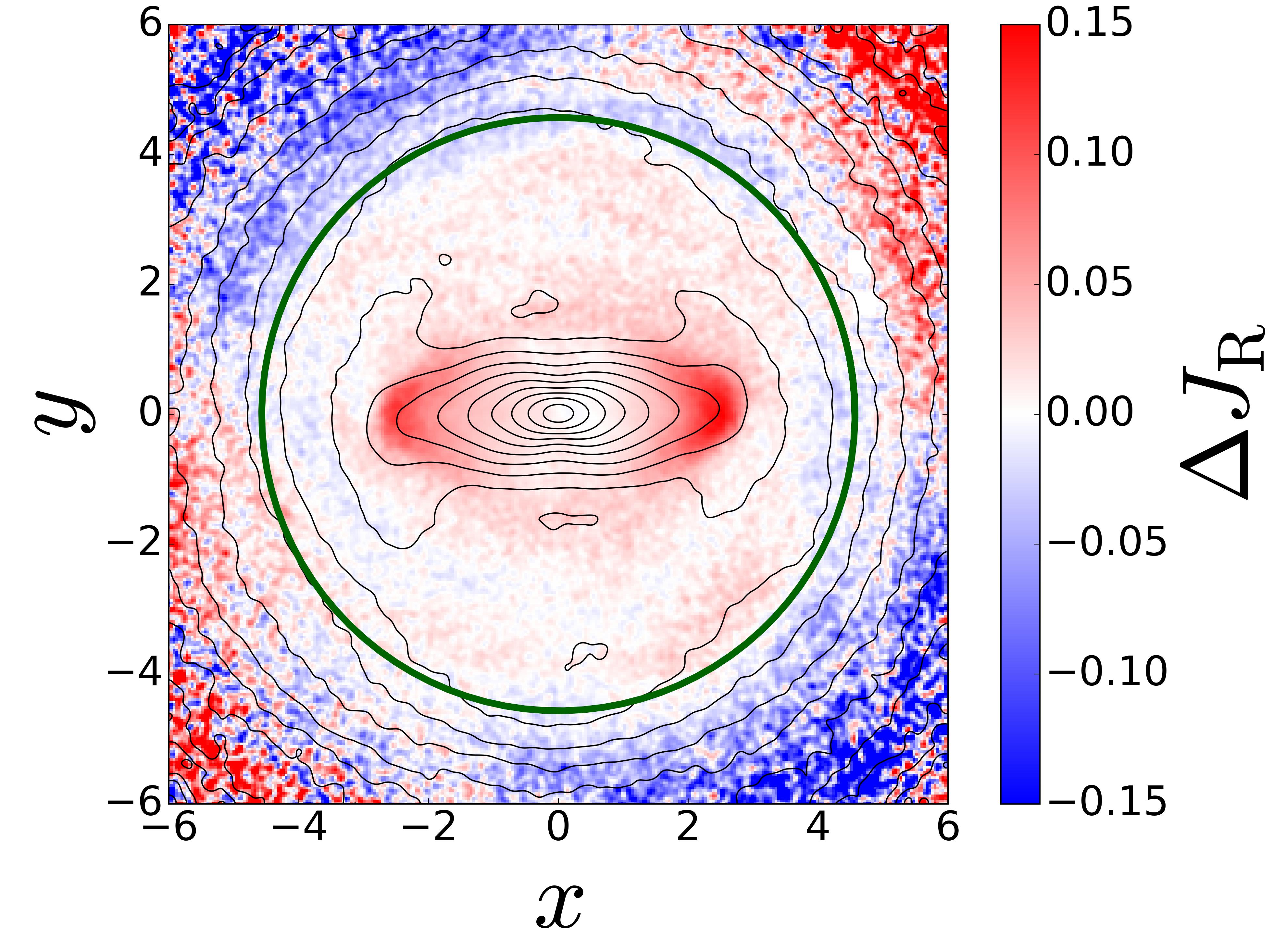}
\caption{2D histograms depicting the changes in angular momentum $\Delta L_z$ and radial actions $\Delta J_\mathrm{R}$ of the orbits from $t=400$ to $t=500$ in bins of size 0.04$\times$0.04. The picture is smoothed with a Gaussian filter with $\sigma=0.08$. Black lines mark isodensity contours, and the green circle shows the location of the corotation.}
\label{fig:momentum_exchange}
\end{figure*}
%%%%%%%%%%%%%%%%%%%%%%%%%%%%%%%%%%%%%%%%%%%%%%%%%%%%%%%
%%%%%%%%%%%%%%%%%%%%%%%%%%%%%%%%%%%%%%%%%%%%%%%%%%%%%%%

%%%%%%%%%%%%%%%%%%%%%%%%%%%%%%%%%%%%%%%%%%%%%%%%%%%%%%%
% Fig. 17
%%%%%%%%%%%%%%%%%%%%%%%%%%%%%%%%%%%%%%%%%%%%%%%%%%%%%%%
\begin{figure}
\begin{center}
\centering   
%\begin{minipage}[t]{0.48\linewidth}
\includegraphics[width=\linewidth]{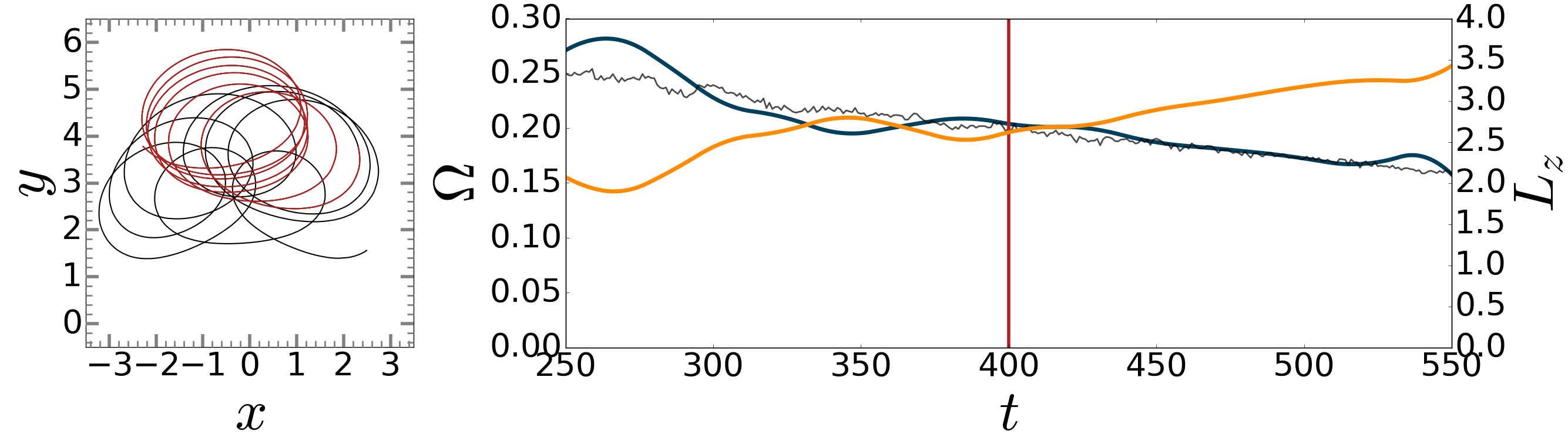}
%\end{minipage}
%\begin{minipage}[t]{0.48\linewidth}
\includegraphics[width=\linewidth]{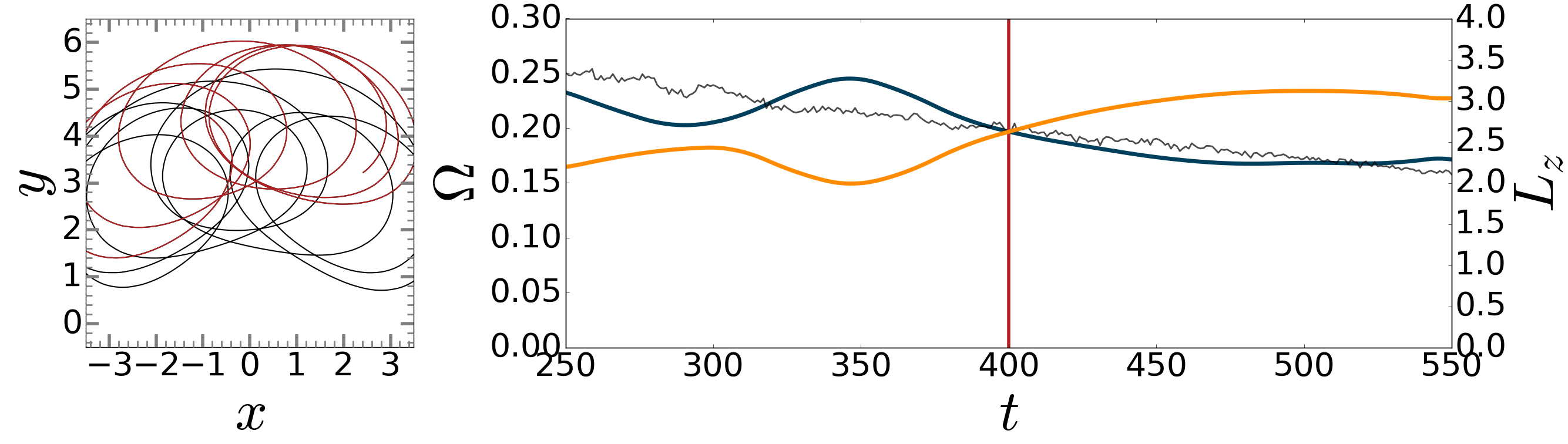}
%\end{minipage}    
\end{center}
%%%%%%%%%%%%%%%%%%%%%%%%%%%%%%%%%%%%%%%%%%%%%%%%%%%%%%%%%%%%%%%%%%%%%%%%%%%%
\caption{\textcolor{black}{\textit{Left:} Orbits in the vicinity of stable Lagrange points on $xy$-plane over the time interval from $t=250$ to $t=550$. \textcolor{black}{\textit{Red} colour marks the orbit at the late time interval from $t=400$ to $t=550$ (to show how the orbits drift outward)}. \textit{Right:} Coevolution of $L_z$ (\textit{orange} line) and \textcolor{black}{angular velocity of the orbit} $\Omega$ (\textit{blue} line). Thin grey line is $\Omega_\mathrm{p}$.} \textcolor{black}{Vertical red line marks the beginning of the late stage of the orbit (after $t=400$).}}
\label{fig:orbits_corotation}
\end{figure}
%%%%%%%%%%%%%%%%%%%%%%%%%%%%%%%%%%%%%%%%%%%%%%%%%%%%%%%
%%%%%%%%%%%%%%%%%%%%%%%%%%%%%%%%%%%%%%%%%%%%%%%%%%%%%%%
One more interesting question is what triggers the secular changes in orbits, i.e. why the orbit starts to lose its angular momentum. Here we should note that alongside the orbit capturing, there are some global processes that also affect the bar as a whole. In particular, many numerical and theoretical studies discussed an angular momentum redistribution which occurs alongside the bar formation and growth~\citep{Hernquist1992,Weinberg2002,Athanassoula2003,Ceverino_Klypin2007, Weinberg2007a, Weinberg2007b, Sellwood2008, Petersen_etal2019}. These studies proved that the dark halo particles play an important role in the bar life, taking the momentum from the bar particles due to the resonance interaction. At the same time, the bar slows down due to the loss of angular momentum. This is exactly what happens with the bar in the model considered here, as was already shown in the right panel of Fig.~\ref{fig:Nbodymodel}. Fig.~\ref{fig:momentum_exchange}, where \textcolor{black}{secular} angular momentum and radial action changes for the disc particles from $t=400$ to $t=500$ are presented, shows this in a more profound way. The maps are constructed by averaging the overall inflow and outflow of the corresponding variables inside the bins of size $0.04\times0.04$ during the mentioned time interval. It can be clearly seen that the bar region as a whole loses the angular momentum, and orbits become more elongated here ($J_\mathrm{R}$ increases, see Fig.~\ref{fig:orbits_evolution}, orbits~2 and 3). It is also interesting that, outside the bar region, the disc is inhomogeneous in terms of angular momentum gain. In particular, there is no clear ring-like structure that can be associated with the corotation\footnote{The location of the CR corresponds to the CR in Fig.~\ref{fig:Nbodymodel_bar}} (green circle), where the recipients of angular momentum should be located~\citep{Athanassoula2003}. The momentum increases in two blobs above and below the bar. \textcolor{black}{But the most prominent increase in $L_z$ occurs at CR near two stable Langrange points. In Fig.~\ref{fig:orbits_corotation} (left column) we present examples of two \textcolor{black}{CR} orbits that absorb $L_z$. The angular velocity \textcolor{black}{$\Omega$} of these orbits practically coincides with $\Omega_\mathrm{p}$ over the entire time interval from $t=250$ to $t=550$ (Fig.~\ref{fig:orbits_corotation}, right column). At the same time, the angular momentum shows a ``stubborn'' secular increase, and the orbits themselves drift outwards (compare black and blue loops of the orbit in the left panels of Fig.~\ref{fig:orbits_corotation}).} 
\par
\textcolor{black}{Our Fig.~\ref{fig:momentum_exchange} differs significantly from figure~14 in \citet{Ceverino_Klypin2007}, and our Fig.~\ref{fig:orbits_corotation} is different from their figure~15. \citet{Ceverino_Klypin2007} found that, over one period of the bar ($\sim150$~Myr), the largest changes in angular momentum occur at CR, where regions of both gain and loss angular momentum are identified. Considering individual orbits, \citet{Ceverino_Klypin2007} showed that particles at the corotation near two stable Lagrange points move between these regions in such a way that losses and gains of angular momentum are compensated. Here, we study longer periods of time ($t=400-500$, i.~e. $\sim1.4$~Gyr) and, thus, see a secular effect. This effect manifests itself both as the average gain in angular momentum by regions near two stable Lagrange points and at the level of individual orbits. The orbits slowly but steadily increase $L_z$, drifting outward and pushing CR to the periphery. At the same time, they maintain their angular velocity equal to the angular velocity of the bar, which gradually decreases.}
\par
\textcolor{black}{\citet{Ceverino_Klypin2007} confirmed their result regarding no net torque over the period time of the bar rotation analysing orbits in an analytical model with a bar rotating with constant speed. And this, the constancy of the bar speed, is the key difference between our results and results by \citet{Ceverino_Klypin2007}. Only a bar with decreasing angular velocity\footnote{\textcolor{black}{The $N$-body bar in \citet{Ceverino_Klypin2007} also slowed down, but the authors looked at the sinks and inflows of angular momentum on a fairly short time scale when the bar angular velocity changes little, and they were unable to capture secular effects. Although the trend towards an increase in $L_z$ for an individual orbit in the $N$-body potential over a time interval of 2 Gyr can be seen in their figure~15.}} turns regions near stable Lagrange points into sites of angular momentum sink.}
%as well as in a spiral-like structure in the right lower corner and a relatively small patch in the right upper corner. Other areas of the disc demonstrate a loss of angular momentum.
%Here we would like to verify the role of the bar itself in its growth.
%-----------------------------------------------------%
%-----------------------------------------------------%
\subsection{Bar slowdown as a condition for maintaining the trap}
\label{sec:bar_slowdown}
% Sec. 6.5.
%-----------------------------------------------------%
%-----------------------------------------------------%

%%%%%%%%%%%%%%%%%%%%%%%%%%%%%%%%%%%%%%%%%%%%%%%%%%%%%%%
% Fig. 18
%%%%%%%%%%%%%%%%%%%%%%%%%%%%%%%%%%%%%%%%%%%%%%%%%%%%%%%
\begin{figure*}
\begin{center}
\centering
    %\includegraphics[width=\textwidth]{img/new/orbits/orbit_evol_1_234.png}  
    %\bigskip
    %\includegraphics[width=\textwidth]{img/new/orbits/orbit_evol_2_16.png}
\begin{minipage}[t]{0.49\linewidth}
\includegraphics[width=\textwidth]{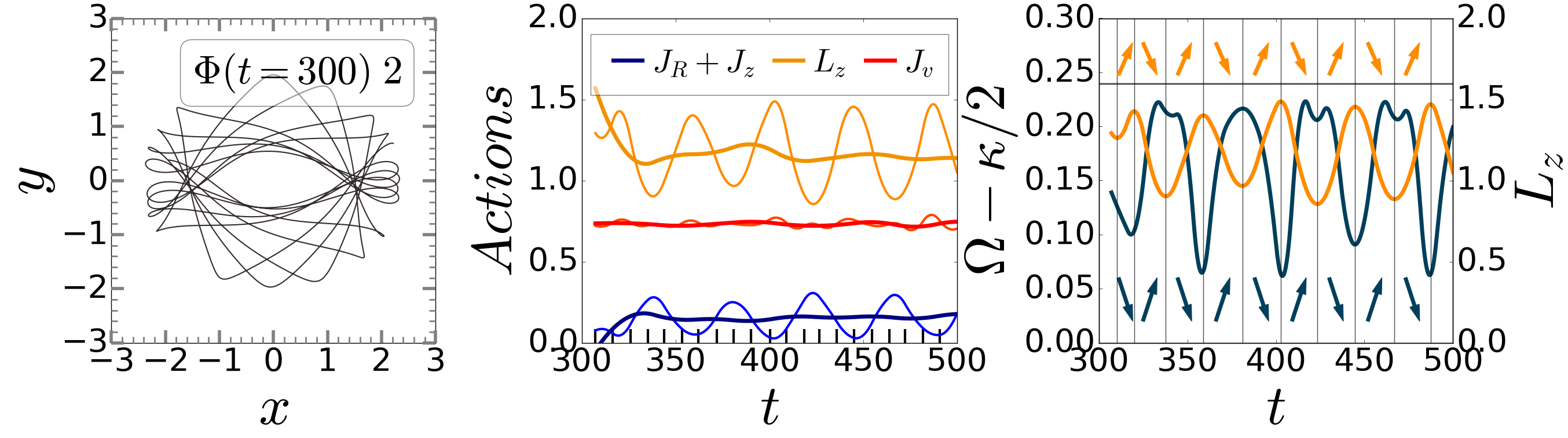}
\end{minipage}
\begin{minipage}[t]{0.49\linewidth}
\includegraphics[width=\textwidth]{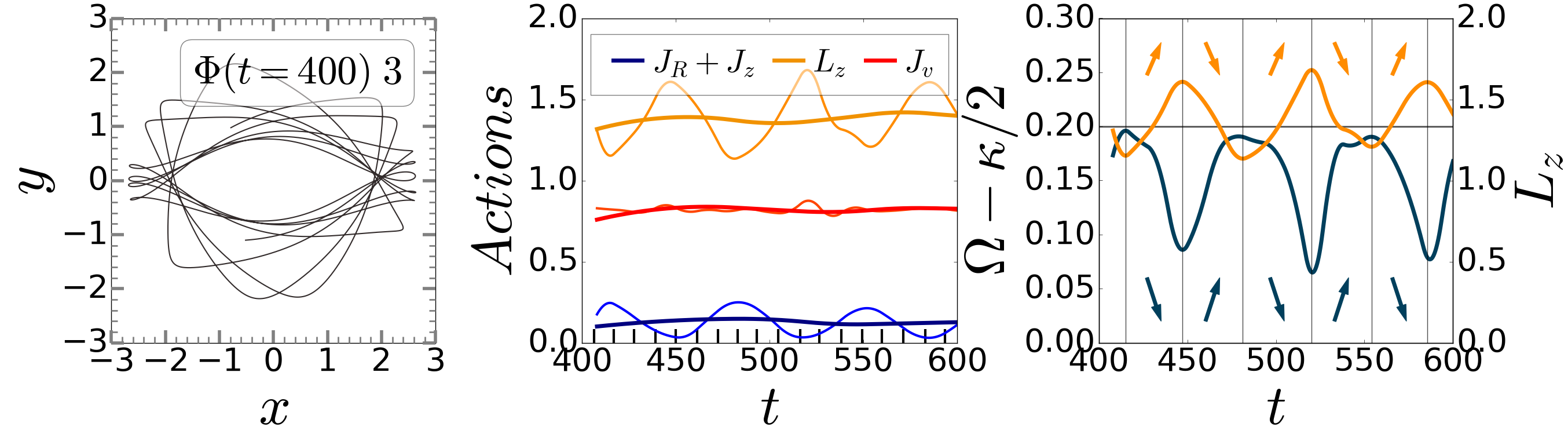}
\end{minipage}
\smallskip
\begin{minipage}[h]{0.49\linewidth}
\includegraphics[width=\textwidth]{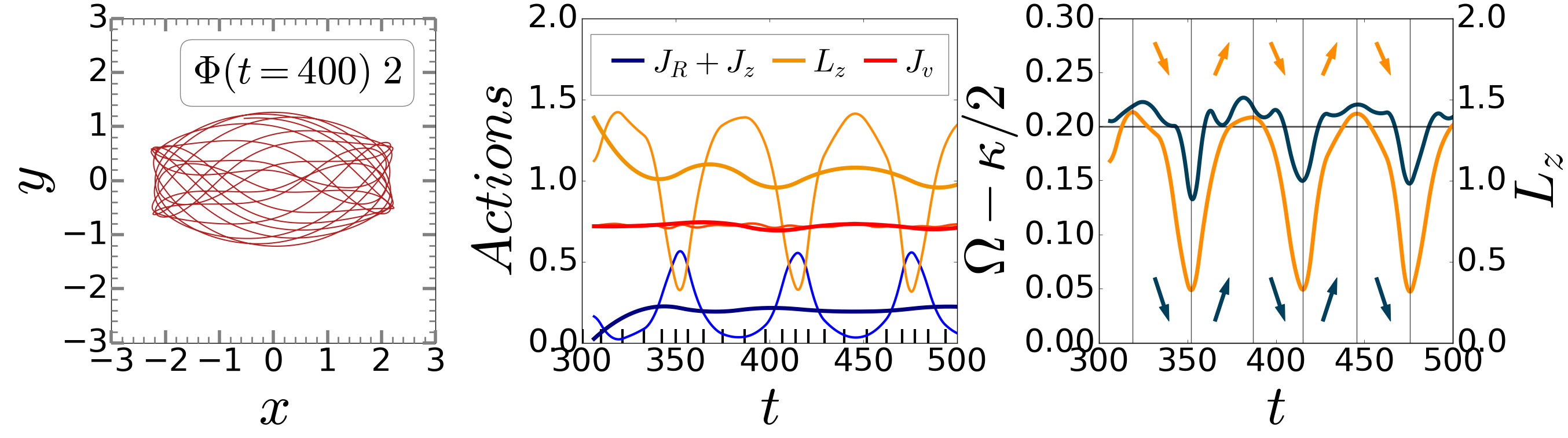}
\end{minipage}
\begin{minipage}[h]{0.49\linewidth}
\includegraphics[width=\textwidth]{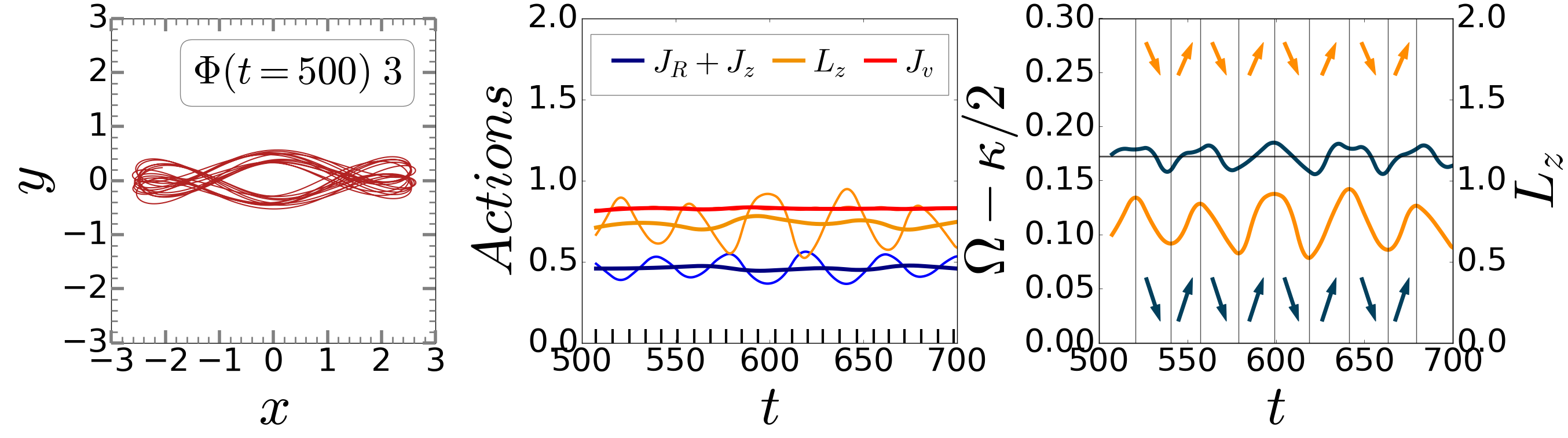}
\end{minipage}
\smallskip
\begin{minipage}[t]{0.49\linewidth}
\includegraphics[width=\textwidth]{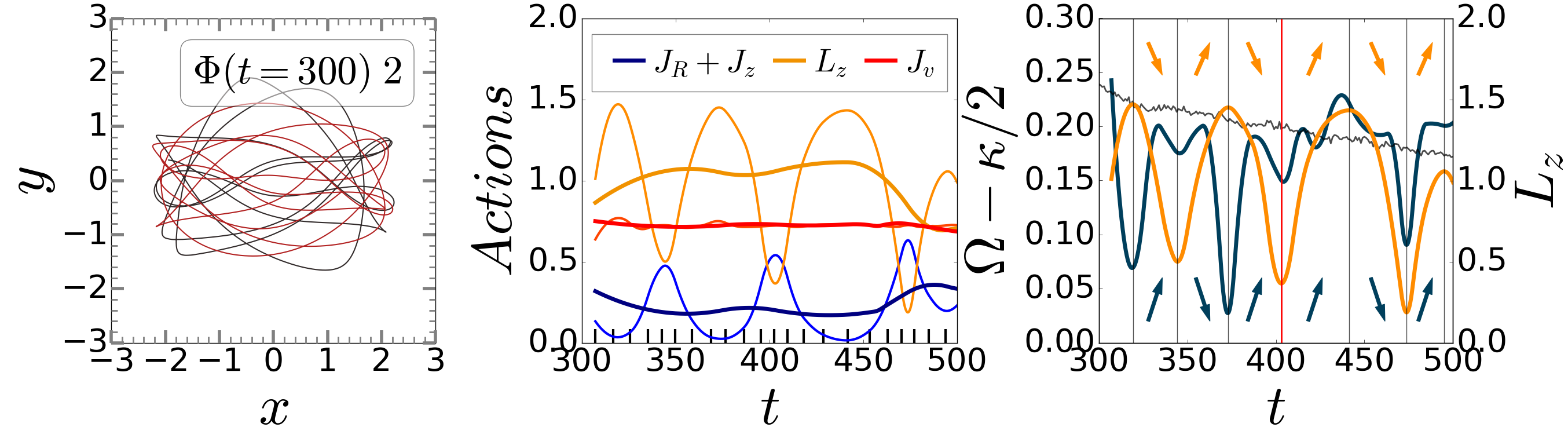}
\end{minipage}
\begin{minipage}[t]{0.49\linewidth}
\includegraphics[width=\textwidth]{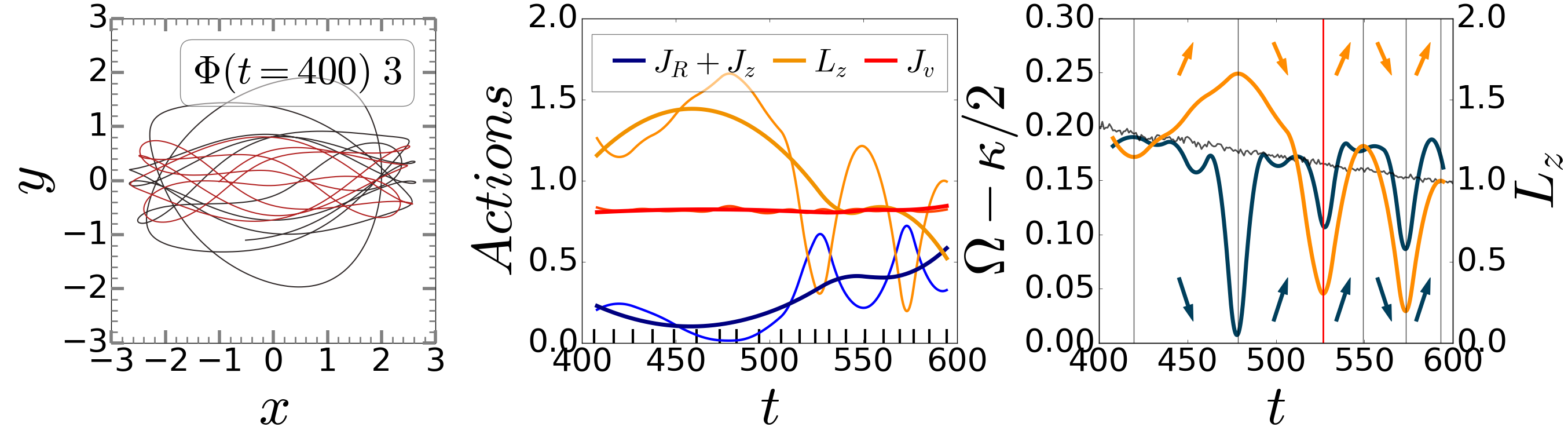}
\end{minipage}
    %\bigskip
    %\includegraphics[width=\textwidth]{img/new/orbits/orbit_evol_4_35.png}
\end{center}
%%%%%%%%%%%%%%%%%%%%%%%%%%%%%%%%%%%%%%%%%%%%%%%%%%%%%%%%%%%%%%%%%%%%%%%%%%%%
\caption{\textcolor{black}{\textit{Top row:} Orbits on the $xy$-plane and evolution of actions and precession rate in the frozen $N$-body potential with constant $\Omega_\mathrm{p}$. Particles are the same as in Fig.~\ref{fig:orbits_evolution}, i.e. orbit 2 (first three plots) and orbit 3 (second three plots). The initial moments of evolution are labelled in the plot with the orbit. The potential and the pattern angular speed are also taken for this time moment. Short marks on the abscissa axis of the plots with the evolution of the actions show the moments of the passage of the apocentres of the orbit. Coloured arrows in plots with the coevolution of $L_z$ (\textit{orange} line) and $\Omega - \kappa/2$ (\textit{blue} line) indicate the direction of change in the corresponding values during the time between two subsequent extrema of $L_z$. \textit{Thin grey} line is the pattern speed. \textit{Middle row:} the same as in the top row but for the time interval after orbit trapping in the bar. \textit{\textcolor{black}{Bottom} row:} the same as in the top row but for a slow downing bar. \textit{Black} and \textit{red} colour in the plots with orbits correspond to negative and positive values of LB-derivative, respectively.}
%\textit{Left:} Orbits on $xy$-plane in a frozen $N$-body potential. The initial moment of evolution is labeled in the plot. \textit{black} and \textit{red} colour correspond to negative and positive value of LB-derivative, respectively. \textit{Centre:} Evolution of orbital actions $(J_\mathrm{R}+J_z, L_z, J_f)$. The short marks at the bottom show the moments of passage of the apocentres of the orbit. \textit{Right:} Coevolution of $L_z$ (\textit{orange} line) and $\Omega - \kappa/2$ (\textit{blue} line). Coloured arrows indicate the direction of change in the value during the time between two adjacent extrema $L_z$. The evolution in a stationary rotating potential was separately traced for orbits number 2 and 3 before and after capture in the bar.
}
\label{fig:orbits_evolution_stationary}
\end{figure*}
%%%%%%%%%%%%%%%%%%%%%%%%%%%%%%%%%%%%%%%%%%%%%%%%%%%%%%%
%%%%%%%%%%%%%%%%%%%%%%%%%%%%%%%%%%%%%%%%%%%%%%%%%%%%%%%

%%%%%%%%%%%%%%%%%%%%%%%%%%%%%%%%%%%%%%%%%%%%%%%%%%%%%%%
% Fig. 15
%%%%%%%%%%%%%%%%%%%%%%%%%%%%%%%%%%%%%%%%%%%%%%%%%%%%%%%
% \begin{figure}
% \centering
% \center{\includegraphics[width=\linewidth]{img/Orbits/xy_orbit_dif_new_Phi300_sm.png}}\\
% \center{\includegraphics[width=\linewidth]{img/Orbits/JR_dif_omega_t_new_Phi300_sm.png}}\\
% \center{\includegraphics[width=\linewidth]{img/Orbits/Lz_dif_omega_t_new_Phi300_sm.png}}\\
% \center{\includegraphics[width=\linewidth]{img/Orbits/Jf_dif_omega_t_new_Phi300_sm.png}}\\
% \caption{Comparison of orbits of the same particle in a stationary potential and a potential where the bar slows down. \textit{Top panels}: orbits of the particle in a face-on view for a set of three different but constant pattern speeds (\textit{left}) and continuously decreasing pattern speed (\textit{right}). \textit{Middle and bottom panels}: Changes in the radial action $J_\mathrm{R}$, the angular momentum $L_z$, and adiabatic invariant $J_f$ for the orbits presented in the top panel. The black lines correspond to the orbit of the same particle taken from $N$-body simulations.}
% \label{fig:orbit_slowdown}
% \end{figure}
%%%%%%%%%%%%%%%%%%%%%%%%%%%%%%%%%%%%%%%%%%%%%%%%%%%%%%%
%%%%%%%%%%%%%%%%%%%%%%%%%%%%%%%%%%%%%%%%%%%%%%%%%%%%%%%
\par 
Is bar slow downing important for orbit capturing? To verify this, we carry out simple yet qualitatively fruitful experiments concerning orbit trapping. We take a probe particle, which is trapped in the bar, and analyse whether the same particle would be trapped if the bar does not slow down, that is, the bar rotates with a fixed pattern speed. \textcolor{black}{Fig.~\ref{fig:orbits_evolution_stationary} (left plots in each of the top four panels, consisting of three plots) demonstrates orbits of the particles 2 and 3 (see Fig.~\ref{fig:orbits_evolution}) starting from initial conditions taken from $N$-body simulations corresponding to different time moments.} 
%for time moments 300, 400, and 500 by different colours. 
Each of the orbits is obtained by integration of equations of motion in a frozen $N$-body potential rotating with a pattern speed that the bar has for each of the corresponding time moments \textcolor{black}{(see legend in the panel where an orbit is presented)}. For each time moment, we use a potential of the $N$-body model (including the dark halo contribution) taken for this specific time moment and approximated via the procedure described in Section~\ref{sec:measuring_actions}. 
%The top right panel of each figure shows the same orbit but the integration is carried out in a potential frozen at $t=300$, and where the bar slows down in the same way as it happens in the $N$-body model, i.e. we change the bar pattern speed in a discrete way according to our measurements for it in $N$-body model. 
As can be seen, if the bar does not change its pattern speed the orbit does not experience any change in morphology.
\par 
\textcolor{black}{In Fig.~\ref{fig:orbits_evolution_stationary} (middle plots in each of the top four panels, consisting of three plots), we also show the evolution of $J_\mathrm{R}$, $L_z$, and \textcolor{black}{$J_v$}, respectively, for each of the case described above (see legend).
%plus for the actual orbit from $N$-body model (black lines). 
It can be seen that if an orbit is not trapped at any specific time moment it does not experience a change in actions and remains untrapped, keeping the precession rate below the pattern speed (plots with the coevolution of $L_z$ and $\Omega - \kappa/2$ in the top row of Fig.~\ref{fig:orbits_evolution_stationary}). In the middle row of Fig.~\ref{fig:orbits_evolution_stationary} the orbit is shown after it has been captured into the bar. Although the speed of the pattern remains constant, the captured orbit no longer escapes the trap and retains its shape. 
Only when the bar slows down, the orbit undergoes a change in morphology and actions and experiences gradual trapping into the bar (Fig.~\ref{fig:orbits_evolution_stationary}, bottom row).} 
\par
\textcolor{black}{In \citet{Villa-Vargas_etal2009}, particle trapping is associated with the outward motion of CR, which, in turn, is only possible if the bar slows down (see also \citealp{Li_etal2022}). 
%with their trapping in the CR in its
Our findings, including those described in Section~\ref{sec:Lz_sink}, are in line with these results.} 
\par
\textcolor{black}{Thus, our experiments clearly show that the slow downing of the bar causes changes in the orbit and may well cause the orbits to become trapped in the bar when the mode of orbital motion changes from normal to abnormal. We also want to emphasise here that in our experiments the orbits are changing in response to the bar potential changes, and not due to the interaction with the dark halo.}
\section{Conclusions}
\label{sec:conclusions}
% Sec. 7.
%-----------------------------------------------------%
%-----------------------------------------------------%
In the present work, we studied the structure of the action space for the typical bar in a simulated galaxy. \textcolor{black}{We used the Lynden-Bell language (action variables, the LB-derivative), but applied it to the later stages of bar evolution. We were interested in how the action variables and frequencies change for each specific orbit of our numerical model on a large time scale. We examined correlations in the behavior of these variables.} The crucial point was to determine whether Lynden-Bell's idea of normal/abnormal \textcolor{black}{regions} can be utilised to explain the bar growth, which occurs after the bar has formed. Those regions are characterised by a different sign of the partial derivative of the precession rates over $z$-projection of the angular momentum for a fixed value of the adiabatic invariant \textcolor{black}{$J_v$}, \lbd. Lynden-Bell's hypothesis was that, depending on the sign of the derivative and the bar pattern speed, orbits will be either trapped along the bar and facilitate its growth or run away from the bar, thereby, contributing to its destruction. \textcolor{black}{Analyzing our $N$-body model, we moved from the concept of normal/abnormal regions to the concept of normal/abnormal orbital modes. In the normal mode, $L_z$ and $\Omega_\mathrm{pr}$ of the orbit change asynchronously, and when transitioning to the abnormal mode, these quantities either increase together or decrease simultaneously.}
\par
\textcolor{black}{To calculate actions and frequencies directly from the orbit in $N$-body simulations, we have developed several procedures that allow us to obtain values of all quantities at different times on short, medium, and secular time scales. The first procedure (Section~\ref{sec:axisym_potentia}) uses the publicly available software implementation of St\"ackel fudge method \citep{Binney2012, Sanders_Binney2016} from {\texttt{AGAMA}} package \citep{agama} and allows one to compute instantaneous actions and frequencies from the coordinates and velocities at one point in time. 
The second procedure aims to extract the medium-term evolution of actions and requires knowledge of the evolution of the orbit over a sufficiently large time interval~(Section~\ref{sec:mean_actions_oscillations}). The medium-term evolution of actions is found either by directly integrating the velocities along the orbit in the evolving $N$-body potential or by smoothing instantaneous actions from {\texttt{AGAMA}} (both approaches give very similar results).
In both cases, we smooth the values between successive apocentres using a mean-preserving spline~\citep{Ruiz_Jose2022}. The secular evolution of actions is found by further smoothing medium-term time series. Medium-term frequencies are found from the time interval between the successive passages of $R$ and $z$ maxima for $\kappa$ and $\omega_z$, respectively, and from the angle between apocentres for $\Omega$. Medium-term and secular frequency evolution is calculated in a similar way to the actions. Most results presented in this paper are based on the analysis of medium-term and secular actions and frequencies calculated \textit{directly in the non-axisymmetric $N$-body potential} via the integration of velocities along the corresponding orbits, while actions and frequencies from {\texttt{AGAMA}} are used for the preliminary analysis of the model.}
% mean value between two neighbouring extrema of their
\par
\textcolor{black}{
%After creating the tool, we can explore our $N$-body model in Lynden-Bell's terms.
Using the described procedures, we were able to study our $N$-body model in variables suggested by~\cite{Lynden-Bell1979}.
We tracked \lbd both for individual particles (Section~\ref{sec:trapping}) and for the ensemble of particles (Section~\ref{sec:trapped_particles}). Our main results are as follows. %related to the Lynden-Bell derivative \lbd can be described as follows.
}
\textcolor{black}{
\begin{itemize}
\item[1.] For the initial time moment, we found that the normal region of orbits, where the LB-derivative is negative, is non-existent while the whole disc is contained within an abnormal region, where the derivative sign is positive~(Fig.~\ref{fig:Jf_Lz}). This contradicts an old idea that the region should be contained within a small central part of a galaxy. A similar fact was found by~\citet{Pol2020a,Pol2020b}. These authors showed that the role of the normal orbits in preventing the bar formation or its destruction was overestimated by Lynden-Bell~(\citeyear{Lynden-Bell1979}) for realistic potentials.
\item[2.] At late stages, the Lynden-Bell zero derivative line can be used to roughly separate the mature $N$-body bar (\lbd$>0$) and disc (\lbd$<0$) at any given time~(Fig.~\ref{fig:density_regions}). Such a division of the system allows one to distinguish most of the bar particles relatively easily using procedures from \texttt{AGAMA} software package. This is a good practical result, which can be used in future dynamical studies of bars in $N$-body simulations. We also find that orbits near the Lynden-Bell zero derivative line precess relatively slowly,~i.~e. the difference between their precession rate and the bar pattern speed, divided by the angular velocity, is close to zero (Fig.~\ref{fig:LB-map}).
\item[3. ] However, as we show in Fig.~\ref{fig:LB-map_density}, this division is not the full story, since one essentially ignores a number of orbits that are being trapped by the bar and reside to the right of the zero LB-derivative line. We conclude that the concept of the Lynden-Bell zero derivative line does not provide a complete picture of the secular evolution of the bar and does not allow us to understand exactly how orbits become trapped in the bar.
\item[4.] In addition to Lynden-Bell's ideas, we build on the concept of normal/abnormal orbital modes, considering for each orbit how its behaviour changes over a large time scale in terms of the joint change in $L_z$ and $\Omega_\mathrm{pr}$. In the normal mode, $L_z$ and $\Omega_\mathrm{pr}$ change asynchronously, and when transitioning to the abnormal mode, they either increase together or decrease simultaneously~(Fig.~\ref{fig:orbits_evolution}).
\item[5.] The concept of normal/abnormal modes has shed light on how orbits become trapped in a bar in $N$-body simulations. Trapping occurs when a particle, while in a normal orbital mode, loses $L_z$ on a long-term time scale, while increasing $\Omega_\mathrm{pr}$ until $\Omega_\mathrm{pr}$ becomes almost equal to $\Omega_\mathrm{p}$. After $\Omega_\mathrm{pr}$ reaches values close to $\Omega_\mathrm{p}$, the sign of \lbd changes (becomes $>0$), and $L_z$ begins to librate (and even experience secular changes) synchronously with $\Omega_\mathrm{pr}$. In abnormal mode, $\Omega_\mathrm{pr}$ no longer goes beyond the current value of $\Omega_\mathrm{p}$, which slowly decreases over time.
\item[6.] Before switching to an abnormal mode and being trapped in a bar, the orbit retains \textcolor{black}{$J_v$}. This means that with a secular decrease in $L_z$, it increases $J_\mathrm{R}$ (stretches out) and $J_z$ (moves out of the mid-plane). The conservation of \textcolor{black}{$J_v$} and secular changes in other actions are characteristic both for individual orbits and for the entire ensemble of orbits trapped by the bar~(Fig.~\ref{fig:hist} and Fig.~\ref{fig:mean_actions}).
\item[7.] The difference between normal/abnormal regions and normal/abnormal orbital modes is clearly seen in orbits with a positive LB-derivative and high $L_z$ values~(Fig.~\ref{fig:orbits_largeLz}). Such orbits have an elliptical shape close to circular. Formally, they are in the normal region, but have already switched to an abnormal mode of behaviour and have a precession speed almost equal to the speed of the pattern. By decreasing $L_z$ in the future, they will become more elongated and may even go out of the mid-plane, but for now, having fallen into the trap of the bar, they outline its outer boundaries and make up the peripheral thin part of the bar.
\item[8.] For the transition to the abnormal mode to be possible, the particle must lose angular momentum on a long-term time scale. This means that there must be particles that gain this momentum. These could be, for example, CR particles in the vicinity of two stable Lagrange points that absorb angular momentum.
%In particular, CR particles in the vicinity of two stable Lagrange points absorb angular momenta. 
These particles gradually drift outward, moving the position of the CR to the periphery. This shows up well on a large time scale~(Fig.~\ref{fig:orbits_corotation}).
\item[9.] We have also established that trapping into the bar can not occur without a secular slowdown of the bar. Experimenting with probe particles, we have found that if one freezes the bar pattern speed no orbit trapping occurs, i.e. a stationary bar does not trap orbits~(Fig.~\ref{fig:orbits_evolution_stationary}).
\end{itemize}
}
\par
\textcolor{black}{Our approach and findings clarify and expand the picture of bar formation and evolution in numerical models. In the future, we plan to use our approach for analysing the evolution of $J_z$ and to provide answers to a number of unresolved questions regarding how the bar acquires its thick part, the B/PS bulge.}
\par
%one can easily calculate the actions and precession rates of the orbits using \texttt{AGAMA} software bar particles can be carried out simply by finding the region, where the partial derivative of precession rate of $z$-projection over angular momentum is close to zero. 

%-----------------------------------------------------%
%-----------------------------------------------------%
\section*{Data availability}
The data underlying this article will be shared on reasonable request to the corresponding author.
%-----------------------------------------------------%
%-----------------------------------------------------%

%-----------------------------------------------------%
%-----------------------------------------------------%
\section*{Acknowledgements}
%without which this work would not be possible.%%%%%%%%
%-----------------------------------------------------%
%-----------------------------------------------------%
We acknowledge financial support from the Russian Science Foundation, grant no. 24-22-00376. 
\par
We thank the anonymous referee for his/her thorough review and highly appreciate the comments and suggestions that significantly contributed to improving the quality of the article.
\par
We acknowledge the use of the~\texttt{AGAMA}~\citep{agama} and ~\texttt{mpsplines}~\citep{Ruiz_Jose2022} \texttt{python} packages, without which the present work would not be possible.

%-----------------------------------------------------%
%-----------------------------------------------------%
\bibliographystyle{mnras}
\bibliography{article7} % if your bibtex file is called example.bib
%-----------------------------------------------------%
%-----------------------------------------------------%

%-----------------------------------------------------%
%-----------------------------------------------------%
\appendix
\section{Lyapunov exponents}
\label{sec:appendix_LE}
%-----------------------------------------------------%
%-----------------------------------------------------%
To obtain an additional argument supporting the dominating role of the regular orbits, we supplemented our analysis of orbital chaoticity by calculating the Lyapunov exponents. We consider the subsequent results as auxiliary, since they are based on the tools of the {\tt AGAMA} package and are obtained by using only the spherically symmetric part of the evolving potential. 
%But they fit into the general outline of the conclusions that the proportion of regular orbits in our $N$-body model is large in the considered time interval.
\par
The {\tt AGAMA} software package has a special subroutine for the calculation of the Lyapunov exponent for the orbit at a given potential. For this study, we used the $N$-body rotating potential of our model at $t=400$. We then apply the standard expansion of the model potential into spherical functions \textcolor{black}{(with $l_\mathrm{max} = 12$)}. This was done via the corresponding subroutine from the {\tt AGAMA} package. Then, we traced the evolution of all model particles in the spherically symmetric part of the potential over 100 time units \textcolor{black}{(we study the particle trapping in this time interval in Section~\ref{sec:trapped_particles})}. 
%A finite-time estimate of Lyapunov exponent is one of the quantities that can be obtained when calculating the evolution of an orbit in the AGAMA package. 
An estimate of the Lyapunov exponent over a finite time is one of the quantities that can be obtained by calculating the evolution of an orbit. In the case of {\tt AGAMA}, its subroutine returns zero for a non-chaotic orbit and an estimate of the Lyapunov exponent for a chaotic orbit.
\par
\begin{figure}
\begin{minipage}[t]{0.29\linewidth}%
\includegraphics[width=1\linewidth]{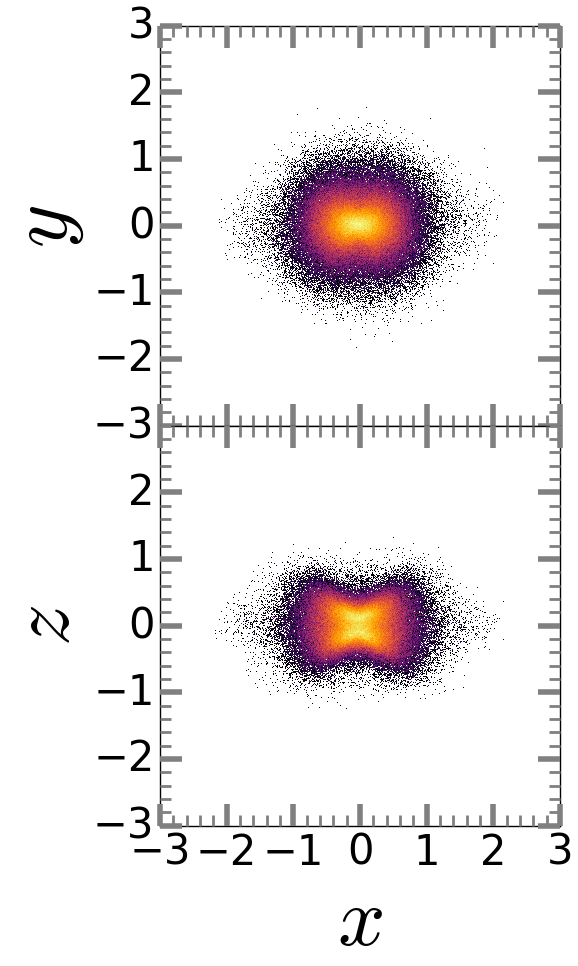}  
\end{minipage}
\begin{minipage}[t]{0.68\linewidth}%
\includegraphics[width=1\linewidth]{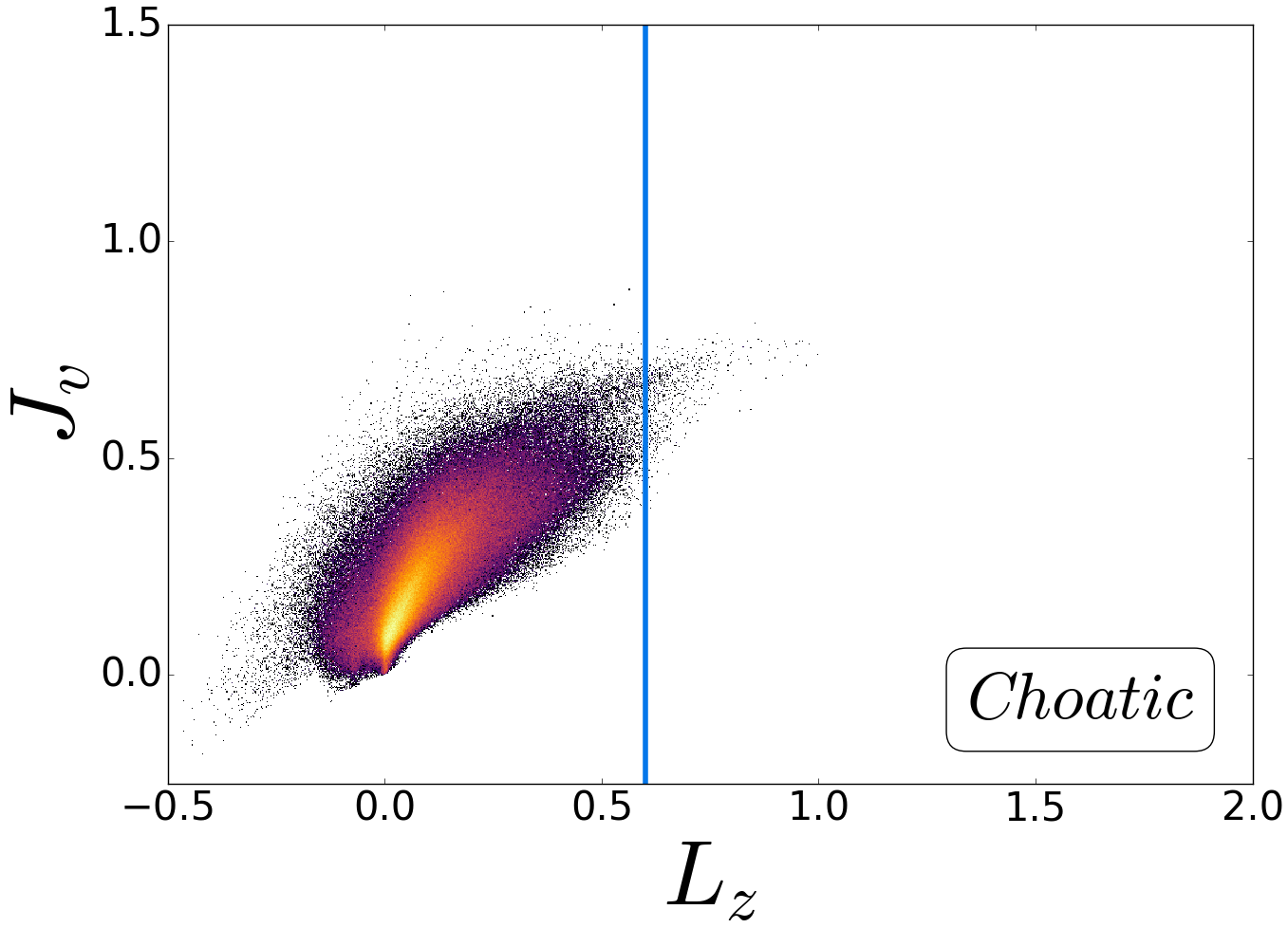}  
\end{minipage}\\
% \begin{minipage}[t]{0.29\linewidth}%
% \includegraphics[width=1\linewidth]{img/new/xyxz_LB<0.png}  
% \end{minipage}
% \begin{minipage}[t]{0.68\linewidth}%
% \includegraphics[width=1\linewidth]{img/new/Lz_Jf_LB<0.png}  
% \end{minipage}
\caption{\textcolor{black}{
\textit{Left two panels:}
Chaotic particles (with a positive Lyapynov exponent) on the $xy$ (\textit{top left}) and $xz$ (\textit{bottom left}) planes. \textit{Right panels:} Distribution of chaotic particles on the $(L_z, \, J_v)$ plane (\textit{bottom}). 
%\textit{Bottom panels:} The same as in upper panels but for particles in \textit{normal mode}. 
A vertical straight blue line separates central normal particles on the $L_z-J_v$ plane for $t=400$ (see Section~\ref{sec:trapped_particles} and Appendix~\ref{sec:appendix_normal_abnormal}).}
}
\label{fig:haos_part}
\end{figure}

We plot all the chaotic particles 
%\textcolor{black}{(which are chaotic in at least one of the two expansions)} 
on the $(xy)$, $(xz)$, and $(L_z, J_v)$ planes in Fig.~\ref{fig:haos_part}. The total number of chaotic orbits is found to be approximately 400 thousand out of more than two million bar particles (i.e. about 20\% as was also found in~\citealt{Valluri_etal2016}), and most of them are concentrated in the central regions.

%%%%%%%%%%%%%%%%%%%%%%%%%%%%%%%%%%%%%%%%%%%%%%%%%%%%%%%
% Fig. A1
%%%%%%%%%%%%%%%%%%%%%%%%%%%%%%%%%%%%%%%%%%%%%%%%%%%%%%%

%%%%%%%%%%%%%%%%%%%%%%%%%%%%%%%%%%%%%%%%%%%%%%%%%%%%%%%
%%%%%%%%%%%%%%%%%%%%%%%%%%%%%%%%%%%%%%%%%%%%%%%%%%%%%%%

%-----------------------------------------------------%
%-----------------------------------------------------%
\section{Complete adiabatic invariant}
\label{sec:appendix_invariant}
%-----------------------------------------------------%
%-----------------------------------------------------%

%%%%%%%%%%%%%%%%%%%%%%%%%%%%%%%%%%%%%%%%%%%%%%%%%%%%%%%
% Fig. B1
%%%%%%%%%%%%%%%%%%%%%%%%%%%%%%%%%%%%%%%%%%%%%%%%%%%%%%%
\textcolor{black}{One of the classical results of orbital dynamics is that a particle, which moves in $xy$-plane and is located in the ILR $(\Omega - \kappa/2 = \Omega_\mathrm{p})$, preserves the integral of motion $J_f = J_\mathrm{R} + L_z/2$. Lynden-Bell used this property to formulate his criterion for the capturing of orbits by the bar \citep{Lynden-Bell1979}. Lynden-Bell considered the initial stages of the axisymmetric disc evolution and, thus, only planar orbits. In this appendix, we provide a rationale for which adiabatic invariant should be used in the case of a 3D bar. We are talking about the invariant\footnote{\textcolor{black}{We thank the reviewer for drawing our attention to the fact that the adiabatic invariant $J_v$ can be found analytically by introducing certain fast and slow variables. We are also grateful for his outline of how to do this.}} $J_v$ that we use.}
\par
\textcolor{black}{
From the point of view of Hamiltonian mechanics, a galactic bar is a perturbation of an axisymmetric disk, which rotates with angular velocity $\Omega_\mathrm{p}$. In this case, the Hamiltonian has the following form:
\begin{equation}
    H(\mathbf{p}, R, z, \phi) = \dfrac{1}{2}\mathbf{p^2} - \Omega_\mathrm{p} L_z + \Phi(R,z) + \delta \Phi(R,z,\phi) 
\end{equation}
Here $\phi = \varphi - \Omega_\mathrm{p} t$ is an azimuthal angle in a rotating reference frame. If we rewrite this Hamiltonian in terms of unperturbed action-angle variables $\bm{J}=(J_\mathrm{R}, \,J_z, \,L_z)$ and $\bm{\theta}=(\theta_\mathrm{R}, \,\theta_z,\, \theta_\phi)$ (frequencies associated with them are $\bm{\Omega} = (\kappa,\, \omega_z, \,\Omega - \Omega_\mathrm{p})$), we obtain the classical Hamiltonian of the perturbation theory:
\begin{equation}
\label{eq:pert_H}
    H(\bm{J}, \bm{\theta}) = H_0(\bm{J}) + H_1(\bm{J}, \bm{\theta})
\end{equation}
Note, for the Hamiltonian of the axisymmetric system $\theta_\phi=(\Omega-\Omega_\mathrm{p})(t-t_0)$ is not exactly equal $\phi$, but close to it and coincides with it at the apocenters and periacenters of the orbit. }
\par
\textcolor{black}{Bar particles and particles, that are in the process of capture into the bar, are close to inner Lindblad resonance $\Omega - \kappa/2 \approx \Omega_\mathrm{p}$. It means that $\theta_\phi - \theta_\mathrm{R}/2$ is slow variable compared to $\theta_\mathrm{R}$. In addition, the works \citet{Parul_etal2020} and \citet{Tikhonenko_2021}, which investigated the spectral dynamics of the bar,  showed that, for most particles in the ILR, ratio of vertical-to-in-plane frequencies $f_z/f_x \approx \omega_z / (\Omega - \Omega_\mathrm{p}) \approx 2 \omega_z / \kappa$ lies in the range from about $1.6$ to $2.1$. Thus, we can write the following relationship between frequencies $\omega_z - \kappa \ll \kappa$. Hence, $\theta_z - \theta_\mathrm{R}$ is slow variable compared to $\theta_\mathrm{R}$. As a result, most of the bar particles simultaneously lie near two resonances: azimuthal (ILR) and vertical (the so-called vertical ILR or vILR). }
\par
\textcolor{black}{ One can make a canonical transformation to resonant angle-action variables $(\bm{J}, \bm{\theta}) \rightarrow (\bm{J}_\mathrm{res}, \bm{\theta}_\mathrm{res})$, here $\bm{J}_\mathrm{res} = (J_0,\,J_1,\,J_2)$, $\bm{\theta}_\mathrm{res} = (\theta_\mathrm{R},\,  \theta_1 = \theta_z - \theta_\mathrm{R},\, \theta_2 = \theta_\phi - \theta_\mathrm{R} /2)$. Following the recommendation of an anonymous reviewer, we use the following generating function:
\begin{equation}
   S(\bm{J}_\mathrm{res}, \bm{\theta})  =  J_0 \theta_\mathrm{R} + J_1 \theta_1 + J_2 \theta_2 = (J_0 - J_1 - J_2/2) \theta_\mathrm{R} + J_1 \theta_z + J_2 \theta_\phi.
\end{equation}
Then, from $\bm{\theta}_\mathrm{res} = \partial S / \partial \bm{J}_\mathrm{res}$ and from the expression $\bm{J} = \partial S / \partial \bm{\theta}$, we get the relationship between the variables $(J_\mathrm{R} = J_0 - J_1 - J_1/2, J_z = J_1, L_z = J_2)$. Thus, we introduce a new invariant $J_v = J_0 = J_\mathrm{R} + J_z + L_z/2$. Canonical transformation to resonant variables allows us to move on to the averaged equations of motion and get rid of the fast variable $\theta_\mathrm{R}$.}
\par
\textcolor{black}{We move on to equations of motion containing only slow variables. To do this, we introduce the averaged Hamiltonian of the following form:
\begin{equation}
    \overline{H}(\bm{J}_{\mathrm{res}}, \theta_1, \theta_2) = \dfrac{1}{2\pi} \int_0^{2\pi} H(\bm{J}_{\mathrm{res}}, \theta_\mathrm{R}, \theta_1, \theta_2) d\theta_\mathrm{R}
\end{equation}
Hence, $\dot{J_v} = - \partial \overline{H} / \partial \theta_\mathrm{R} = 0$, which means that $J_v = \mathrm{const}$ and is the complete adiabatic invariant for most particles of the bar and those close to the bar.  }
%\par
%\textcolor{black}{We thank the reviewer for drawing our attention to the fact that the adiabatic invariant $J_v$ can be found analytically by introducing certain fast and slow variables.}

%%%%%%%%%%%%%%%%%%%%%%%%%%%%%%%%%%%%%%%%%%%%%%%%%%%%%%%
%%%%%%%%%%%%%%%%%%%%%%%%%%%%%%%%%%%%%%%%%%%%%%%%%%%%%%%

%-----------------------------------------------------%
%-----------------------------------------------------%
\section{Normal/abnormal orbits}
\label{sec:appendix_normal_abnormal}
%-----------------------------------------------------%
%-----------------------------------------------------%

%%%%%%%%%%%%%%%%%%%%%%%%%%%%%%%%%%%%%%%%%%%%%%%%%%%%%%%
% Fig. C1
%%%%%%%%%%%%%%%%%%%%%%%%%%%%%%%%%%%%%%%%%%%%%%%%%%%%%%%
\begin{figure}
\begin{minipage}[t]{0.98\linewidth}
\includegraphics[width=1\linewidth]{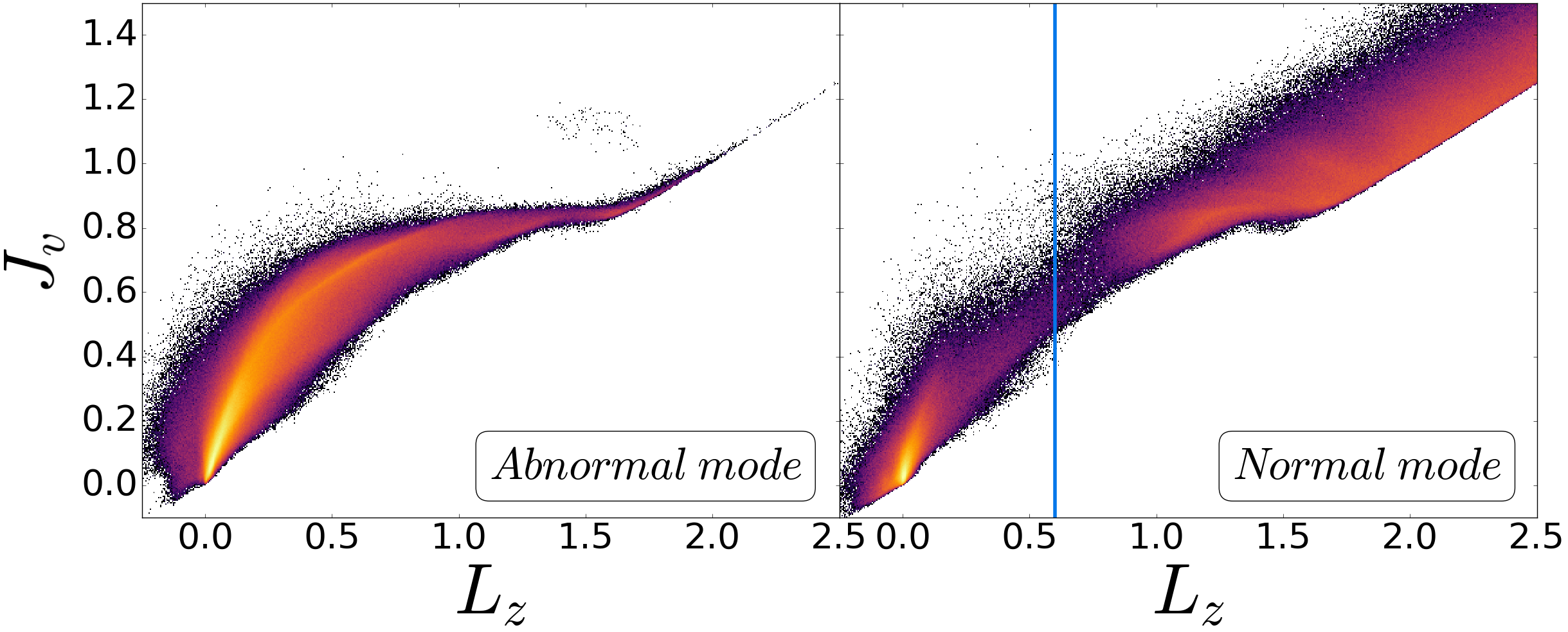}  
\end{minipage}\\
\begin{minipage}[t]{0.98\linewidth}
\includegraphics[width=1\linewidth]{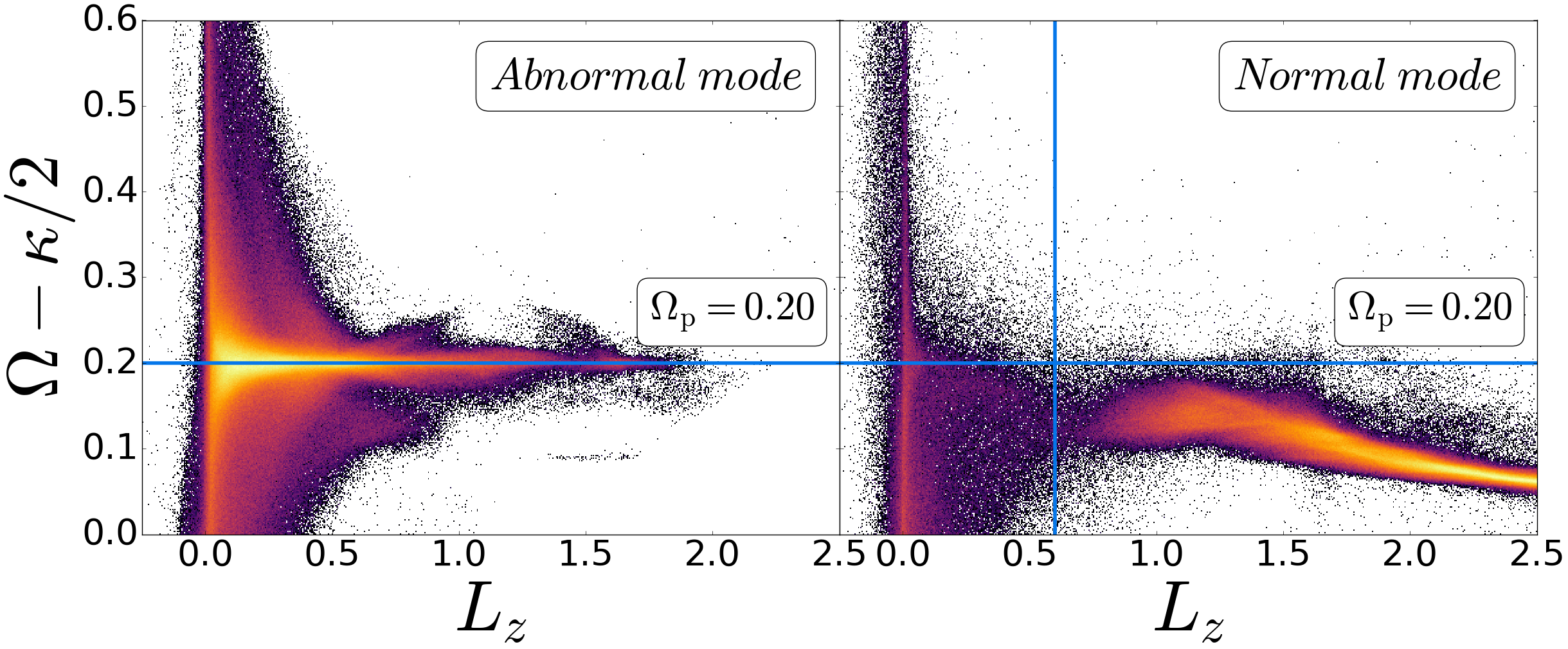}  
\end{minipage}\\
%\begin{minipage}[t]
%{0.98\linewidth}
%\includegraphics[width=1\linewidth]{img/new/xyxz_normal_abnormal_t400_0.png}  
%\end{minipage}
\caption{\textcolor{black}
{\textit{Upper row}: 
The 2D density distribution of particles in the $N$-body model over $L_z$--\textcolor{black}{$J_v$} plane for particles in \textit{abnormal} (\textit{left}) and \textit{normal} modes (\textit{right}) at $t=400$. Actions are averaged over librations. \textit{Bottom row}: Distributions of particle precession rates depending on the angular momentum $z$-projection $L_z$. A straight blue line corresponds to the bar pattern speed with the value indicated in the annotating textbox. A vertical straight blue line ($L_z=0.6$) separates central orbits in a normal mode.}}
\label{fig:normabnorm_part1}
\end{figure}
%%%%%%%%%%%%%%%%%%%%%%%%%%%%%%%%%%%%%%%%%%%%%%%%%%%%%%%
%%%%%%%%%%%%%%%%%%%%%%%%%%%%%%%%%%%%%%%%%%%%%%%%%%%%%%%

%%%%%%%%%%%%%%%%%%%%%%%%%%%%%%%%%%%%%%%%%%%%%%%%%%%%%%%
% Fig. B2
%%%%%%%%%%%%%%%%%%%%%%%%%%%%%%%%%%%%%%%%%%%%%%%%%%%%%%%
\begin{figure}
%\begin{minipage}[t]{0.98\linewidth}
%\includegraphics[width=1\linewidth]{img/new/Lz_Jf_normabnorm_400.png}  
%\end{minipage}\\
%\begin{minipage}[t]{0.98\linewidth}
%\includegraphics[width=1\linewidth]{img/new/Lz_omega-kappa:2_normabnorm_400.png}  
%\end{minipage}\\
\begin{minipage}[t]
{0.98\linewidth}
\includegraphics[width=1\linewidth]{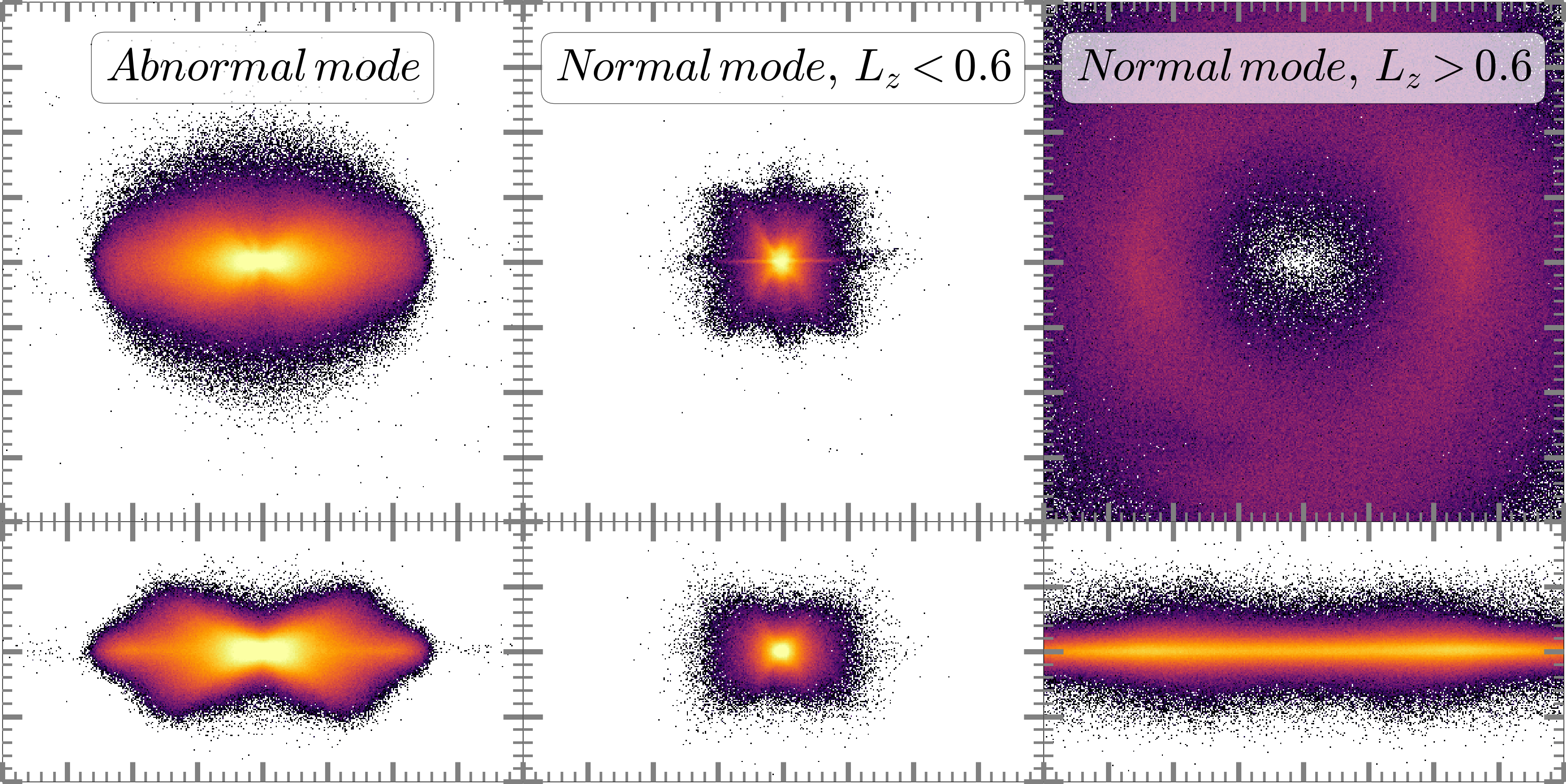}  
\end{minipage}
\caption{\textcolor{black}
{The 2D density distribution of particles in the $N$-body model on the $(xy)$ (\textit{upper plots}) and $(xz)$ (\textit{bottom plots}) planes for particles in \textit{abnormal} (\textit{left}) and \textit{normal} modes. %Actions are averaged over librations. 
Orbits in a normal mode are divided into two parts by the conditions $L_z<0.6$ (central orbits) or $L_z>0.6$ (disc orbits).}}
\label{fig:normabnorm_part2}
\end{figure}
%%%%%%%%%%%%%%%%%%%%%%%%%%%%%%%%%%%%%%%%%%%%%%%%%%%%%%%
%%%%%%%%%%%%%%%%%%%%%%%%%%%%%%%%%%%%%%%%%%%%%%%%%%%%%%%

\textcolor{black}{In Section~\ref{sec:trapped_particles}, when identifying an ensemble of orbits that are being trapped in a bar in the time interval from $t=400$ to $t=500$, we encountered the fact that chaotic orbits in the central regions of the model change the sign of the LB-derivative in an arbitrary manner. If we select all the orbits in the normal mode at the moment $t=400$, we will see that, in addition to the orbits in the disc, some of the central orbits will also be in this mode (Fig.~\ref{fig:normabnorm_part1}, top right plot). Orbits in the abnormal mode occupy a wide range of $L_z$ from $L_z\approx0$ (in central regions) to $L_z\approx1.3$ with a small admixture of orbits with even larger $L_z$ values, which are adjacent to the line of almost circular orbits (Fig.~\ref{fig:normabnorm_part1}, top left plot). The precession speed of the orbits in the abnormal mode is practically glued to the angular speed of the pattern (Fig.~\ref{fig:normabnorm_part1}, bottom left plot). The precession speed of the disc orbits in the normal mode is lower than the angular speed of the pattern, as is the precession speed of the central orbits in the normal mode (Fig.~\ref{fig:normabnorm_part1}, bottom right plot). The central orbits in the normal mode coincide in their position on the $L_z$--\textcolor{black}{$J_v$} plane with the chaotic orbits in Fig.~\ref{fig:haos_part}. About half of them are chaotic. In further analysis we do not consider them, separating them from bar particles by the condition $L_z<0.6$.}
\par
\textcolor{black}{Fig.~\ref{fig:normabnorm_part2} shows the face-on and edge-on views of three subsystems. On the left, there is a bar assembled from orbits in an abnormal mode at $t=400$. In the middle, there is a central subsystem, assembled from orbits in the normal mode with $L_z<0.6$, and these are predominantly chaotic orbits. On the right, there is a disc identified by the condition $L_z>0.6$ and negative LB-derivative.}

%-----------------------------------------------------%
%-----------------------------------------------------%
\label{lastpage}
\end{document}